\documentclass[fleqn,usenatbib,useAMS]{mnras}

\usepackage{graphicx}	
\usepackage{amsmath}	
\usepackage{amssymb}	
\usepackage{multicol}        
\usepackage{bm}		
\usepackage{pdflscape}	
\usepackage{verbatim}
\usepackage[flushleft]{threeparttable}
\usepackage[dvipsnames]{xcolor}
\usepackage{ulem}

\newcommand{\kms}{\,km\,s$^{-1}$} 

\newcommand{\zabs}{\ensuremath{z_{\rm abs}}}

\newcommand{\HH}{\mbox{H$\rm _2$}}

\newcommand{\dla}{damped Lyman-$\alpha$}

\newcommand{\lya}{\mbox{${\rm Ly}\alpha$}}

\newcommand{\HI}{H\,{\sc i}}
\newcommand{\CI}{C\,{\sc i}}
\newcommand{\CII}{C\,{\sc ii}}
\newcommand{\CrII}{Cr\,{\sc ii}}
\newcommand{\FeII}{Fe\,{\sc ii}}

\newcommand{\MnII}{Mn\,{\sc i}}

\newcommand{\NiII}{Ni\,{\sc ii}}
\newcommand{\OI}{O\,{\sc i}}
\newcommand{\SII}{S\,{\sc ii}}
\newcommand{\SiII}{Si\,{\sc ii}}
\newcommand{\TiII}{Ti\,{\sc ii}}
\newcommand{\ZnII}{Zn\,{\sc ii}}

\definecolor{green}{rgb}{0,0.4,0}

\newcommand{\unige}{Department of Astronomy, University of Geneva, Chemin Pegasi 51, 1290 Versoix, Switzerland}
\newcommand{\iap}{Institut d'Astrophysique de Paris, CNRS-SU, UMR\,7095, 98bis bd Arago, 75014 Paris, France}
\newcommand{\ioffe}{Ioffe Institute, {Politekhnicheskaya 26}, 194021 Saint Petersburg, Russia}

\newcommand{\kassi}{Korea Astronomy and Space Science Institute, 776, Daedeokdae-ro, Yuseong-gu, Daejeon, 34055, Korea}

\usepackage[T1]{fontenc}
\usepackage{ae,aecompl}

\usepackage{newtxtext,newtxmath}
\graphicspath{{./}{figures/}}

\title[Seven new ESDLAs]{Extremely strong DLAs at high redshift: Gas cooling and H$_2$ formation}

\author[K. N. Telikova et al.]{K. N. Telikova$^{1}$\thanks{Contact e-mail: \href{mailto:telikova.astro@mail.ioffe.ru}{telikova.astro@mail.ioffe.ru}}, 
S. A. Balashev$^{1}$, P. Noterdaeme$^{2}$, 
J.-K. Krogager$^{2,3}$, A. Ranjan$^{2,4}$
\\
$^{1}$ \ioffe \\
$^{2}$ \iap \\
$^{3}$ \unige \\
$^{4}$ \kassi
}

\date{Last updated 2015 May 22; in original form 2013 September 5}

\pubyear{2015}

\begin{document}
\label{firstpage}
\pagerange{\pageref{firstpage}--\pageref{lastpage}}
\maketitle

\begin{abstract}
We present a spectroscopic investigation with VLT/X-shooter of seven candidate extremely strong \dla\ absorption systems
(ESDLAs, $N$(H\,{\sc i})$\ge 5\times 10^{21}$~cm$^{-2}$)
observed along quasar sightlines. 
We confirm the extremely high column densities, albeit slightly (0.1~dex) lower than the original ESDLA definition for four systems. We measured  
low-ionisation metal abundances and dust extinction for all systems. 
For two systems we also found strong associated \HH\ absorption $\log N(\HH)\text{[cm$^{-2}$]}=18.16\pm0.03$ and $19.28\pm0.06$ at $z=3.26$ and 2.25 towards J\,2205+1021 and J\,2359+1354, respectively), while for the remaining five we measured conservative upper limits on the \HH\ column densities of typically $\log N(\HH)\text{[cm$^{-2}$]}<17.3$. The increased H$_2$ detection rate ($10-55$\% at 68\% confidence level) at high \HI\ column density compared to the overall \dla\ population ($\sim 5-10$\%) confirms previous works. 
We find that these seven ESDLAs have similar observed properties as those previously studied towards quasars and gamma-ray burst afterglows, suggesting they probe inner regions of galaxies. We use the abundance of ionised carbon in excited fine-structure level to calculate the cooling rates through the \CII~$\lambda$158$\mu$m emission, and compare them with the cooling rates from \dla\ systems in the literature. We find that the cooling rates distribution of ESDLAs also presents the same bimodality as previously observed for the general (mostly lower \HI\ column density) \dla\ population. 
\end{abstract}




\begingroup
\let\clearpage\relax
\endgroup
\newpage

\section{Introduction}\label{sec:Introduction}
The analysis of the interstellar medium (ISM) of high-redshift galaxies in emission is hampered by the sensitivity limit and spatial resolution of the instruments.
Another widely-used method for the investigation of the gaseous content of galaxies at high redshifts is to analyse the absorption lines they imprint in the spectra of bright background sources such as $\gamma$-ray burst (GRB) afterglows and quasars. In particular, atomic hydrogen manifests itself as characteristic strong Lyman-$\alpha$ \HI\ absorption. Observations of this line with ground-based telescopes are limited to a minimum redshift of $z\approx1.6$ due to the atmospheric cut-off at $\sim$3200~\AA. For systems with column density $N(\text{\HI})\gtrsim2\times10^{20}$\,cm$^{-2}$, so-called  \dla\ systems (DLAs, see review by \citealt{Wolfe2005}), hydrogen is predominantly neutral. DLAs have therefore long been considered as regions where star formation processes may potentially occur. Besides \HI\ absorption lines, numerous associated absorption lines from metals at different ionization states are usually detected in DLAs. Moreover, in a small fraction of them, molecules are also found: mostly \HH\, (first detected by \citealt{Levshakov1985}), but also HD (first detected by \citealt{Varshalovich2001}) and CO (first detected by \citealt{Srianand2008}).
It was shown in simulations \citep[e.g.]{Rahmati2014} as well as in observations \citep{Noterdaeme2012b,Noterdaeme2014,Ranjan2018,Ranjan2020} that DLA systems with $N(\text{\HI})\gtrsim 5\times10^{21}$ cm$^{-2}$, so-called extremely strong \dla\ (ESDLA) systems, are most likely associated with galaxies at small impact parameters, typically $<3$~kpc, while the rest of the DLA population statistically probes the neutral gas at larger impact parameters \citep{Krogager2017}, most likely associated with the circumgalactic medium. Therefore, investigating ESDLA systems may provide more direct information on galaxy evolution, in  particular the physical and chemical conditions and a closer link to star formation in the associated high-redshift galaxies.
However, statistical analysis of ESDLA systems is complicated by their small cross sections, seen in the steep power-law nature of the DLA \HI\ column density distribution: less than 1 per cent of the DLAs detected in the Sloan Digital Sky Survey (SDSS) ($\sim 100$ systems) have \HI\ column densities larger than $5\times10^{21}$ cm$^{-2}$ \citep{Noterdaeme2014}. Thereby, detailed analysis of ESDLAs at medium/high spectral resolution has been obtained so far for only about twenty systems toward quasars \citep[e.g.][and references therein]{Ranjan2020} and about a dozen systems toward GRB afterglows \citep{Bolmer2019}, where 
the line of sight pin-points regions of star-formation in galaxies, where the burst occurs.

In this work we present the analysis of seven additional ESDLA systems at $z=2-3$ from the same selection as \citet{Ranjan2020}, and also obtained with the intermediate-resolution spectrograph X-shooter on the Very Large Telescope (VLT)~\citep{Vernet2011}. This paper is organised as follows. In section~\ref{sec:Data} we describe the observations and data reduction. In section~\ref{sec:spectroscopic_analysis} we present the analysis of the collected spectra. 
Lastly, we discuss the results and present our summary in sections~\ref{sec:Discussion} and \ref{sec:conclusions}, respectively.

\section{Observations and data reduction}\label{sec:Data}

The ESDLAs candidates were taken from the parent SDSS-DR14 DLA sample with the conditions to be observable from Paranal in a reasonable amount of observing time. A first set of data was published by \citet{Ranjan2020} and the remaining were prepared for observations based on best observability during the allocated period.
Observations were carried out in 2018 under Program ID 0101.A-0891(A) (PI: Ranjan) in service mode with the intermediate-resolution multi-wavelength spectrograph X-shooter mounted on the European Southern Observatory Very Large Telescope Unit 2 (Kueyen). Observations were obtained with the nodding mode using a nod throw of 4 arcsec and slit widths of 1.6, 0.9 and 1.2 arcsec for the UVB, VIS and NIR arms, respectively. The UVB and VIS detectors were read out using 1$\times$2 binning. 
The data were processed using the esorex pipeline version 2.6.8 specifically designed to combine exposures with nodding offsets along the slit \citep{Goldoni2006, Modigliani2010}. All spectra were flux-calibrated using observations of a spectroscopic standard star taken on the same night as the respective science target. For 4 targets, we obtained two separate observations that were individually processed and subsequently combined using an inverse-variance weighted combination. The 1-dimensional spectra were then extracted from the combined observations following the steps of the optimal extraction algorithm \citep{Horne1986}. Wavelengths were converted to vacuum and expressed in the heliocentric rest-frame. Lastly, the spectra were corrected for Galactic extinction using the dust maps by \citet{Schlafly2011ApJ}. Since we do not analyze any features in regions of strong telluric absorption, the spectra have not been corrected for telluric absorption.

Since the observations were performed under good seeing conditions, throughout the analysis, we have used a nominal spectral resolution in the UVB, VIS, and NIR arms of 5400, 8900 and 5600, respectively, corresponding to the smaller slit widths, which matches the average seeing.
Observational details are presented in Table~\ref{tab:log}.

\begin{table*}
\centering
\caption{Log of X-shooter observations}
\label{tab:log}
\addtolength{\tabcolsep}{-2pt}
\begin{tabular}{cccccc}
\hline 
Quasar & Date  &  Airmass   & Seeing$^\star$  & Exposure time$\dagger$  &CNR$\dagger$ \\ 
       &                     &        & (arcsec) &s&\\
\hline
J\,002409.3-072554.8  & 2018-08-16 &  1.357 & 0.83 &1490, 1440, 300 & 22, 12, 8\\
\hline
J\,123815.9+162043.9 & 2018-04-09 &1.323 & 0.67& 1490, 1440, 300 &  30, 15, 13\\ 
J\,123815.9+162043.9 & 2018-04-16 &  1.328 & 1.01 & \ 1490, 1440, 300 & 30, 15, 13\\
\hline
J\,135316.9+095634.8 & 2018-04-19 &   1.215 & 0.85 & 1490, 1440, 300 & 20, 14, 11\\
\hline
J\,141801.9+071846.3 & 2018-04-23 &   1.221 & 0.69 & 1490, 1440, 300 & 21, 12, 8\\
\hline
J\,220525.7+102119.9 & 2018-07-09 &  1.232 & 0.94 & 1490, 1440, 300 & 18, 7, 6 \\
J\,220525.7+102119.9 & 2018-07-09 &  1.233 & 0.89 & 1490, 1440, 300 & 20, 8, 6 \\
\hline
J\,235124.8-063916.4 & 2018-08-11 &  1.337 & 1.05 & 1490, 1440, 300 & 14, 9, 5\\
J\,235124.8-063916.4 & 2018-08-14 & 1.348 & 0.83 & 1490, 1440, 300 & 15, 9, 5\\
\hline
J\,235916.6+135445.0 & 2018-08-12 &  1.374 & 0.78 & 1490, 1440, 300 & 18, 13, 13\\
J\,235916.6+135445.0 & 2018-08-15 &  1.313 & 1.01 & 1490, 1440, 300 & 15, 10, 13\\
\hline 
\end{tabular}
\addtolength{\tabcolsep}{2pt}
\begin{tablenotes}
\item $\star$ Corrected by airmass
\item Values for airmass and seeing correspond to the start of the respective exposures. 
\item $\dagger$ Exposure times and typical continuum-to-noise ratios (CNRs) are given for the UVB, VIS, and NIR arms, respectively.
\end{tablenotes}
\end{table*}

\section{Spectroscopic analysis}\label{sec:spectroscopic_analysis}

We analysed the absorption lines using standard multi-component Voigt-profile fitting procedures \citep[for the details see][]{Balashev2019}. A summary of the main properties of the ESDLAs analysed here is presented in Table~\ref{tab:fit_results}.  Details on the absorption-line analysis are provided in the sub-sections below with the corresponding table parameters and figures presented in the appendix. 

\renewcommand\arraystretch{1.2}
\begin{table*}
\centering
\caption{Main derived properties of the ESDLAs.}
\label{tab:fit_results}
\begin{tabular}{cllcccllcrcc}
\hline 
Quasar & $z_{\rm em}^{\star}$&$z_{\rm abs}$  & $\log N(\text{\HI})$& $\log N(\text{\HI})^{\rm SDSS}$  & $\log N(\rm H_2)$ & [X/H] & [Fe/X] & $\log N(\rm Fe^{dust})$&$A_{\rm V}^\dagger$ \\ 
& & & [cm$^{-2}$]& [cm$^{-2}$]& [cm$^{-2}$]& & & [cm$^{-2}$]& \\
\hline
J\,0024$-$0725  &$2.87$& $2.68120$ & $21.81\pm0.01$ & 21.8&$<17.20$ & $-1.77^{+0.06}_{-0.09}$ &  $-0.56^{+0.10}_{-0.09}$ &$15.40^{+0.12}_{-0.18}$& $<0.1$   \\
J\,1238$+$1620  &$3.43$&$3.20907$ &  $21.60\pm0.01$ & 21.7&$<17.25$ & $-1.01^{+0.05}_{-0.05}$ & $-0.43^{+0.06}_{-0.05}$&$15.89^{+0.13}_{-0.16}$& $<0.1$   \\
J\,1353$+$0956 &$3.61$&$3.33326$ & $21.61\pm0.01$ & 21.7&$<17.30$ & $-1.61^{+0.13}_{-0.20}$ &  $-0.46^{+0.22}_{-0.15}$ &$15.32^{+0.22}_{-0.61}$& $<0.1$ \\
J\,1418$+$0718 &$2.58$&$2.39211$ & $21.59\pm0.02$ & 21.8&$<17.20$ & $-1.50^{+0.05}_{-0.04}$ & $-0.36^{+0.05}_{-0.06}$ &$15.34^{+0.13}_{-0.15}$& $<0.1$   \\
J\,2205$+$1021 &$3.41$&$3.25516$ & $21.61\pm0.02$ & 21.7&$18.16\pm0.03$  & $-0.93^{+0.05}_{-0.05}$ & $-0.87^{+0.09}_{-0.07}$ &$16.12^{+0.09}_{-0.10}$& $<0.1$   \\
J\,2351$-$0639&$2.91$&$2.55744$& $21.90\pm0.01$ & 21.9&$<17.75$ & $-1.58^{+0.05}_{-0.05}$ & $-0.46^{+0.11}_{-0.11}$ & $15.64^{+0.13}_{-0.17}$&$<0.1$  \\
J\,2359$+$1354&$2.77$&$2.2499$ & $21.96\pm0.02$ & 22.0&$19.28\pm0.06$ & $-0.47^{+0.03}_{-0.03}$ & $-0.94^{+0.03}_{-0.03}$ &$16.94^{+0.23}_{-0.37}$& $0.3$   \\
\hline
\end{tabular}
\begin{tablenotes}
        \item $\star$ Quasar redshift.
        \item $\dagger$ The upper limits on $A_{\rm V}$ correspond to the systems, for which obtained extinction less than the systematic uncertainty ($\lesssim0.1$) due to dispersion of intrinsic spectral shapes of quasars. 
        \item X refers to a volatile element, which is zinc for all ESDLAs but that towards J\,1418$+$0718, for which X is sulphur, see text.
\end{tablenotes}
\end{table*}

\subsection{Neutral atomic hydrogen}\label{subsec:HI_fitting}

\begin{figure}
\centering
    \renewcommand{\arraystretch}{0}
    \begin{tabular}{c}
    \includegraphics[trim={0.0cm 0.0cm 0.0cm 0.0cm},clip,width=1.\columnwidth]{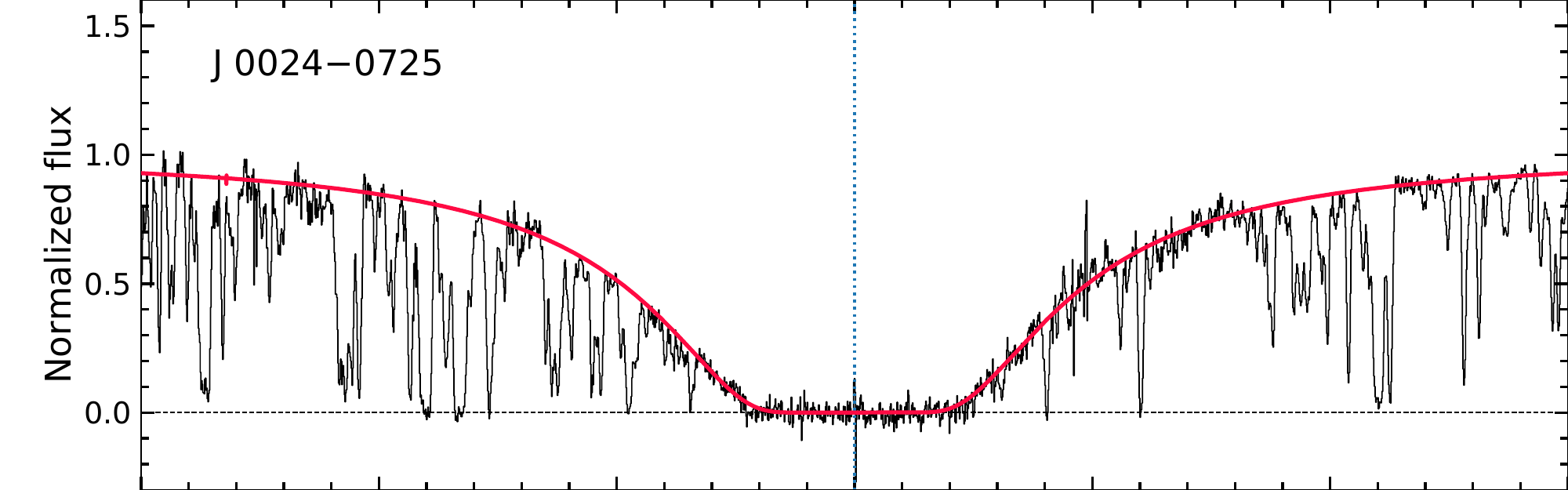}
    \\
    \includegraphics[trim={0.0cm 0.0cm 0.0cm 0.0cm},clip,width=1.\columnwidth]{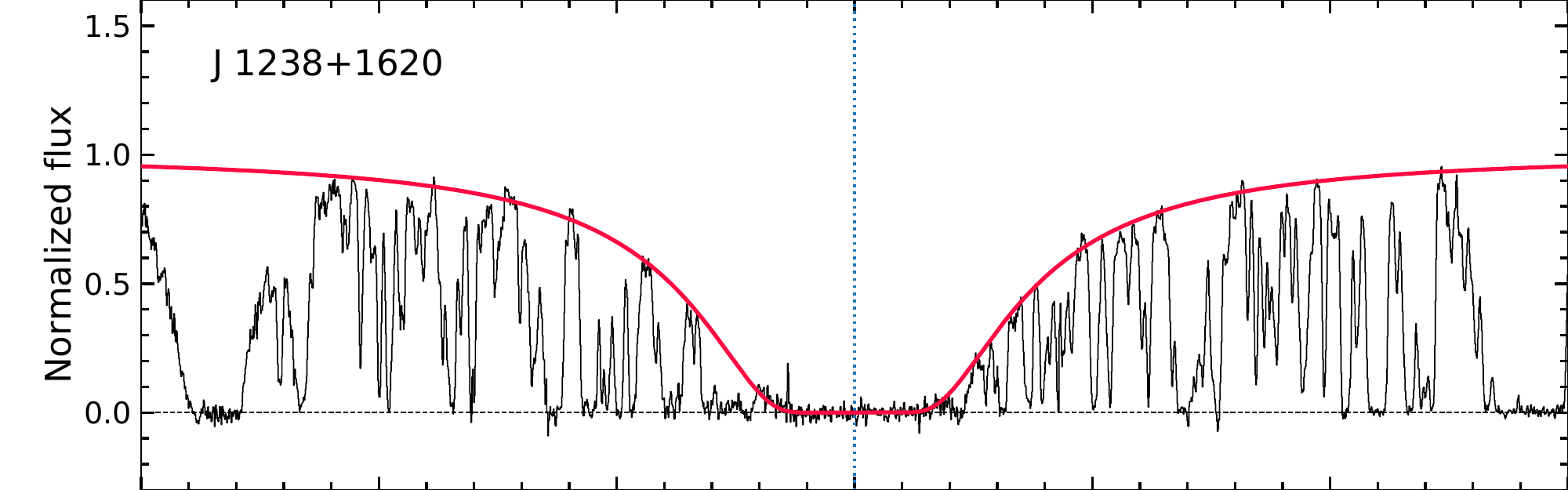}
    \\
    \includegraphics[trim={0.0cm 0.0cm 0.0cm 0.0cm},clip,width=1.\columnwidth]{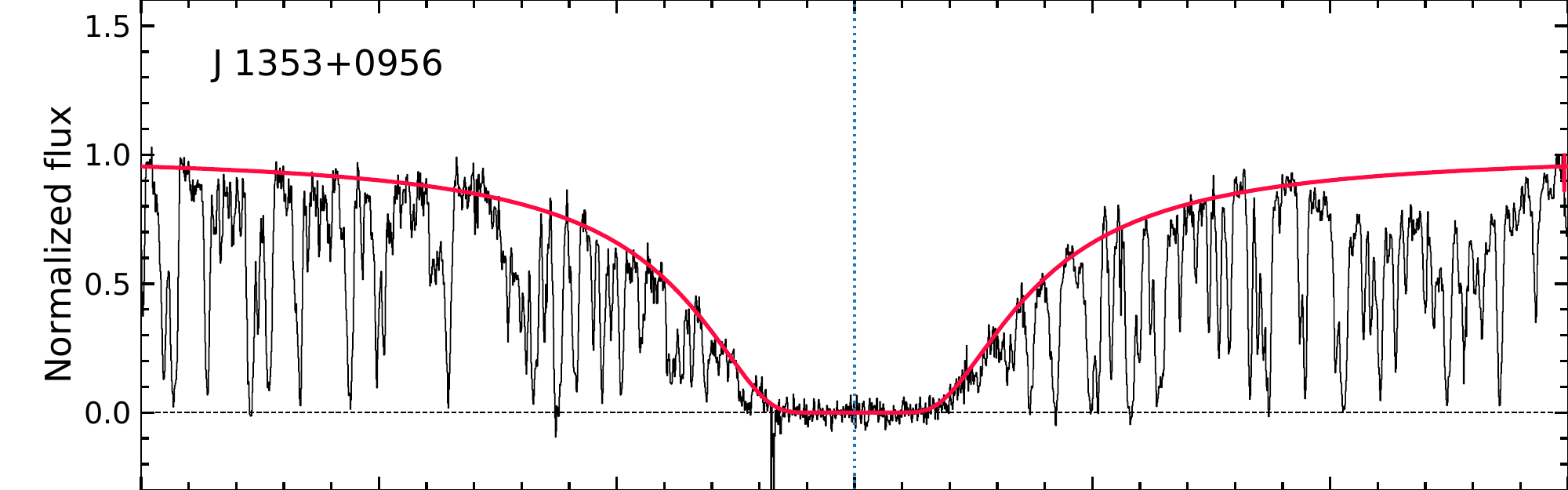}
    \\
    \includegraphics[trim={0.0cm 0.0cm 0.0cm 0.0cm},clip,width=1.\columnwidth]{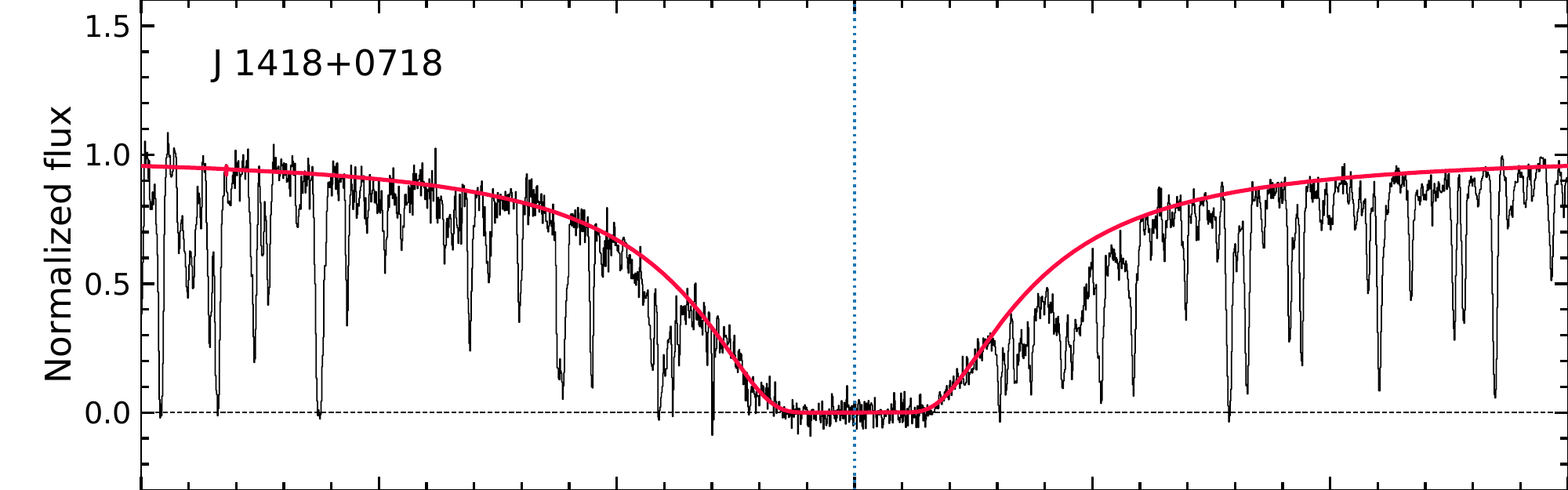}
    \\
    \includegraphics[trim={0.0cm 0.0cm 0.0cm 0.0cm},clip,width=1.\columnwidth]{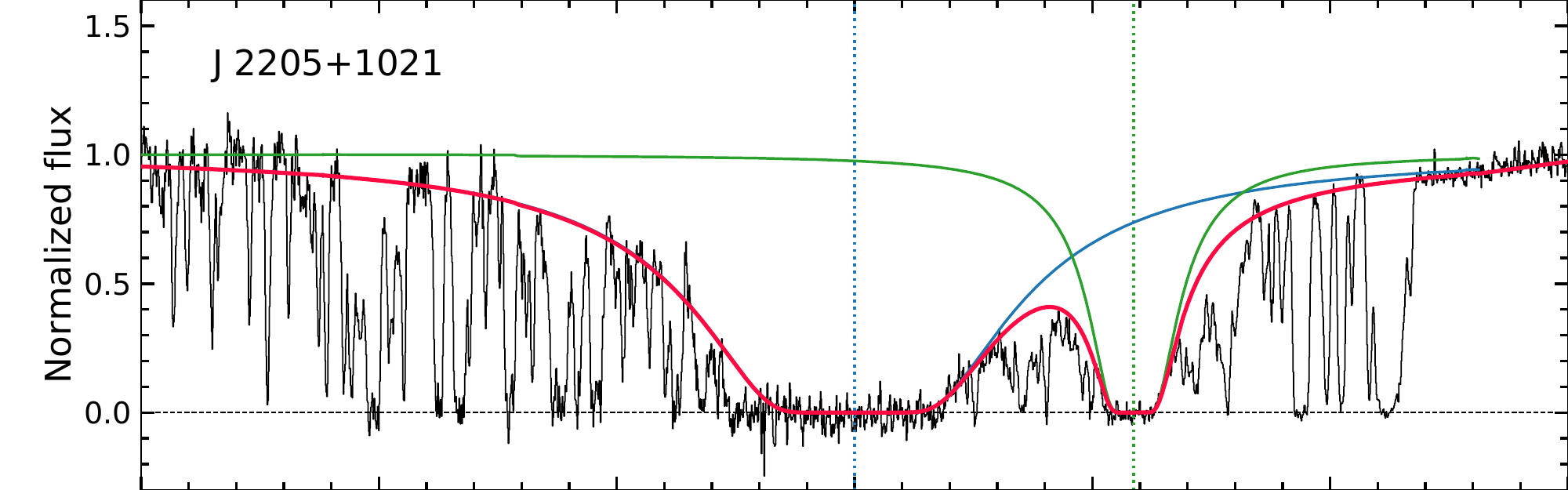}
    \\
    \includegraphics[trim={0.0cm 0.0cm 0.0cm 0.0cm},clip,width=1.\columnwidth]{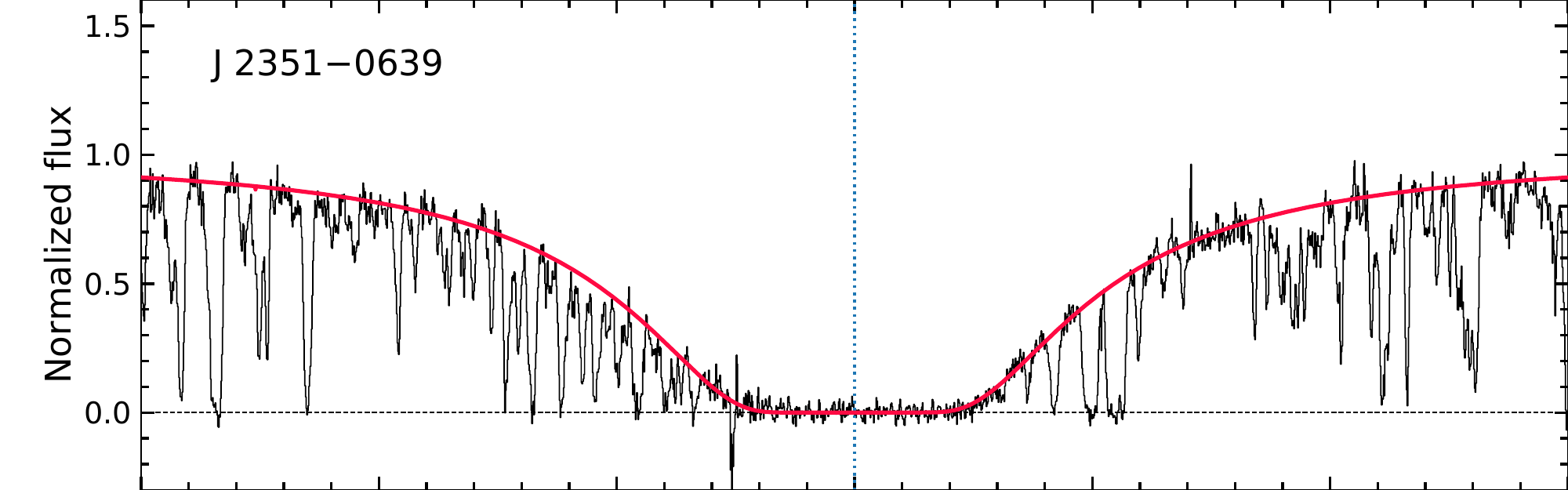}
    \\
    \includegraphics[trim={0.0cm 0.0cm 0.0cm 0.0cm},clip,width=1.\columnwidth]{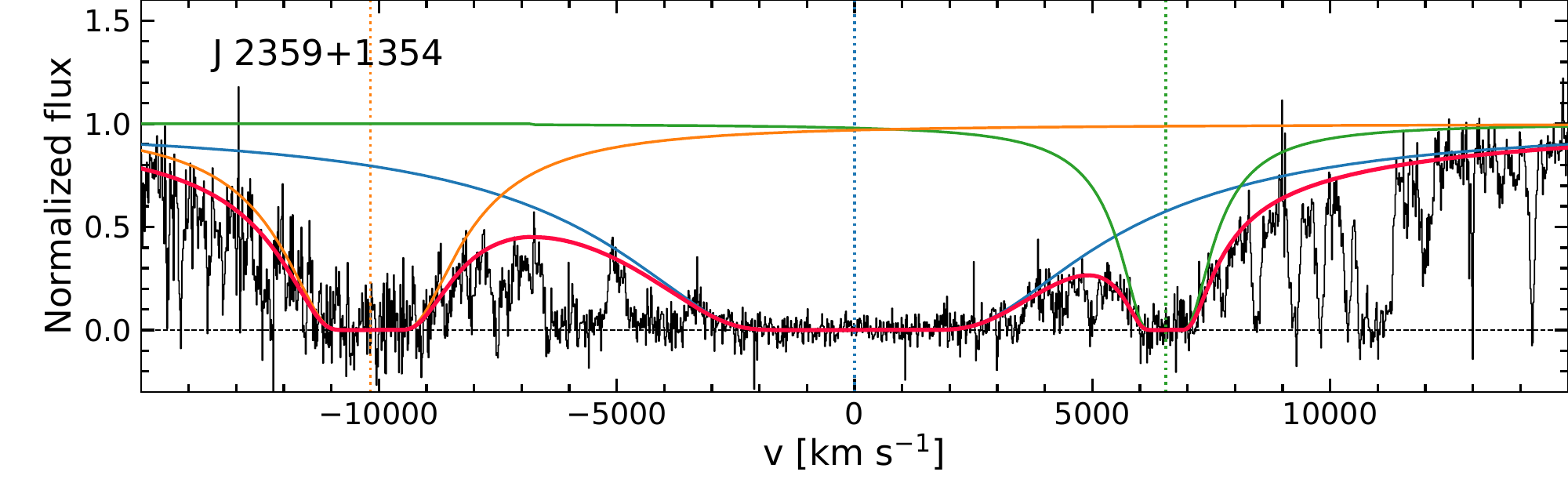}
    \\
    \end{tabular}
    \renewcommand{\arraystretch}{0}
\caption{Fit to the \HI\ Ly$\alpha$ profiles of ESDLAs in the sample. Black and red lines show the observed spectrum and total fit profiles, respectively. Blue solid line indicates the profile of studied ESDLA, while the green and orange lines correspond to additional fitted damped Ly$\alpha$ transitions. The x-axis indicates the relative velocity with respect to the ESDLA redshifts.
\label{fig:HI}}
\end{figure}

We estimated the \HI\ column densities through fitting Ly-series lines (in most cases Ly$\alpha$ and Ly$\beta$) with a one-component Voigt-profile model. 
Since extremely strong damped Ly$\alpha$ lines absorb the quasar continuum over a wide wavelength range, the widely-used method of continuum reconstruction by eye can be significantly biased. In turn, we modeled the continuum at the location of Ly$\alpha$ lines using Chebyshev polynomials\footnote{The order of polynomial was defined based on the wavelength range and usually was $6-8$.}, whose parameters were fitted simultaneously with profiles of \HI\ absorption lines. As an initial guess of the unabsorbed quasar continuum, we first used the visual estimation aided by a B-spline interpolation. In the high column density regime, the derived column densities are actually insensitive to the choice of the Doppler parameter $b$. Therefore, we fixed $b$ with a value chosen to roughly reproduce the width of the observed high Lyman series lines. Fit profiles to the \HI\ Ly$\alpha$ absorption lines are shown in Fig.~\ref{fig:HI}.
Although the estimated \HI\ column densities are consistent with the values measured from the low signal-to-noise ratio, low resolution spectra from the Sloan Digital Sky Survey \citep{Noterdaeme2012a,Noterdaeme2014}, four systems (J\,1238$+$1612, J\,1353$+$0956, J\,1418$+$0718 and J\,2205$+$1021) have \HI\ column densities $\sim0.1$~dex below the threshold originally used to define ESDLAs ($N(\text{\HI})\ge 5\times10^{21}$ cm$^{-2}$). 
For the sake of simplicity, we continue calling them ESDLAs in the following.

\subsection{Molecular hydrogen}\label{subsec:molecular_hydrogen}

\begin{figure}
\includegraphics [width=\columnwidth]{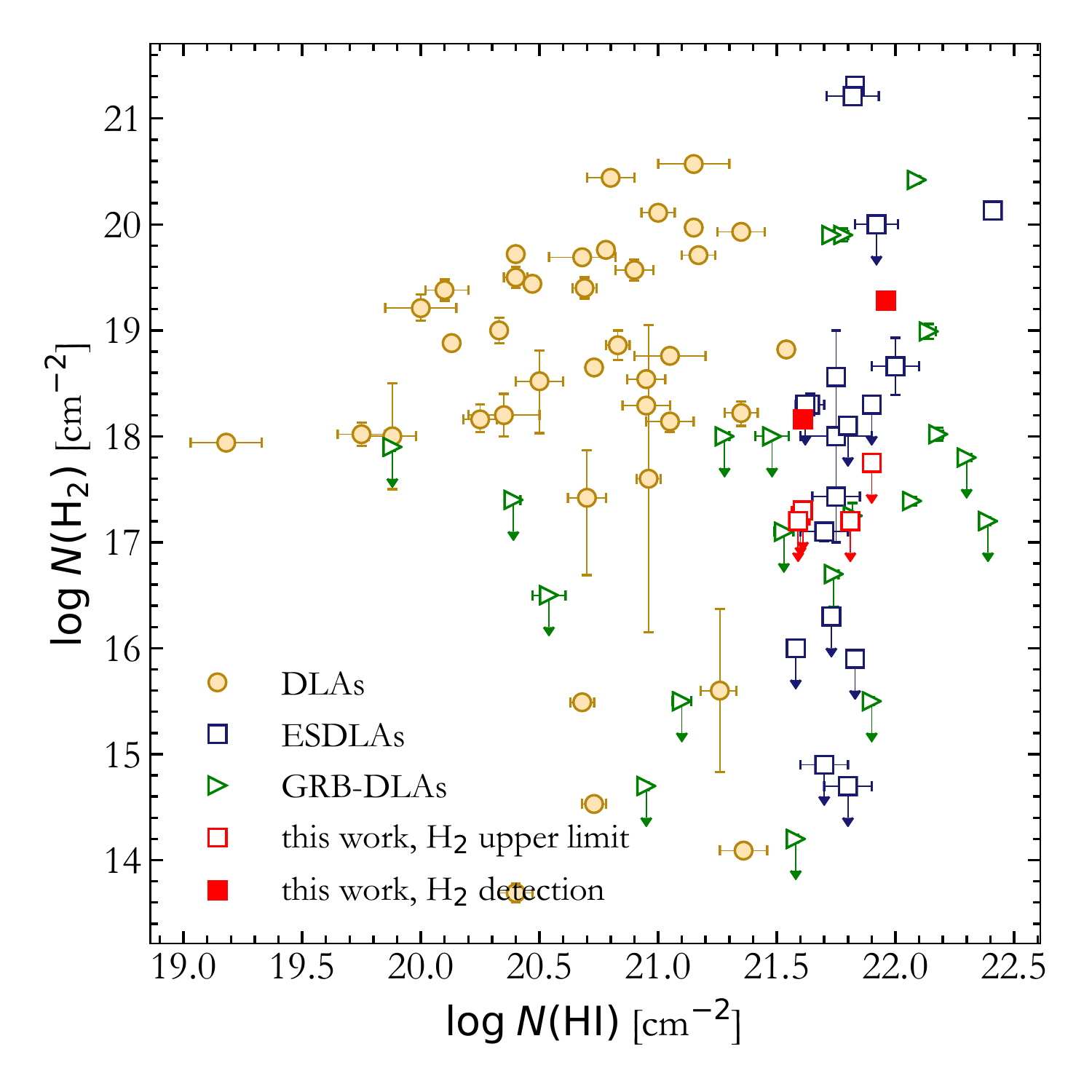}
\caption{\HI\ vs \HH\ column densities. ESDLAs from this work are shown by open (\HH\ upper limits) and filled (\HH\ detections) red squares. Blue squares and brown circles are ESDLAs and DLAs from \citet{Ranjan2018,Ranjan2020,Noterdaeme2007a,Noterdaeme2008,Noterdaeme2008b,Noterdaeme2010,Noterdaeme2014,Noterdaeme2015,Noterdaeme2018,Guimaraes2012,Fynbo2011,Rahmani2013,Ellison2001,Klimenko2015,Balashev2010,Balashev2011,Balashev2017,Balashev2019,Albornoz2014,Carswell2011,Srianand2010,Jorgenson2010,Milutinovic2010,Krogager2016}. GRB-DLAs from \citet{Bolmer2019} are shown by triangles.
}
\label{fig:HI-H2}
\end{figure}

We searched for resonant rest-frame UV \HH\ absorption lines, i.e. lines in the Lyman and Werner bands. For two systems, towards J\,2205$+$1021 and J\,2359$+$1354, we firmly detected strong \HH\ absorption with total column densities $\log N(\HH)\text{[cm$^{-2}$]} = 18.16\pm0.03$ and $\log N(\HH)\text{[cm$^{-2}$]} = 19.28\pm0.06$, respectively. 
For the remaining five systems, there are no obvious \HH\ signatures detectable in the X-shooter spectra. Therefore, we estimated conservative upper limits on \HH\ column densities, focusing on the $J=0$ and $J=1$ rotational levels that are known to contain most of the total column density. We used a method similar to that described by \citet{Ranjan2020}. We fixed the Doppler parameter $b$ to 1~\kms\ and the $T_{01}$ temperature to 100\,K, and created synthetic profiles at redshift varied within the range that corresponds to the observed velocity extent of the metal lines (typically  $100-200$\kms). Due to presence of \lya\ forest in the \HH\ absorption region we focused only on the positive residuals. The upper limit of the $N(\rm H_2)$ is obtained as the largest column density for which the synthetic profile remains consistent with the observed spectrum with corresponding fraction of positive residuals above 2 $\sigma$ level less than 5\%.
Constraints on the total $N(\rm H_2)$ are presented in Table~\ref{tab:fit_results}. In Fig.~\ref{fig:HI-H2} 
 we provide \HI$-$\HH\ diagram for the DLAs detected at $z\gtrsim2$. Our measurements of \HH\ column densities 
 result in an incidence rate of \HH\ of $10-55$\% based on the binomial proportion interval at 68\% confidence level in agreement with that obtained from ESDLAs toward quasars, $\sim 40$\% \citep{Balashev2018,Ranjan2020}, as well as from GRB afterglows, $\sim 30$\% \citep{Bolmer2019}. Although here we have only upper limits on \HH\ column densities for five out of seven system, our results confirm a higher \HH\ incidence rate for high \HI\ column density systems compared to that of the overall DLA population, $\sim 5-10$\% \citep{Balashev2018}.

\subsection{Metal column densities and dust depletion}\label{subsec:metals_and_depletion}

The main purpose of the current work is the derivation of the chemical and physical conditions in the neutral phase of the absorbing gas. Consequently, we focus on the low-ionization metal species (with ionization potential $<13.6$\,eV) such as \SiII, \ZnII, \CrII, \SII, \FeII, \MnII, \NiII, \TiII, \OI, \CI, \CII. All species, except \CI, which is more likely associated with molecular gas, were fitted using a common $b$ parameter and velocity structure. Throughout the paper we calculate metal abundances with respect to solar ones \citep[from][]{Asplund2009} as\footnote{For our seven ESDLAs $N({\rm H})=N(\text{\HI})+2N(\HH)\approx N(\text{\HI})$, since 
$\log N(\HH)\text{[cm$^{-2}$]} < 19.3$.}:
\begin{equation}
    [{\rm X/H}] = \log \left( \frac{N({\rm X})}{N({\rm \text{H}})} \right) - \log \left(\frac{{\rm X}}{{\rm H}} \right)_\odot
\end{equation}
and
\begin{equation}
    [{\rm Fe/X}] = \log \left( \frac{N({\rm \text{\FeII}})}{N({\rm X})} \right) - \log \left(\frac{{\rm Fe}}{{\rm X}} \right)_\odot.
\end{equation}

Both sulphur and zinc are known to be volatile species that deplete little into dust \citep[e.g.]{Vladilo2000}. 
We hence estimated the metallicities and dust depletion factors using zinc-to-hydrogen and iron-to-zinc ratios for all systems except for J\,1418$+$0718, for which we used sulphur instead of zinc, 
owing to the better constraint on \SII\ than on \ZnII\ for this system.
The sixth column of Table~\ref{tab:fit_results} presents these metallicities only, but detailed abundances are provided for each system in the appendix.

Since the production and shielding of \HH\ 
is intimately linked to the dust content, 
we assume an intrinsic solar abundance pattern and estimate the 
column density of iron locked into the dust from the observed column densities as

\begin{equation}
    N({\rm Fe^{dust}}) = (1-10^{\rm[Fe/X]}) N({\rm X})\left(\frac{{\rm Fe}}{\rm{X}}\right)_\odot,
\end{equation}
where X again refers to zinc (sulphur for J\,1418$+$0718).

\subsection{[\CII] cooling rate}\label{subsec:cooling_rate}
\begin{figure}
\includegraphics [width=\columnwidth]{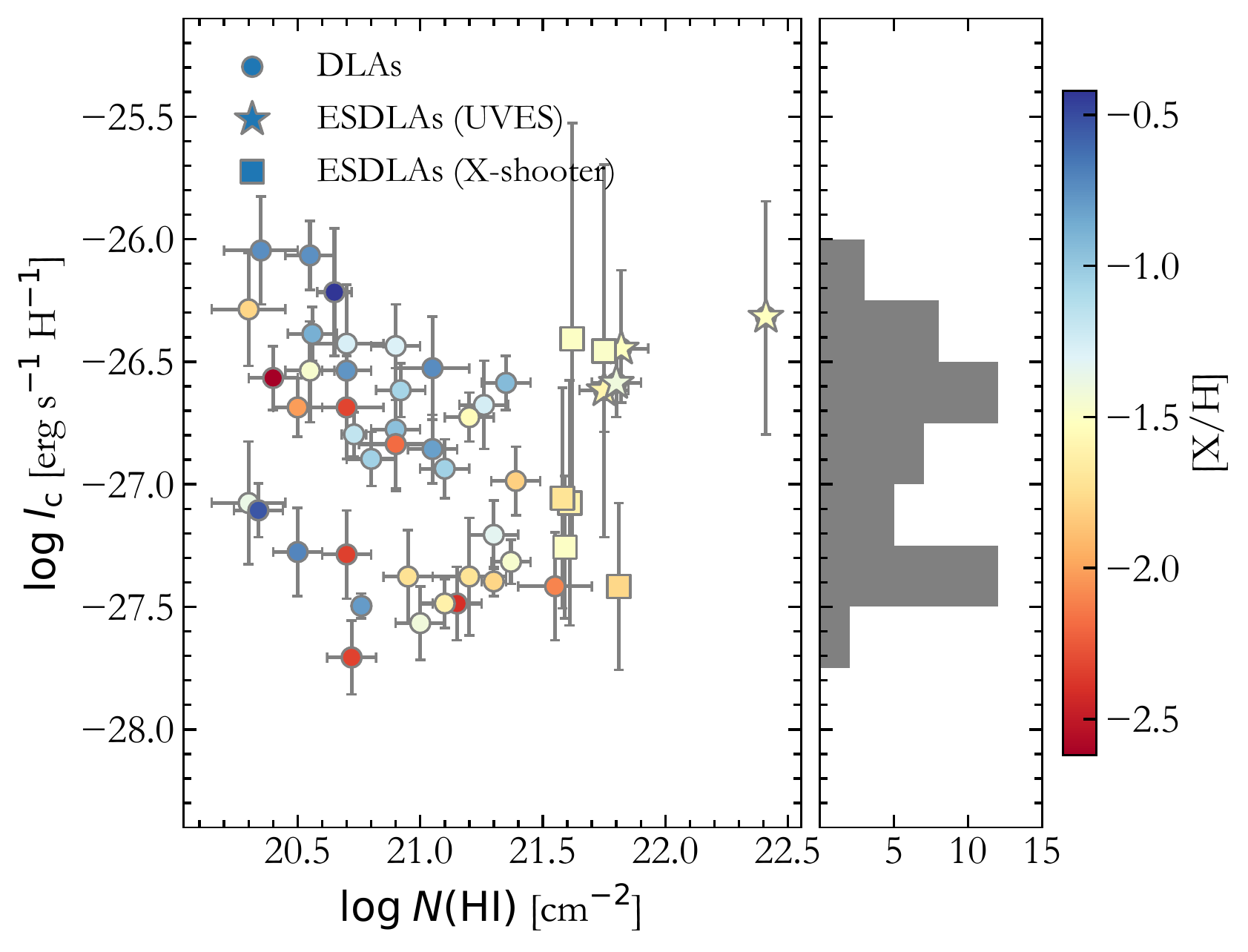}
\caption{[\CII] cooling rates vs \HI\ column densities. Circles correspond to the DLAs from \citet[][and references therein]{Wolfe2008} and \citet{Dutta2014}. Stars and squares are the ESDLAs from this work, obtained with UVES and X-shooter, respectively. Colors encode the metallicity.
Histogram corresponds to the cooling rate distribution, marginalised over the \HI\ column densities.}
\label{fig:cooling_rate}
\end{figure}

The fine-structure [\CII]~$\lambda$158$\mu$m transition from collisionally excited $^2P^{\circ}_{3/2}$ level \citep[e.g.][]{Goldsmith2012} is a key contributor to the cooling of the cold ISM \citep[e.g.][]{Lagache2018}. Instead of direct [\CII]~$\lambda$158$\mu$m emission, one can estimate the ISM cooling rate $l_{\rm c}$ locally (per unit hydrogen atom) in absorption, by measuring the column density of ionised carbon in its first excited level $N($\CII*) through its electronic transition at 1335\AA\ 
\citep{Pottasch1979, Wolfe2003}:

\begin{equation}\label{eq:cooling_rate}
    l_{\rm c} = \frac{N(\text{\CII*}) h\nu_{ul}A_{ul}}{N(\text{\HI})}{\rm~erg~s^{-1}~H^{-1}},
\end{equation}
where $A_{ul}=2.29 \times 10^{-6}$~s$^{-1}$ is the Einstein coefficient for the spontaneous decay $^2P^{\circ}_{3/2}\to ^2P^{\circ}_{1/2}$ and $h\nu_{ul}$ is the energy of this transition. 
It was possible to constrain the \CII*\ column densities for three systems in our X-shooter sample (towards J\,0024$-$0725, J\,1353$+$0956 and J\,1418$+$0718), while the \CII* absorption is strongly saturated for the remaining four. 
To enrich this sample, 
we also estimated \CII*\ column densities in the ESDLAs sample from \citet{Ranjan2020}. In total, we derived the [\CII] cooling rates for seven ESDLAs obtained with X-shooter and for four ESDLAs obtained with UVES~\citep{Dekker2000} 
(see Table~\ref{tab:CII*}). For ESDLAs detected in X-shooter spectra we can not exclude saturation of the \CII* absorption, since the velocity structure is likely unresolved. Therefore we obtained a conservative interval estimate on \CII* column density using two limiting cases. As a lower limit of $\log N$(\CII*) we used measurement obtained by the simultaneous fit of \CII* absorption line with other metal lines in the system. These lower limits correspond to the optically thin or intermediate saturation of \CII*\ lines. To obtain conservative upper limits on $N$(\CII*) we fit \CII* lines using one-component profile with Doppler parameter $b=3$\kms as more robustly derived at high resolution. 
As a point estimate for further calculations (except J\,1349$+$0448, for which we provide only a lower limit to $N$(\CII*)) we use the mean column density of the obtained $N$(\CII*) range. 
The quoted uncertainties in Table~\ref{tab:CII*} correspond to this conservative range.
In Fig.~\ref{fig:cooling_rate} we compare the derived [\CII] cooling rates in ESDLAs with those of DLAs from \citet{Wolfe2008}\footnote{For several systems in Table 1 from \citet{Wolfe2008}, which have zero uncertainties on $\log\, N$(\HI), we use a value of 0.1 dex for uncertainty on $\log\, N$(\HI). 
These values correspond to the uncertainties typically measured using high-resolution spectra and are consistent with the error bars shown in Fig. 1 in \citet{Wolfe2008}.} and \citet{Dutta2014}. 
Our results tend to support the bimodality in the [\CII] cooling rate distribution previously observed by \citet{Wolfe2008} with the following two groups of DLAs: {\it high} $\log l_{\rm c}[\rm erg\,s^{-1}\,H^{-1}] > -27$ and {\it low} $\log l_{\rm c}[\rm erg\,s^{-1}\,H^{-1}] \lesssim -27$. Indeed, four ESDLAs in our sample have {\it low} cooling rates of $\log l_{\rm c} [\rm erg\,s^{-1}\,H^{-1}] \sim -27.2$, 
while six have {\it high} cooling rates of $\log l_{\rm c}[\rm erg\,s^{-1}\,H^{-1}] \sim -26.5$. To quantitatively test if our systems 
do not contradict with the statistical properties previously obtained for the overall DLAs, we use a Gaussian mixture algorithm with two components, as done by \citet{Wolfe2008}, but applying it to the whole DLA plus ESDLA sample. 
We obtained two groups of cooling rates with median $\log l_{\rm c}[\rm erg\,s^{-1}\,H^{-1}] = -26.6$ and $-27.3$ with corresponding fraction of systems of 0.6 and 0.4, respectively.

For all ESDLA systems in our sample, observed with UVES, that is, where the velocity structure is easily resolved and saturation of the lines can be easily assessed, we measure {\it high} cooling rates, while four of six X-shooter ESDLAs fall in the {\it low} cooling rate group. 
It is legitimate to wonder whether this difference {in ESDLA cooling rates} arises from the use of two instruments. 
However, even our conservative procedure for the $N(\text{\CII*})$ upper limit rules out high saturation of the \CII* lines (see Table~\ref{tab:CII*}). Therefore, the provided interval estimation of the $N(\text{\CII*})$ for X-shooter systems is quite robust. In addition, the difference in the cooling rates of the X-shooter and UVES ESDLAs in our sample is supported by the coincidentally higher \HH\ abundance in the ESDLAs studied with UVES than in those studied with X-shooter.

\citet{Wolfe2008}, under the assumption that 
heating through photoelectric effect equates [\CII] cooling, suggested that the bimodality in [\CII] cooling rates originates from a mass-transition in a galaxy formation scenario and corresponds to different star-formation regimes. 
Recently, \citet{Roy2017} argued that inferring the star-formation rate from \CII* is not straightforward due to the presence of \CII* in both the warm neutral medium (WNM) and the cold neutral medium (CNM). 
The authors stressed the absence of evident correlation between $N({\rm \text{\CII*}})$ and the \HI\ column density in each of these two phases in the Galaxy. This means that the two-phase origin of \CII* absorption must be taken into account.
Moreover, [\CII]~$\lambda$158$\mu$m emission is the main coolant only for the cold phase of neutral ISM, while the cooling of the warm phase is dominated by Ly$\alpha$ emission, which sets 
the WNM temperature of $\sim10^4$\,K (as robustly derived in several DLAs, see e.g. \citealt[][]{Carswell2012,Cooke2015,Noterdaeme2012c,Noterdaeme2021b}). 
Therefore, assuming dominant [\CII] cooling and thermal balance to derive the heating rate and hence the star-formation is only possible when most of \CII*\ arises from the cold phase with temperatures $< \rm few \times 100$\,K.

Interestingly, the consistent bimodality in the $l_{\rm c}$ distribution observed in our sample of ESDLAs does not comply with the explanation from \citet{Wolfe2008} about its origin in DLA samples. 
\citet{Wolfe2008} argued that the bimodality arises from a difference in star-formation regimes for two distinct galaxy populations: more massive metal-rich and less massive metal-poor systems. However, our CII*-selected ESDLA sample is quite homogeneous in terms of metallicities ($\rm [X/H]\sim-1.5$, see Table~\ref{tab:CII*}) and \SiII~$\lambda1526$ equivalent widths (which was used by \citeauthor{Wolfe2008} as mass indicator). The selection of ESDLAs is also more homogeneous in terms of the galactic environment that the line of sight traverses compared to that of DLAs, since ESDLAs are more likely to probe galaxies at low impact parameters, while DLAs probe predominantly the interface between interstellar and circumgalactic gas at larger impact parameters.

As an alternative explanation, we suggest that the $l_{\rm c}$ bimodality is more likely due to the phase separation of the neutral ISM rather than a bimodality in the star-formation regimes. We address this issue quantitatively in the work by \citet{Balashev2021} and show that the bimodality originates in the \CII*/\CII--metallicity plane and can be naturally explained by the phase separation in the ISM and the dependence of their characteristic number densities on metallicity.

\begin{table*}
\centering
\caption{Fine-structure excitation of ionised carbon and derived cooling rates.}\label{tab:CII*}
\begin{tabular}{ccccccccc}
\hline
Quasar & $\log N$(\HI) & $\log N(\HH)$&$\log f^\dagger$&[X/H]&$\log N$(\CII*)& $\log l_{\rm c} $& instrument & Refs$^\star$\\
&[cm$^{-2}$] &[cm$^{-2}$] & & & [cm$^{-2}$]& [erg~s$^{-1}$~H$^{-1}$] & & \\
\hline
 J\,0024$-$0725 & $21.81\pm0.01$ &$<17.20$&$<-4.30$ &$-1.77^{+0.06}_{-0.09}$ &$13.93\pm0.35$ & $ -27.42\pm0.35$ & XS &1\\
 J\,1353$+$0956  & $21.61\pm0.01$ &$<17.30$&$<-4.00$&$-1.61^{+0.13}_{-0.20}$ &$14.07\pm0.50$ &  $-27.08\pm0.50$ & XS&1\\
 J\,1418$+$0718  & $21.59\pm0.02$ &$<17.20$&$<-4.07$&$-1.50^{+0.05}_{-0.04}$ &$13.87\pm0.29$ &  $-27.26\pm0.29$& XS&1\\
J\,0017$+$1307 & $21.62\pm0.03$ &$<18.3$&$<-2.99$& $-1.50\pm0.09$&$14.76\pm0.88$  &  $-26.40\pm0.88$& XS&2\\
J\,1349$+$0448  & $21.80\pm0.01$ &$<18.1$&$<-3.39$ & $-1.35\pm0.06$&$>14.16$ &  $>-27.17$& XS&2\\
J\,2232$+$1242  &$21.75\pm0.03$  &$18.56\pm0.02$& $-2.89\pm0.03$& $ -1.48\pm0.05$&$14.83\pm0.76$ & $-26.46\pm0.76$& XS&2\\
J\,2322$+$0033  &$21.58\pm0.03$  &$<16.0$& $<-5.25$&$-1.71\pm0.13$ &$14.06\pm0.45$ &  $-27.06\pm0.45$& XS&2\\
 
HE\,0027$-$1836 &$21.75\pm0.10$  &$17.43\pm0.02$& $-4.02^{+0.10}_{-0.04}$ &$ -1.59\pm0.10$ &$14.67^{+0.13}_{-0.09}$ &  $-26.62^{+0.16}_{-0.13}$  & UVES & 4\\       
J\,0843$+$0221& $21.82\pm0.11$ &$21.21\pm0.02$& $-0.48^{+0.07}_{-0.08}$& $-1.52^{+0.08}_{-0.10}$&$14.91^{+0.28}_{-0.17}$  & $-26.45^{+0.30}_{-0.19}$ & UVES & 3\\
Q\,1157$+$0128& $21.80\pm0.10$ &$<14.75$&$<-6.65$ & $-1.40\pm0.10$&$14.7\pm0.02$  & $-26.59\pm0.10$ & UVES & 5\\
J\,2140$-$0321  & $22.41\pm0.03$ &$20.13\pm0.07$& $-1.98\pm0.06$&$-1.52\pm0.08$ &$15.63^{+0.46}_{-0.47}$  &  $-26.32^{+0.46}_{-0.46}$ & UVES & 2\\

\hline
\end{tabular}
\begin{tablenotes}
\item
$\dagger$ $f=2N$(\HH)/($2N$(\HH)$+N$(\HI)) is molecular fraction.
\item 
$\star$ References correspond to the \HI\ and \HH\ column densities measurements: (1)~this work, (2)~\citet{Ranjan2020}, (3)~\citet{Balashev2017}, (4)~\citet{Noterdaeme2007b}, (5)~\citet{Noterdaeme2008}.
\end{tablenotes}
\end{table*}
\renewcommand\arraystretch{1}

\subsection{Electron density}
\label{sec:eden}
\begin{figure}
\includegraphics [width=\columnwidth]{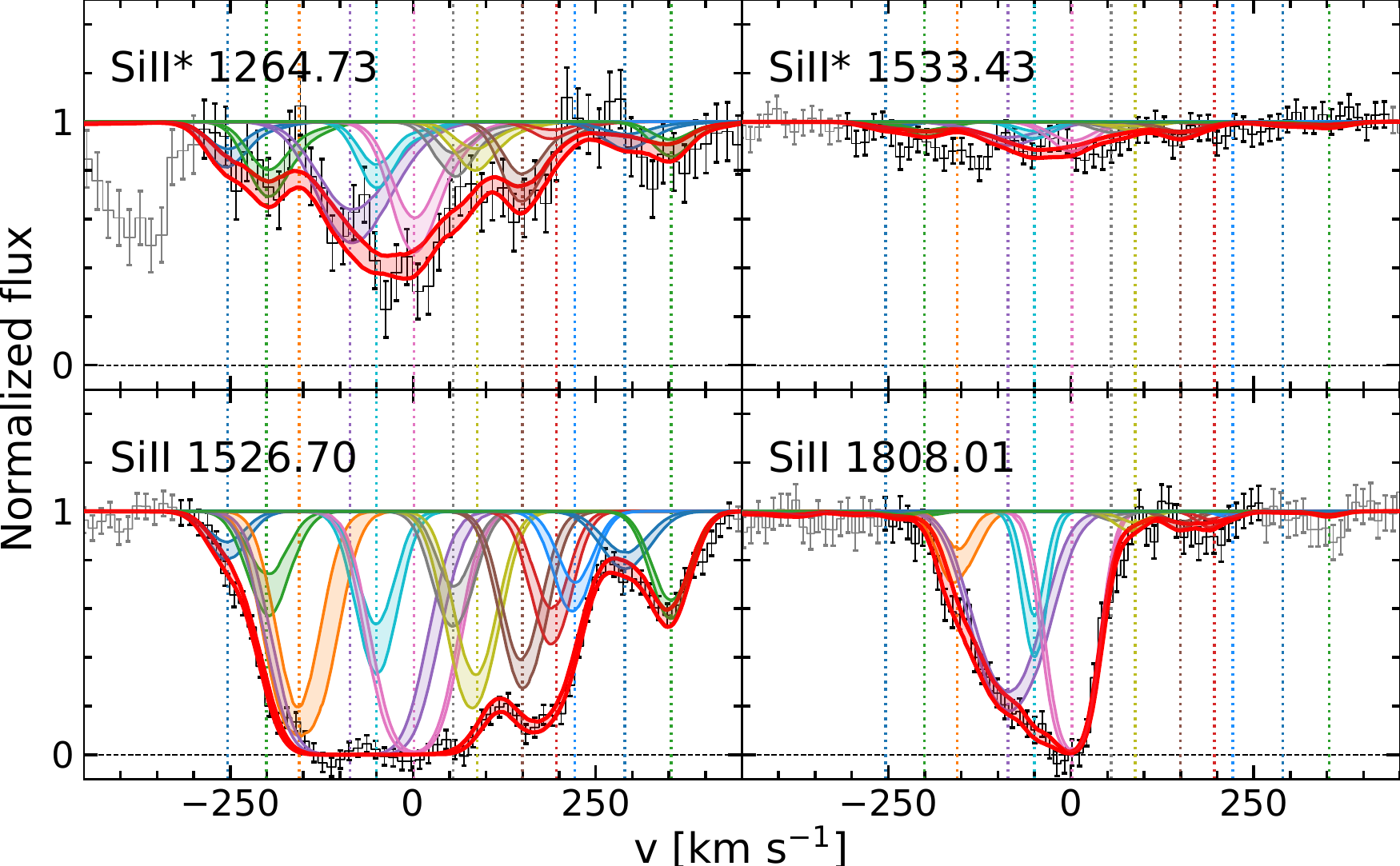}
\caption{Fit to \SiII\ and \SiII* lines in J\,2359$+$1354. The red  regions indicate the total Voigt profiles with corresponding 1$\sigma$ uncertainties. The vertical dashed lines correspond to the relative positions of the individual fit components.}
\label{fig:J2359_SiII}
\end{figure}

Far from ionizing sources, the collisional excitation of \CII*\ and \SiII* dominates over the radiative excitation and
\CII*\ and \SiII* column densities may be used to estimate the electron density assuming equilibrium between collisional excitation and spontaneous radiative de-excitation \citep{Srianand2000}.
For the system at $z=2.2486$ toward J\,2359$+$1354 with the highest \HI\ column density, we detected \SiII* absorption via \SiII*~$\lambda$1264\AA\ and \SiII*~$\lambda$1533\AA\ transitions, which were fitted simultaneously with \SiII\ and other available metal species (see Fig.~\ref{fig:J2359_me}). The line profiles of \SiII\ and \SiII* transitions are shown in Fig.~\ref{fig:J2359_SiII}.
Although these lines are partially blended we obtained a total $\log N({\rm \text{\SiII*}})\text{[cm$^{-2}$]}=13.84^{+0.04}_{-0.03}$ that translates into $N$(\SiII*)/$N$(\SiII)$=\left(1.0^{+0.1}_{-0.2}\right)\times10^{-3}$. This value is in agreement with what was measured towards J\,1135$-$0010 by \citet{Kulkarni2012}, but ten times higher than what was measured in two other ESDLAs \citep{Noterdaeme2015, Balashev2017}. Excited \SiII\ fine-structure levels can be populated either by collisions with atomic (and molecular) hydrogen in a mostly neutral medium, or by collisions with electrons in an ionized phase surrounding the pockets of neutral gas. However, since the measured $N$(\SiII*)/$N$(\SiII) ratio is relatively high, it would require very high thermal pressures of $\sim10^6$\,K\,cm$^{-3}$ if the collisions were dominated by neutral species for both warm atomic and cold molecular phases. Therefore we suggest that \SiII* is mostly associated with the ionized gas, where it is efficiently populated by electrons. In case of the equilibrium between collisional excitation and spontaneous radiative de-excitation for \SiII, the electron number density $n_{\rm e}$ can be expressed as function of the gas temperature as
\begin{equation}
    n_{\rm e} = \frac{A_{21}}{C_{12}(T)} \frac{N(\text{\SiII*})}{N(\text{\SiII})},
\end{equation}
where $A_{21} = 2.13\times 10^{-4}$\,s$^{-1}$ is the Einstein coefficient for the spontaneous decay for \SiII* and $C_{12} = 3.32\times10^{-5}\,T^{-0.5}\,\exp(-413.4/T)$~cm$^3$s$^{-1}$ is the temperature dependent collisional excitation rate for \SiII\ \citep[e.g.][]{Srianand2000}. Assuming a temperature of $\sim10^4$~K, we obtain an electron number density of $n_{\rm e}\sim0.7$~cm$^{-3}$. This should be considered as a lower limit to the electron density in the ionized medium, since we do not know how much of the observed \SiII\ column density is associated with this phase. 
Hence the actual $N$(\SiII*)/$N$(\SiII) ratio in the ionized medium can be higher \citep[see also][]{Noterdaeme2021}. 

\subsection{Neutral Carbon \label{subsec:neutral_carbon}}

\begin{figure}
\includegraphics [width=\columnwidth]{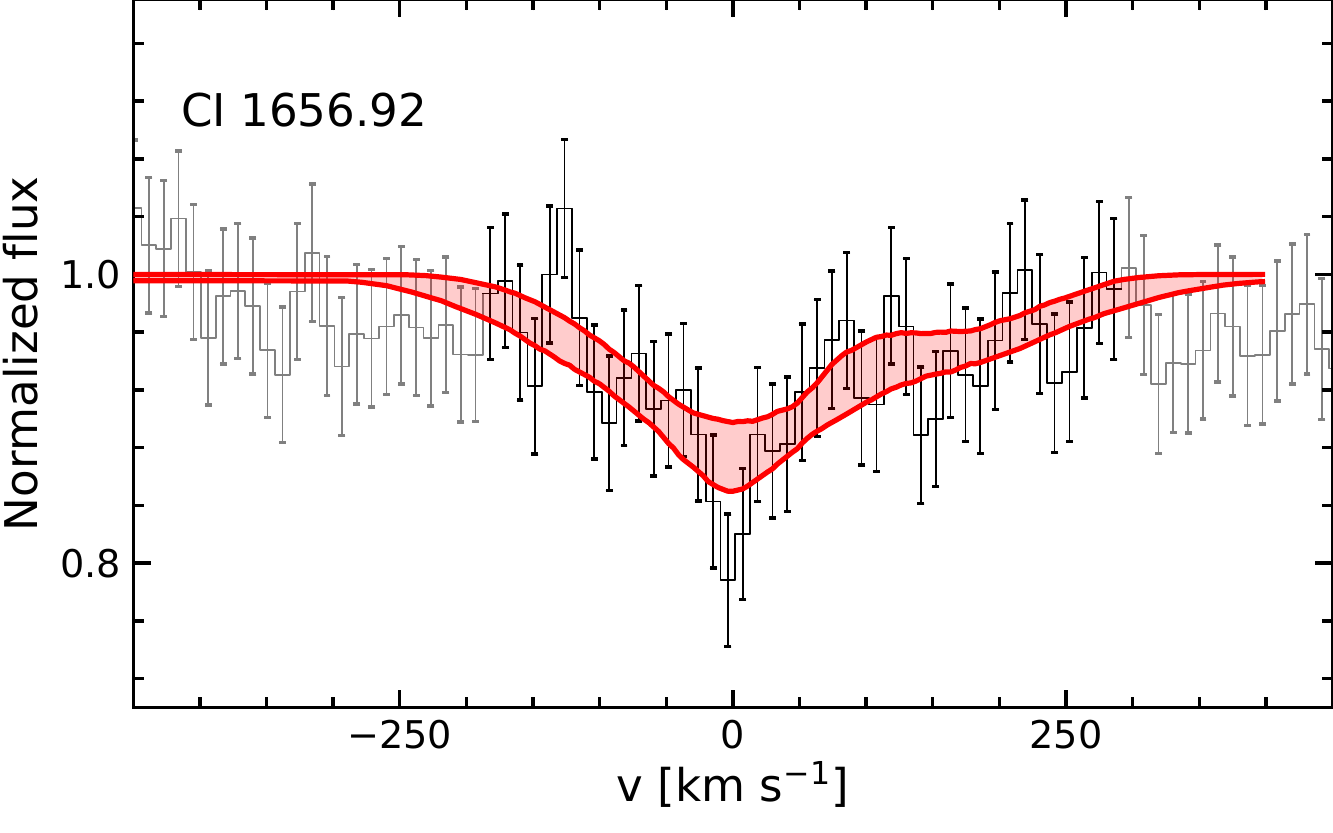}
\caption{Fit to \CI\ lines in J\,2359$+$1354. The red region indicates the joint profile with corresponded 1$\sigma$ uncertainty of the mutually blended \CI\ transitions around $1656$\,\AA\ from the three fine-structure levels. The x-axis shows the relative velocity with respect to the \CI\ $J=0$ transition.}
\label{fig:J2359_CI}
\end{figure}

We searched for the presence of neutral carbon \CI\ lines in our X-shooter spectra of ESDLAs since this species is known to be a good tracer of molecular hydrogen \citep[e.g.][]{srianand2005, Noterdaeme2018}. We detected \CI\ absorption lines only in the system towards J\,2359$+$1354, i.e. the absorber with the highest \HH\ column density and dust extinction. The Doppler parameter and redshift for the \CI\ lines were varied during the one-component fit independently from the low-ionization metals. We measured $\log N(\rm \text{\CI}^{\rm tot})\text{[cm$^{-2}$]} = 13.72\pm0.06$ at $z=2.2485$ from fitting the transitions \CI, \CI* and \CI** at $\sim\lambda$1656\AA\ (unfortunately, other lines are strongly blended), see Fig.~\ref{fig:J2359_CI}. This total column density is similar to what is seen in  other \HH-bearing DLAs  \citep[e.g.][]{Klimenko2020}. 

\subsection{Dust extinction \label{subsec:extinction_measurement}}

We estimated the extinction due to the presence of dust in the intervening absorption system by fitting the observed spectrum with a quasar template spectrum shifted to the respective quasar emission redshift and reddened with the empirical extinction law for the Small Magellanic Cloud \citep{Gordon2003} applied at $z_{\rm abs}$ \citep[see e.g.][]{Srianand2008b}. We used the X-shooter composite spectrum by \citet{Selsing2016} as a reference for the unabsorbed quasar spectrum (except of J\,1353$+$0956 for which we used a composite spectrum from HST data by \citealt{Telfer2002}). We caution that systematic uncertainties prevail over the statistical ones due to the unknown intrinsic shape of the quasar continuum. Therefore, we present in this paper only systematic errors of the $A_{\rm V}$, estimated from the typical dispersion of intrinsic shapes of quasars \cite[see e.g.][]{Noterdaeme2017,Ranjan2020}. 
For six out of seven quasars the measured extinction is compatible with the systematic uncertainties. However, for the system with the highest metallicity and \HH\ column density, namely J\,2359$+$1354, the estimated extinction is statistically significant $A_{\rm V}\approx0.3$. This value is in agreement with the measurements of dust extinction for \CI-selected systems from \citep{Zou2018} with mean $A_{\rm V}\approx0.2$, as well as with overall $A_{\rm v}-N({\rm Fe^{dust}})$ trend, obtained for both local and high-redshift ISM \citep[e.g.][]{Zou2018}.
Additionally, we estimated the extinction by scaling with metallicity the gas-to-dust ratio, obtained for the Milky Way from X-ray absorption of the GRB afterglows \citep{Watson2011}: 
\begin{equation}\label{eq:N(A_v)}
    A^{\rm scale}_{\rm V} = 3.0\times 10^{-22} \text{cm}^{2} N(\rm \text{\HI}) 10^{[\rm Zn/H]}(1-10^{\rm[Fe/Zn]}).
\end{equation}

In Eq.~\ref{eq:N(A_v)} we implicitly take into account that metallicity from \citet{Anders1989}, used to derived relation by \citealt{Watson2011}, is about $\sim1.5$ times higher than that from \citet{Asplund2009}, used in this paper. Applying this equation to our sample we get typical values of $A^{\rm scale}_{\rm V} \lesssim 0.1$, i.e. below detection limit and hence in agreement with spectral measurements, with the exception of the ESDLA toward J\,2359$+$1354, where we obtain $A^{\rm scale}_{\rm V} \sim 0.8$. This value is higher than the extinction estimated directly from the spectrum, $A_{\rm V} \approx 0.3$. The reason of such disagreement can be caused by a difference in gas-to-dust ratio for the Small Magellanic Cloud and the Milky Way. 

\begin{figure*}
\includegraphics[trim=0.5cm 1.0cm 0.5cm 0.5cm,width=\textwidth]{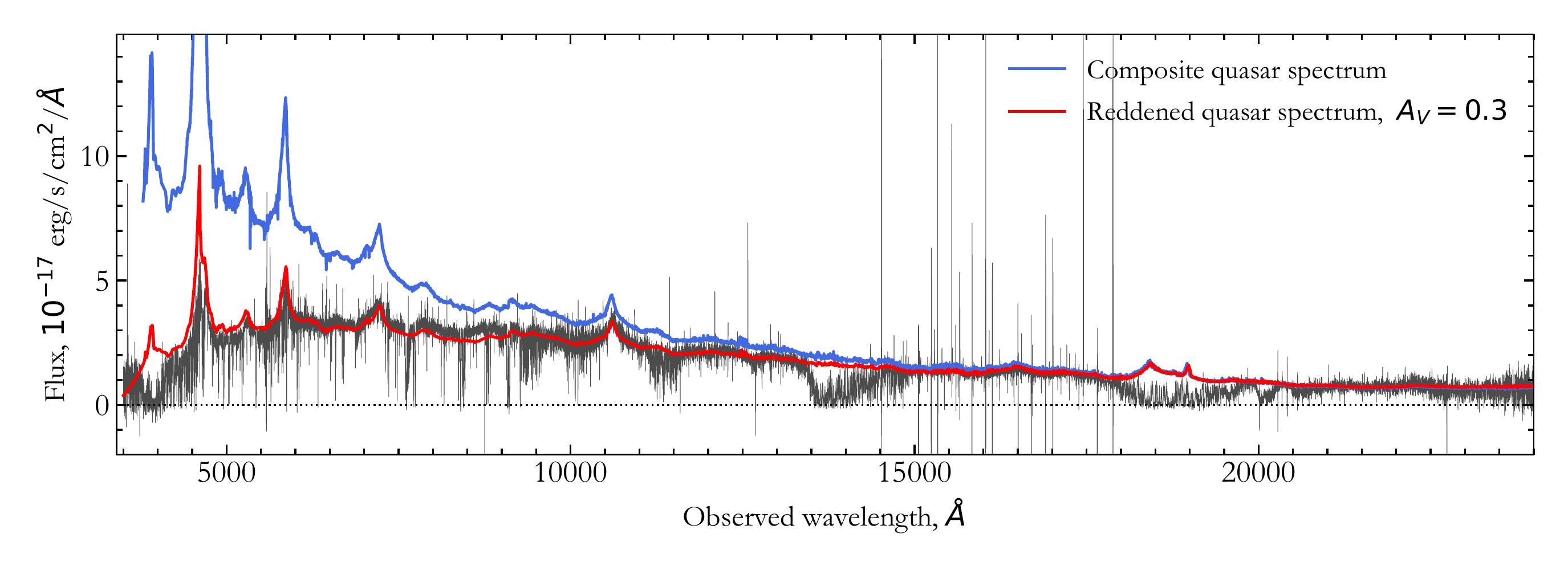}
\caption{Measurement of dust extinction towards the quasar J\,2359$+$1354. The observed X-shooter spectrum is shown in black and matched using a quasar composite spectrum from \citep{Selsing2016} reddened with SMC extinction law at $\zabs=2.25$ with best-fit value $\rm A_V = 0.3$ (red). For comparison, the unreddened quasar composite spectrum is shown in blue.}
\label{fig:J2359_av}
\end{figure*}

\section{Emission lines}

Since ESDLAs are likely arising from gas at small galacto-centric radii \citep{Ranjan2020}, we have searched for their emission counterparts. We searched for 
[\ion{O}{ii}]\,$\lambda$3727 and [\ion{O}{iii}]\,$\lambda$5007. We did not include H$\alpha$ 
as it is either not covered by our NIR spectra or located in the very noisy end of our spectra. 
We also only considered the final combined spectra as the data were obtained using nodding along the slit. Hence, all spatial information along the slit is lost and we only consider the region close to the quasar within 7 pixels above and below the spectral trace (corresponding to $\pm$1.5~arcsec).
For each line, we first subtracted the spectral point spread function (SPSF) of the quasar in order to search for faint emission lines from the foreground absorbing galaxy.  
To this end, we created an empirical SPSF model by median combining the spatial profile of the spectral trace in two regions\footnote{In total, we averaged over relative velocities from 800--2400~km~$^{-1}$ on either side.} on both sides of the expected position of the given emission line.

Next we median filtered and fitted a third order polynomial to the peak flux of the observed spatial profile as a function of wavelength. This provided us with a smooth function for the peak flux density as function of wavelength, $I_{\rm max}(\lambda)$. The empirical SPSF generated above was then rescaled to the peak flux density for a given wavelength, determined by $I_{\rm max}(\lambda)$ before being subtracted from the observed spatial profile at the given wavelength.

\begin{figure}
\includegraphics[trim=0.5cm 1.0cm 0.5cm 0.5cm,width=\columnwidth]{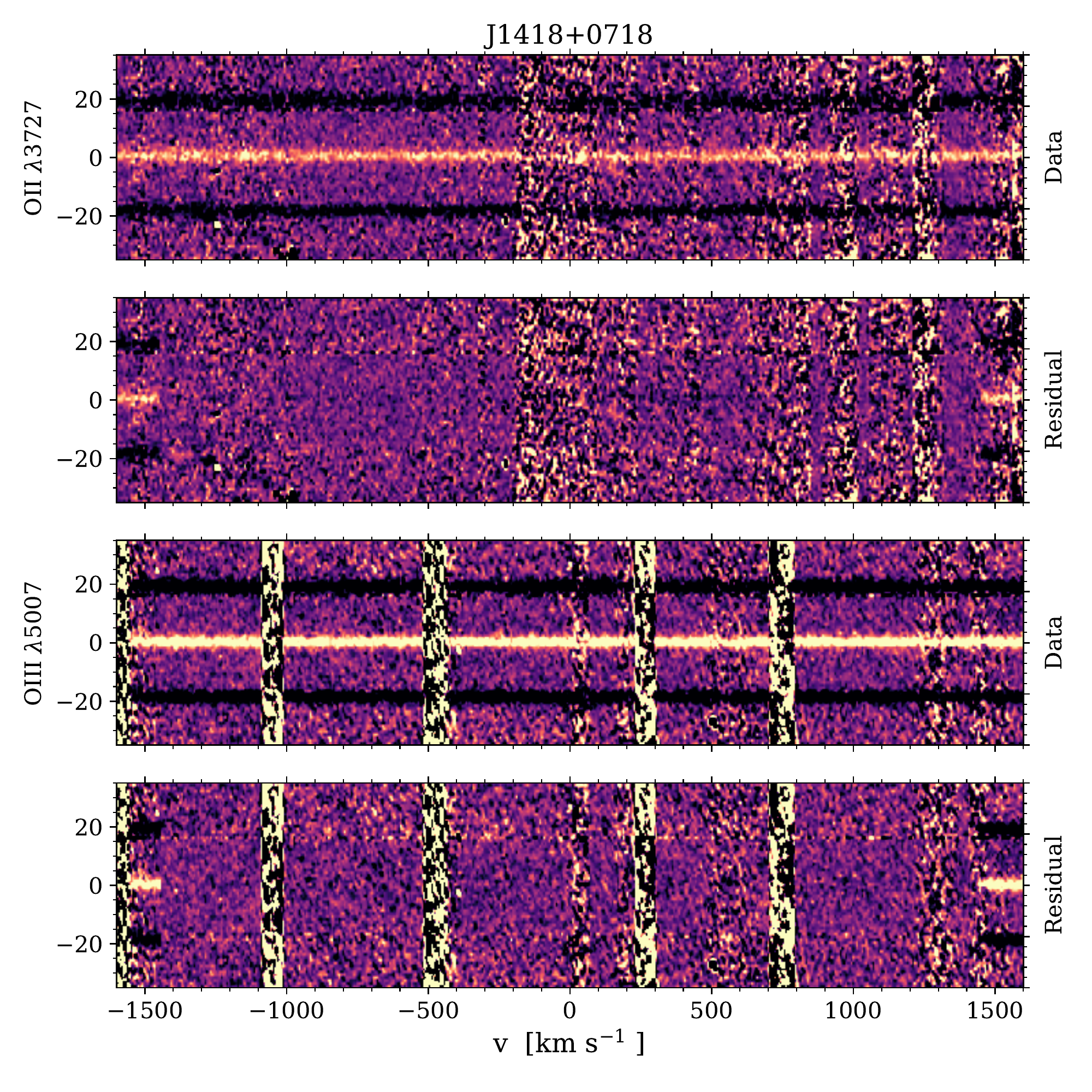}
\caption{Cutouts of  \ion{O}{ii}\,$\lambda$3727 and \ion{O}{iii}\,$\lambda$5007 emission regions from 2D spectrum of system towards J1418$+$0718 without (data) and with (residual) subtracted quasar trace. }
\label{fig:J1418_emission}
\end{figure}

The result is a quasar-subtracted cutout around each expected emission line. One example is shown in Fig.~\ref{fig:J1418_emission}. We do not observe any clear emission lines at the expected relative velocities for \ion{O}{ii}\,$\lambda$3727 and \ion{O}{iii}\,$\lambda$5007. For J\,1418$+$0718 we see some tentative evidence for a faint emission peak at $\sim90$~\kms. However, the line is in a region of rather bright sky emission lines, and should be considered highly uncertain.
Both emission lines for J\,0024$-$0725 fall in regions of very strong and saturated telluric absorption lines, which are not possible to correct, and will therefore not be considered further.

We hence measured upper limits on the emission line fluxes using 500 random apertures displaced along the dispersion axis within $\pm250$~\kms and within $\pm$3 pixels along the spatial axis. Each aperture is a rectangular box 400~\kms wide and $2\times {\rm FWHM_{spatial}}$ high. The spatial FWHM is measured from the model SPSF itself. For our targets the spatial FWHM is in the range of 2--3 pixels (at a scale of 0.21~arcsec per pixel). The upper limits are quoted at 3$\sigma$ using the standard deviation of the distribution of individual aperture fluxes. The limits on line fluxes and estimated star-formation rates based on O\,{\sc ii}\,$\lambda$3727 are summarized in Table~\ref{tab:emission_lines}. The constraints on the star-formation rates are in agreement with those of the ESDLAs with detected emission counterparts \citep{Moller2004,Kulkarni2012,Noterdaeme2012b,Ranjan2020}.

\begin{table}
\caption{Summary of line flux measurements.}
\label{tab:emission_lines}
\begin{tabular}{cccc}
 \hline

Quasar & F([O\,{\sc ii}]\,$\lambda$3727) & SFR$^{(a)}$ & F([O\,{\sc iii}]\,$\lambda$5007) \\
 & erg~s$^{-1}$~cm$^{-2}$ & M$_{\odot}$~yr$^{-1}$ & erg~s$^{-1}$~cm$^{-2}$ \\
 \hline
J\,0024$-$0725 & $ < 4.4 \times 10^{-17}$ & $ < 24 $ & $ < 9.7 \times 10^{-17}$ \\
J\,1238$+$1620 & $ < 8.7 \times 10^{-18}$ & $ < 7  $ & $ < 1.7 \times 10^{-16}$ \\
J\,1353$+$0956 & $ < 2.3 \times 10^{-17}$ & $ < 21 $ & $ < 1.2 \times 10^{-16}$ \\
J\,1418$+$0718 & $ < 1.7 \times 10^{-16}$ & $ < 71 $ & $ < 2.6 \times 10^{-17}$ \\
J\,2205$+$1021 & $ < 1.5 \times 10^{-17}$ & $ < 13 $ & $ < 2.9 \times 10^{-17}$ \\
J\,2351$-$0639 & $ < 5.5 \times 10^{-17}$ & $ < 26 $ & $ < 4.2 \times 10^{-16}$ \\
J\,2359$+$1254 & $ < 4.1 \times 10^{-17}$ & $ < 15 $ & $ < 2.2 \times 10^{-17}$ \\
 \hline
\end{tabular}

{\flushleft
$^{(a)}$ Star-formation rate estimated based on the line luminosity of [O\,{\sc ii}]\,$\lambda$3727 following the calibration by \citet{Kennicutt1998}.
}
\end{table}

\section{Discussion}\label{sec:Discussion}

\begin{figure}
    \includegraphics [width=\columnwidth]{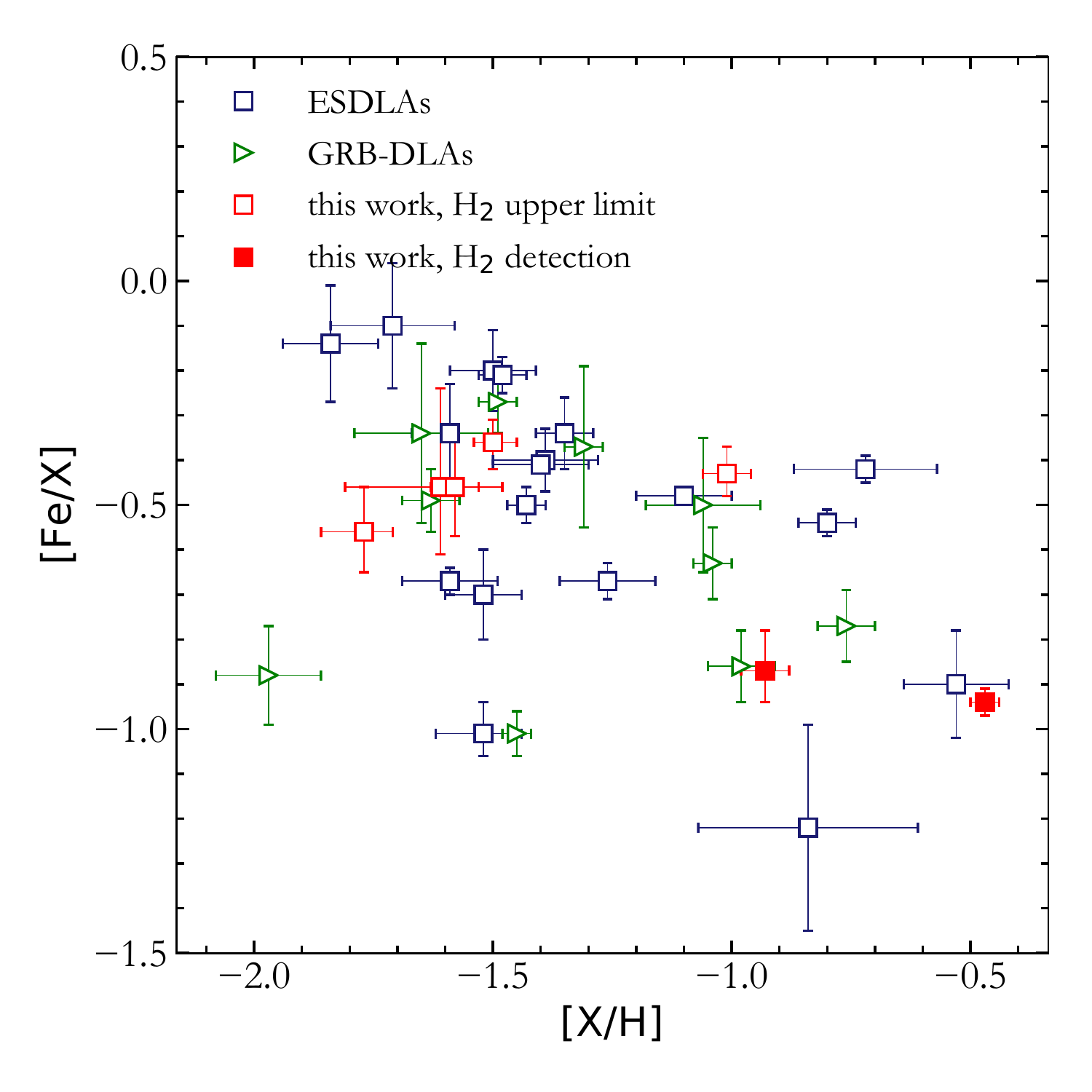}
    \caption{Metallicity vs iron depletion. `X' corresponds to the volatile element Zn or S, depending on availability for each system. ESDLAs from this work are shown by open (\HH\ upper limits) and filled (\HH\ detections) red squares. Blue squares correspond to ESDLAs from \citet{Guimaraes2012,Noterdaeme2015,Ranjan2018,Ranjan2020}. GRB-DLAs from \citet{Bolmer2019} are shown by green triangles. The difference in the adopted solar abundances in the literature samples is within the metallicity uncertainties, so we do not address to it here.}
    \label{fig:depletion-metal}
\end{figure}

\begin{figure}
\includegraphics [width=\columnwidth]{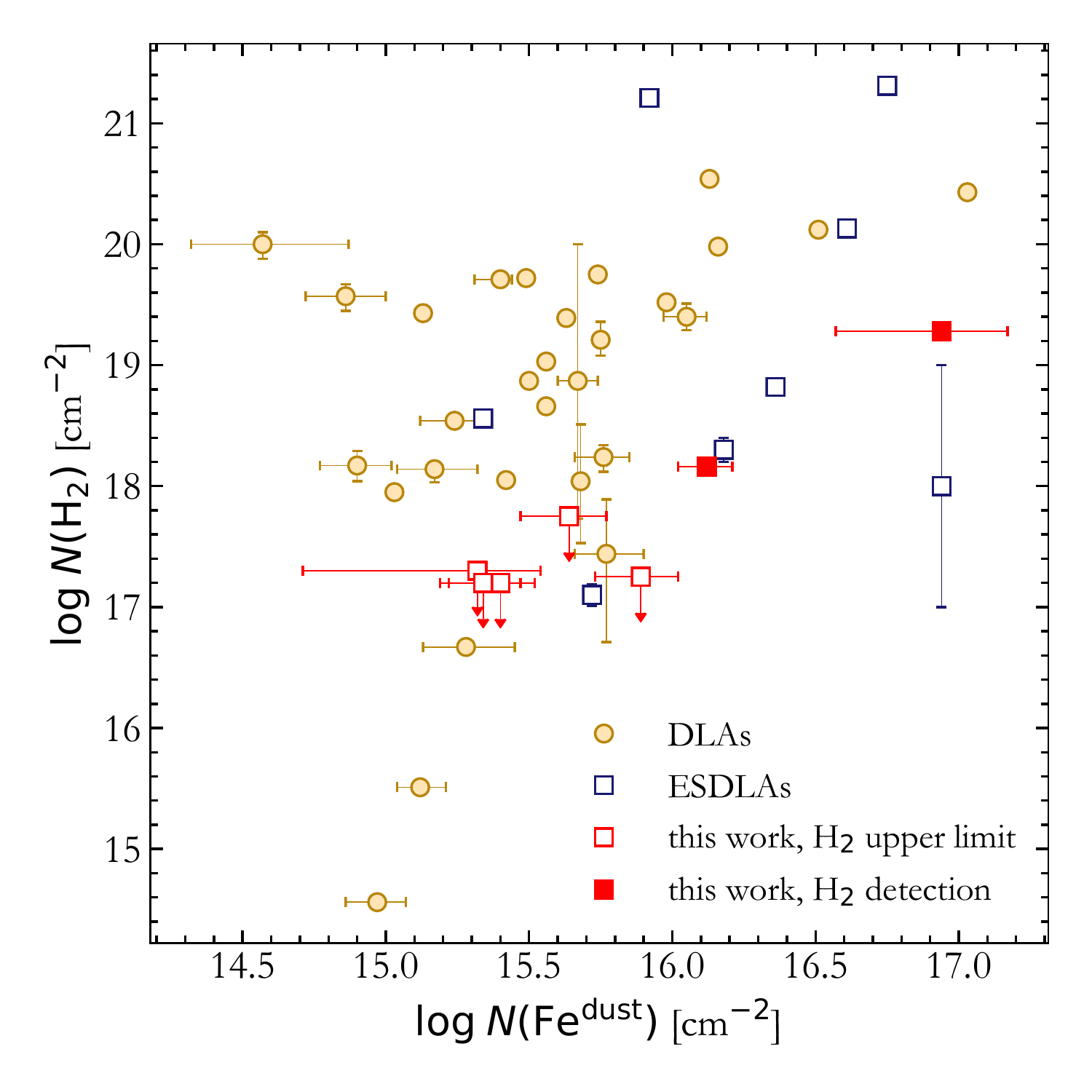}
\caption{Column density of iron depleted into the dust vs \HH\ column density. ESDLAs from this work are shown by red open (\HH\ upper limits) and filled (\HH\ detections) squares. Blue squares and brown circles are ESDLA and DLA systems compiled in  \citet{Balashev2019}, respectively.}
\label{fig:H2-N_Fe_dust}
\end{figure}

From the observed properties of individual ESDLAs toward quasar sightlines, it has been suggested that they arise from similar environments as DLAs observed from GRB afterglow spectra \citep{Noterdaeme2015}. 
Measurements based on about a two dozen of ESDLAs and a dozen of GRB-DLAs further support this idea \citep{Bolmer2019, Ranjan2020}. Therefore, below we compare the properties of our seven systems with a sample of ESDLAs and GRB-DLAs from the literature. Although our sample of ESDLAs is located in the same region in the metallicity--depletion diagram (Fig.~\ref{fig:depletion-metal}\footnote{For ESDLA system toward J\,2246$+$1328 \citep{Ranjan2020} we plot revisited metallicity and depletion from \citet{Ranjan2021submitted}}), three of them have relatively high metallicities ${\rm[Zn/H]} \gtrsim -1$ compared to the average of ESDLAs. 
We detected \HH\ with $\log N(\rm H_2)\text{[cm$^{-2}$]} > 18$ in those two systems with highest depletion factor, which is consistent with study by \citet{Ledoux2003}. 
Since \HH\ formation is connected with the amount of dust, one may expect a correlation between the \HH\ column density with the column density of iron depleted into dust, i.e. the dust fraction \citep{Noterdaeme2008}. In Fig.~\ref{fig:H2-N_Fe_dust} we plot column densities of iron in the dust-phase versus \HH\ column densities. Our sample of ESDLAs are consistent with previous findings, i.e., the incidence rate of \HH\ in DLAs gradually increases for systems with $\log N(\rm Fe^{dust})\text{[cm$^{-2}$]} > 15$ \citep{Noterdaeme2008, Balashev2019}. Indeed, the two ESDLAs from our sample with strong \HH\ absorption have the highest column densities of iron locked into dust, while the other five with upper limits on \HH\  have $\log N(\rm Fe^{dust})\text{[cm$^{-2}$]} < 16$. 
While \HH-bearing systems with $\log N(\HH)\text{[cm$^{-2}$]} > 18$ exhibit a moderate correlation between \HH\ and Fe$^{\rm dust}$ column density (Pearson correlation coefficient $\approx 0.4$), the unknown selection functions does not 
permit a strict conclusion. 
Fig.~\ref{fig:H2-N_Fe_dust} indicates that ESDLAs with \HH\ detection have relatively high $N(\rm Fe^{dust})$ in comparison with other \HH-bearing DLAs. This is evidently explained by the large columns probed by ESDLAs.
ESDLA systems with large $\log N(\rm Fe^{dust})\text{[cm$^{-2}$]} > 16$ have much larger dispersion in the \HH$-$Fe$^{\rm dust}$ diagram, which could be due to
a larger  
number of metal components with different physical conditions along the line of sight, 
whereas \HH\ column densities are most probably determined on smaller scales, i.e. by local physical conditions. 
Since ESDLAs likely probe more central parts of the absorbing galaxies, the corresponding lines of sight are expected to cross regions with more diverse physical conditions 
than regular DLAs that probe predominantly the outskirts of galaxies, resulting in larger dispersion of integrated properties for ESDLAs.

\section{Summary}\label{sec:conclusions}

In this work, we have presented a detailed analysis of seven new ESDLAs. For all systems we have estimated the \HI\ column density, metallicity and dust depletion, dust extinction, and the column density of iron locked up in dust grains. For two out of seven, we have detected strong ($\log N(\text{\HH})\text{[cm$^{-2}$]}>18$) molecular hydrogen absorption. These two systems also demonstrate the highest column density of iron in dust grains, which indicates a significant integrated amount of dust in the corresponding systems. For the ESDLA towards J\,2359$+$1354, which has the strongest \HH\ absorption and highest extinction $A_{\rm V}$, we additionally detected neutral carbon, which is known to be a good tracer of \HH, as well as \SiII* absorption, which may be used to estimate the electron density. Overall, the \HH\ incidence rate of $10-55$\% for ESDLAs in our sample is in agreement with the \HH\ incidence rate observed in GRB-DLAs ($\sim 30$\%, \citealt{Bolmer2019}). This is significantly higher than the incidence rate of the overall population of intervening quasar DLAs ($5-10$\%). 

Combining the sample presented in this work with the sample by \citet{Ranjan2020}, we discuss the measurements of \CII*, which can be used to estimate the [\CII] cooling rate in the ISM. The distribution of the inferred cooling rates for our combined sample of ESDLAs presents the same bimodality as observed in DLA systems \citep{Wolfe2008} with much smaller \HI\ column densities (see Fig.~\ref{fig:cooling_rate}). A more in-depth discussion of the origin of this bimodality is presented by \citet{Balashev2021}.

In order to study the star formation activity in the absorbing galaxy, we have searched for emission lines from [\ion{O}{ii}]\,$\lambda$3727 and [\ion{O}{iii}]\,$\lambda$5007. We do not find any detections of these lines and report only upper limits to the line-fluxes (assuming that the emission counterpart falls inside the slit aperture). The upper limits that we infer are rather high and in full agreement with the overall low star-formation rates observed for ESDLAs with positive detections of the emission counterpart.

\section{Data Availability}
The results of this paper are based on data retrieved from the ESO telescope archive. This data can be shared on reasonable request to the authors.

\section*{Acknowledgements}
This work is partially supported by RFBR grant 18-02-00596. 
KT and SB are supported by the Foundation for the
Advancement of Theoretical Physics and Mathematics ``BASIS''. 
We acknowledge support from the French {\sl Agence Nationale de la Recherche} 
under ANR grant 17-CE31-0011-01 / project ``HIH2'' (PI: Noterdaeme) and from the French-Russian collaborative programme PRC 1845. AR acknowledges support from the National Research Foundation of Korea (NRF) grant funded by the Ministry of Science and ICT (NRF-2019R1C1C1010279). AR also acknowledges the support of the Indo-French Centre for the Promotion of Advanced Research (Centre Franco-Indien pour la Promotion de la Recherche Avanc\'ee) under contract no. 5504-2 (PIs Gupta and Noterdaeme) during the writing of the proposal for the ESO-VLT observations.
JKK acknowledges support from the Swiss National Science Foundation under grant 185692.
The authors are grateful to the European Southern Observatory (ESO) and in particular the Paranal observatory's staff for carrying out their observations in service mode. 

\bibliographystyle{mnras}
\bibliography{references.bib}

\bsp	
\clearpage
\appendix

\section{Spectroscopic analysis}

\subsubsection{J\,0024$-$0725}

\begin{figure}
\includegraphics [width=\columnwidth]{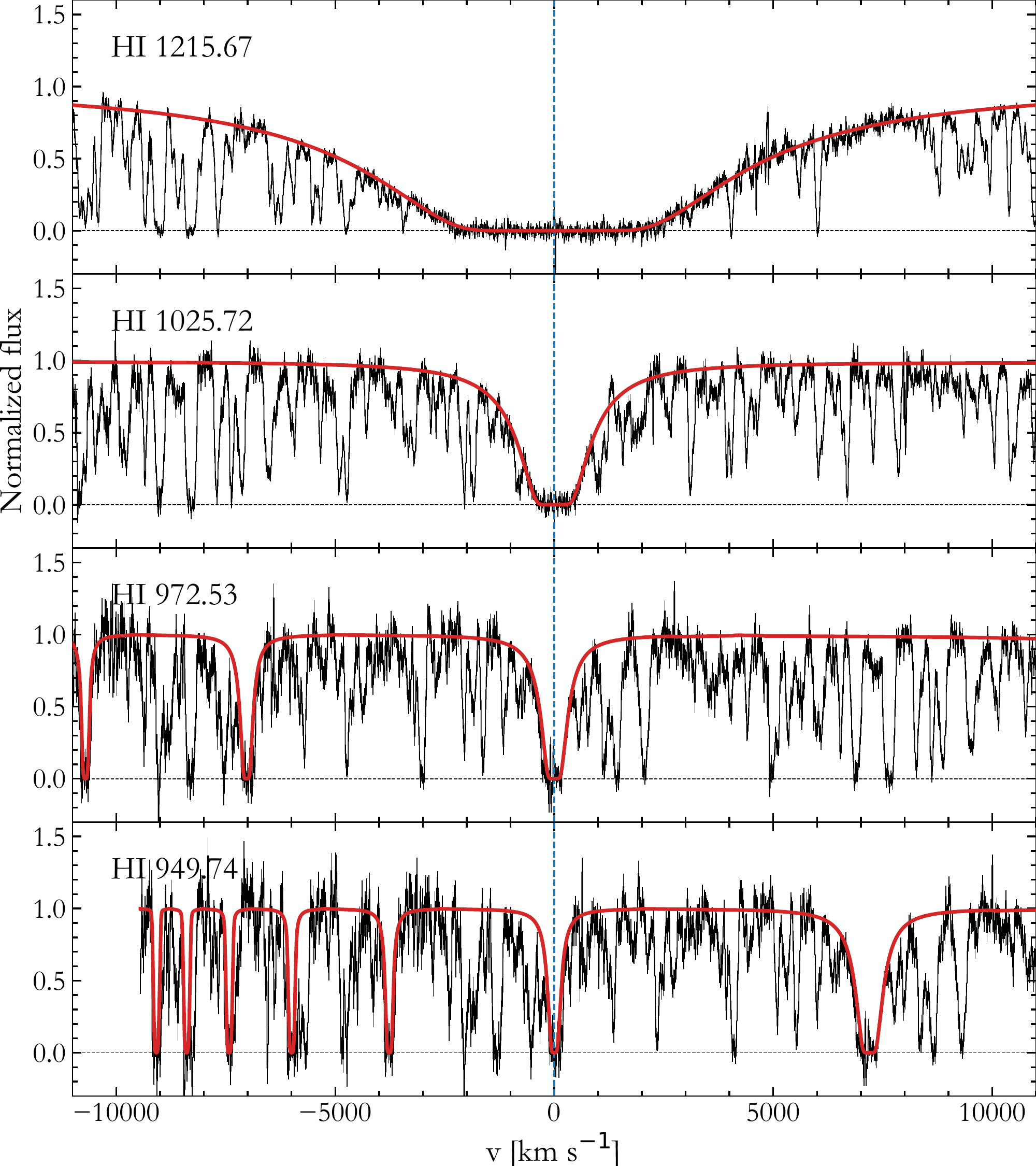}
\caption{Fit to \HI\ absorption lines at z$\sim$2.681 towards J\,0024-0725. The red line presents the profiles of the labeled \HI\ transitions. The vertical dashed line indicates the centers of \HI\ absorption lines.}
\label{fig:J0024_HI}
\end{figure}

\begin{figure*}
\includegraphics [width=0.94\textwidth]{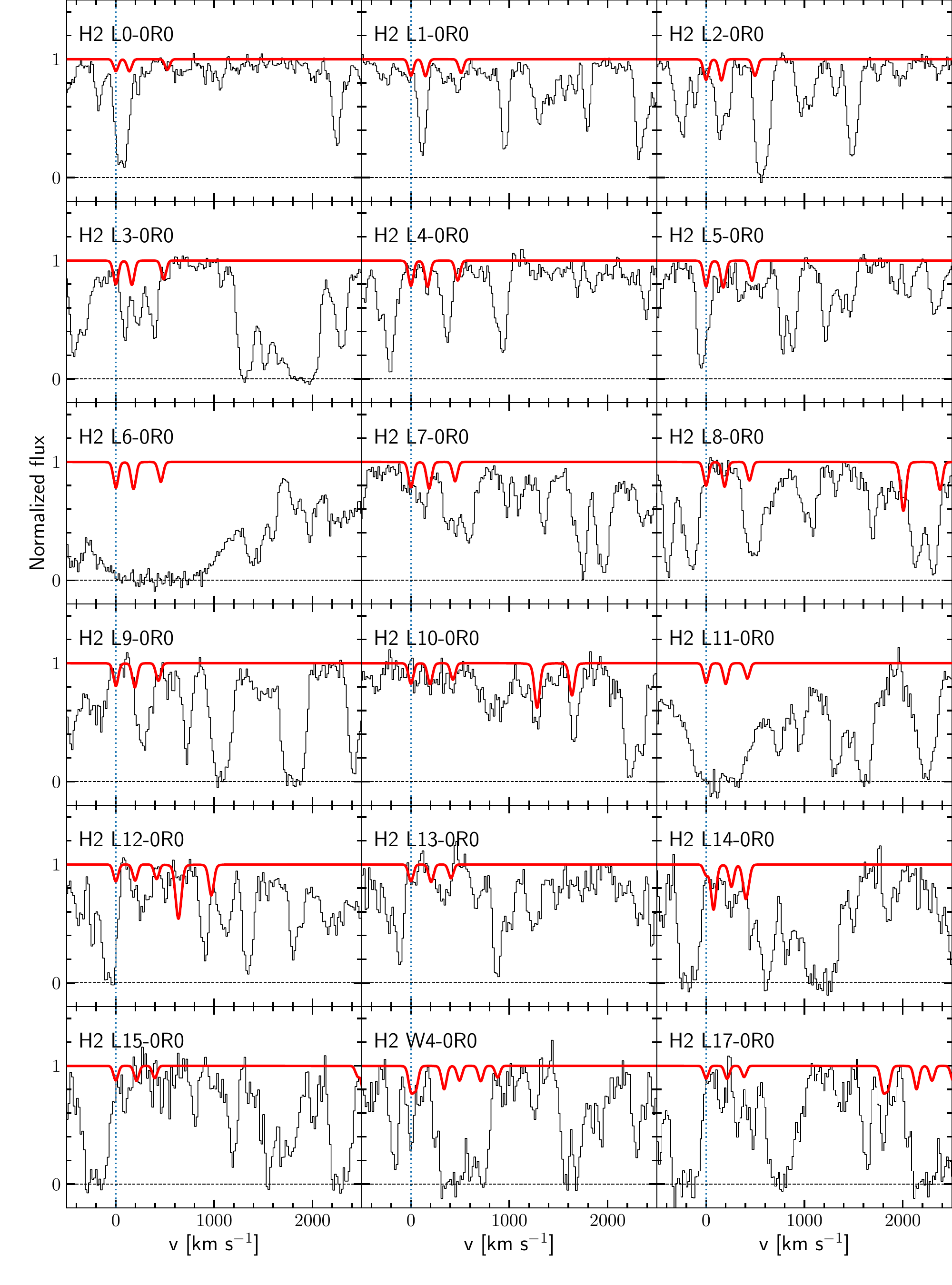}
\caption{The regions of J\,0024$-$0725 spectrum corresponding to the expected position of H$_2$ absorption lines associated with ESDLA at z$\sim$2.681. Each panel corresponds to a particular band of H$_2$ transitions. The red line presents the profile of H$_2$ absorption lines used to obtain an upper limit on the H$_2$ column density. The blue vertical dashed lines indicate the positions of R0 H$_2$ transition, which are in agreement with the positions of ESDLA metal transitions.}
\label{fig:J0024_H2}
\end{figure*}

\begin{figure*}
\includegraphics [width=\textwidth]{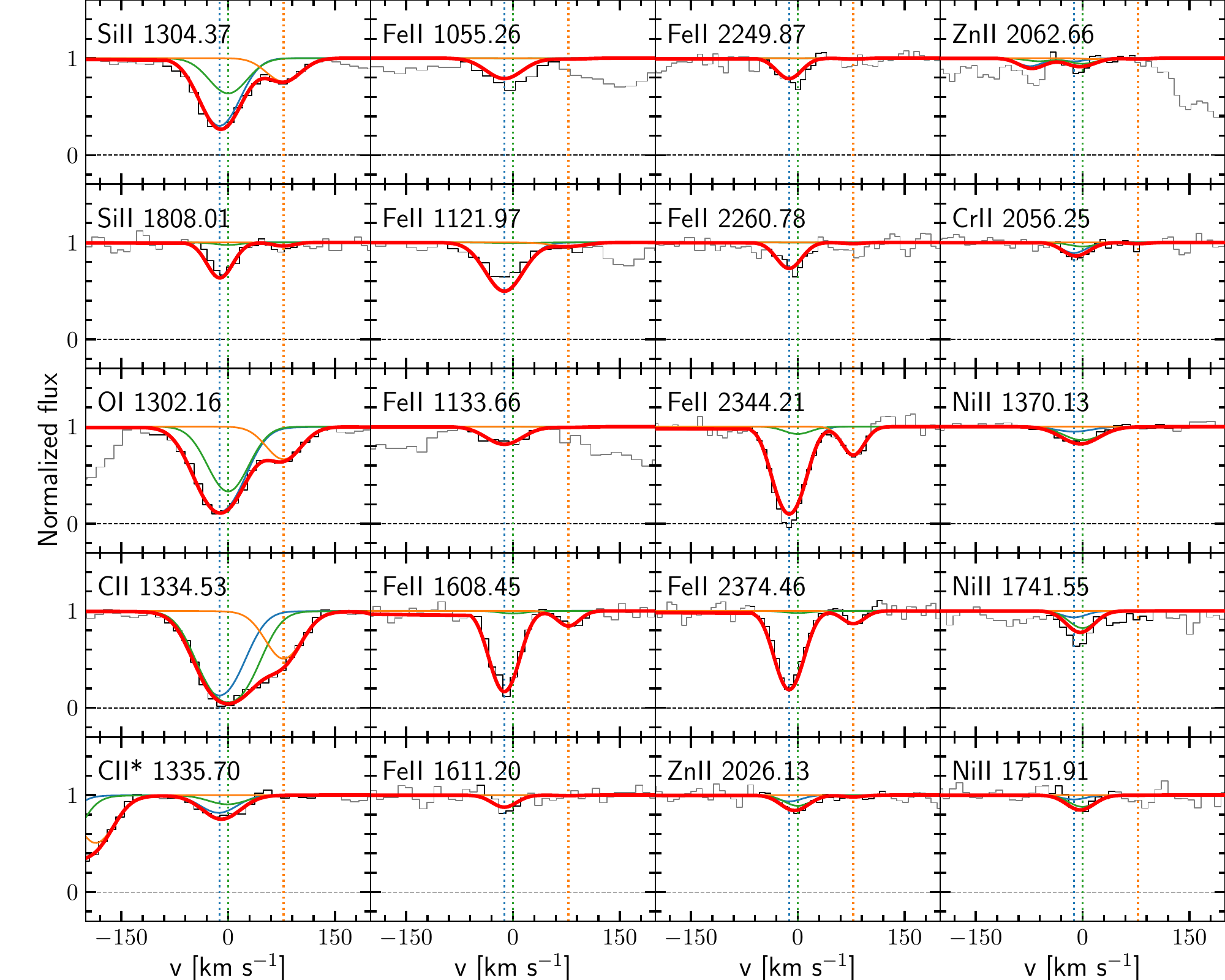}
\caption{Fit to metal absorption lines at z$\sim$2.681 towards J\,0024-0725. The red line presents the total profiles of the labeled metal transitions. The vertical dashed lines indicate the relative positions of individual components of the fit.
}
\label{fig:J0024_me}
\end{figure*}

The fitted \HI\ Ly-series absorptions are shown in Fig.~\ref{fig:J0024_HI}. The \HH\ Voigt profile, corresponded to the upper limit on \HH\ column density, is show in Fig.~\ref{fig:J0024_H2}. The details of the fit to the metal lines are presented in Table~\ref{tab:fit_me_J0024_mcmc}. Point estimates correspond to the mode of the posterior distribution of the parameters and uncertainties correspond to the 68\% credible intervals. The total column densities on \OI\ and \CII\ are the lower limits due to the high saturation of the corresponding transitions. The fitted metal line Voigt-profiles calculated for the best-fit values are shown in Fig.~\ref{fig:J0024_me}.

\begin{table*}
\caption{Fit results for metal lines at z$\sim$2.681 towards  J\,0024$-$0725.}
\label{tab:fit_me_J0024_mcmc}
\begin{tabular}{ccccccc}
\hline 
comp & 1 & 2 & 3 & $\log N_{\rm tot}$ & $\rm [X/H]$ & $\rm [X/ZnII]$ \\
\hline
$z$ & $2.681026(^{+5}_{-8})$ & $2.681148(^{+31}_{-22})$ & $2.682114(^{+14}_{-18})$ &  &  &  \\
$b$, km~s$^{-1}$& $13.2^{+0.5}_{-0.5}$ & $15.1^{+1.9}_{-1.6}$ & $5.0^{+0.7}_{-0.9}$ &  &  &  \\
$\log N$(SiII) & $15.39^{+0.06}_{-0.08}$ & $13.97^{+0.41}_{-0.95}$ & $14.12^{+0.35}_{-0.13}$ & $15.42^{+0.09}_{-0.05}$ & $-1.90^{+0.09}_{-0.05}$ & $-0.13^{+0.12}_{-0.08}$ \\
$\log N$(FeII) & $14.94^{+0.07}_{-0.04}$ & $11.18^{+1.06}_{-0.18}$ & $13.53^{+0.10}_{-0.14}$ & $14.98^{+0.04}_{-0.06}$ & $-2.33^{+0.04}_{-0.06}$ & $-0.56^{+0.10}_{-0.09}$ \\
$\log N$(CrII) & $13.09^{+0.20}_{-0.47}$ & $12.99^{+0.24}_{-0.73}$ & $11.21^{+0.94}_{-0.11}$ & $13.20^{+0.10}_{-0.13}$ & $-2.25^{+0.10}_{-0.13}$ & $-0.48^{+0.13}_{-0.15}$ \\
$\log N$(ZnII) & $12.20^{+0.28}_{-0.62}$ & $12.48^{+0.11}_{-0.25}$ & $11.21^{+0.67}_{-0.09}$ & $12.60^{+0.06}_{-0.09}$ & $-1.77^{+0.06}_{-0.09}$ & $0$ \\
$\log N$(OI) & $17.37^{+0.25}_{-0.23}$ & $15.10^{+0.48}_{-1.48}$ & $15.34^{+0.55}_{-0.50}$ & $17.37^{+0.26}_{-0.20}$ & $-1.13^{+0.26}_{-0.20}$ & $0.64^{+0.28}_{-0.21}$ \\
$\log N$(CII) & $16.77^{+0.45}_{-0.54}$ & $17.63^{+0.39}_{-0.73}$ & $17.11^{+0.25}_{-0.62}$ & $17.61^{+0.33}_{-0.26}$ & $-0.63^{+0.33}_{-0.26}$ & $1.14^{+0.34}_{-0.27}$ \\
$\log N$(CII*) & $13.65^{+0.09}_{-0.15}$ & $13.27^{+0.32}_{-1.31}$ & $10.05^{+0.47}_{-0.05}$ & $13.66^{+0.06}_{-0.05}$ & $-4.58^{+0.06}_{-0.05}$ & $-2.81^{+0.10}_{-0.08}$ \\
$\log N$(NiII) & $13.71^{+0.14}_{-0.52}$ & $13.75^{+0.13}_{-0.38}$ & $12.83^{+0.21}_{-1.93}$ & $13.86^{+0.05}_{-0.05}$ & $-2.17^{+0.05}_{-0.06}$ & $-0.40^{+0.10}_{-0.08}$ \\
\hline
\end{tabular}
\end{table*}

\subsubsection{J\,1238$+$1620}
\begin{figure}
\includegraphics [width=\columnwidth]{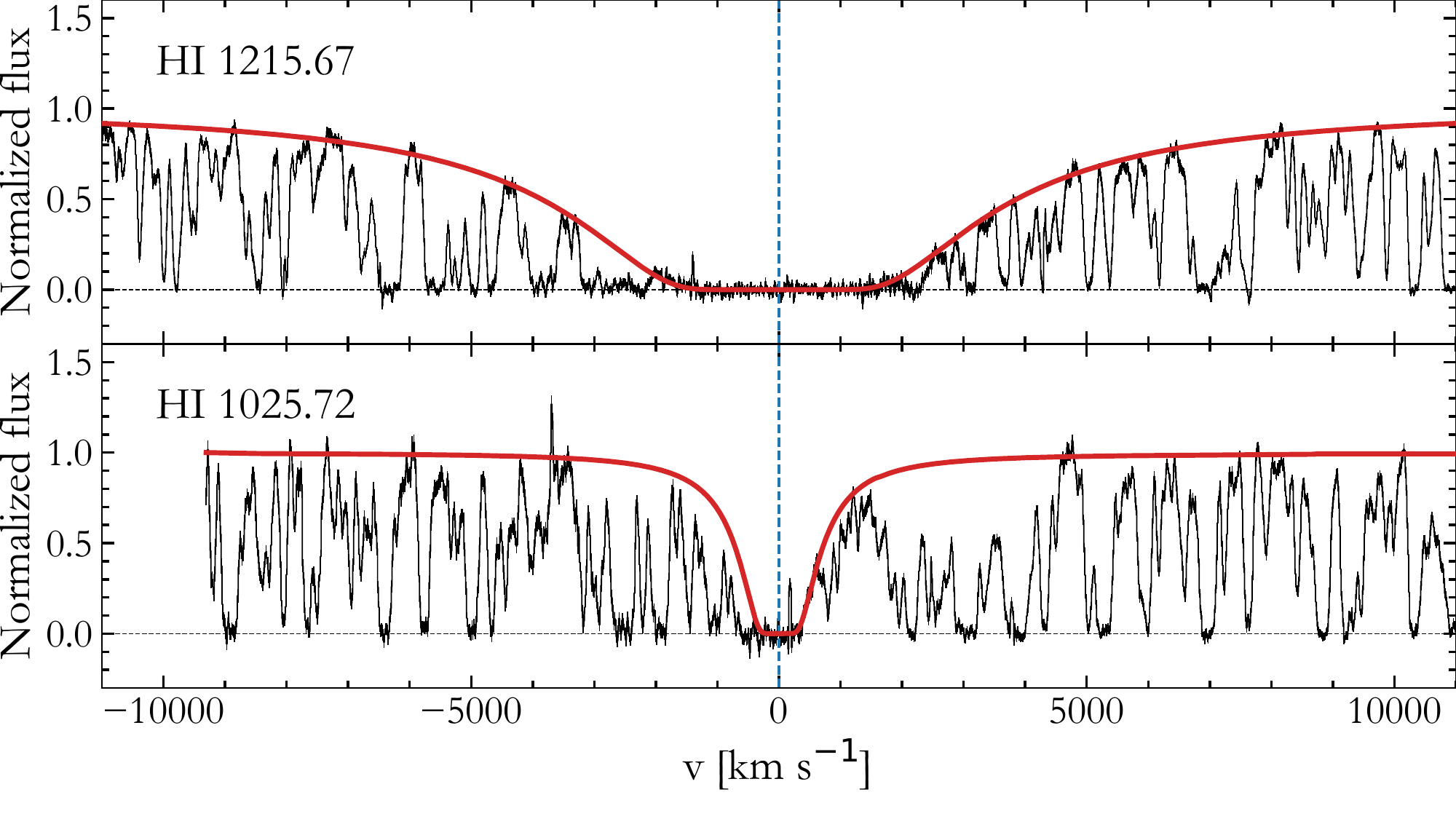}
\caption{Fit to \HI\ absorption lines at z$\sim$3.209 towards J\,1238+1620. The red line presents the profiles of the labeled \HI\ transitions. The vertical dashes line indicates the centers of \HI\ absorption lines.}
\label{fig:J1238_HI}
\end{figure}

\begin{figure*}
\includegraphics [width=\textwidth]{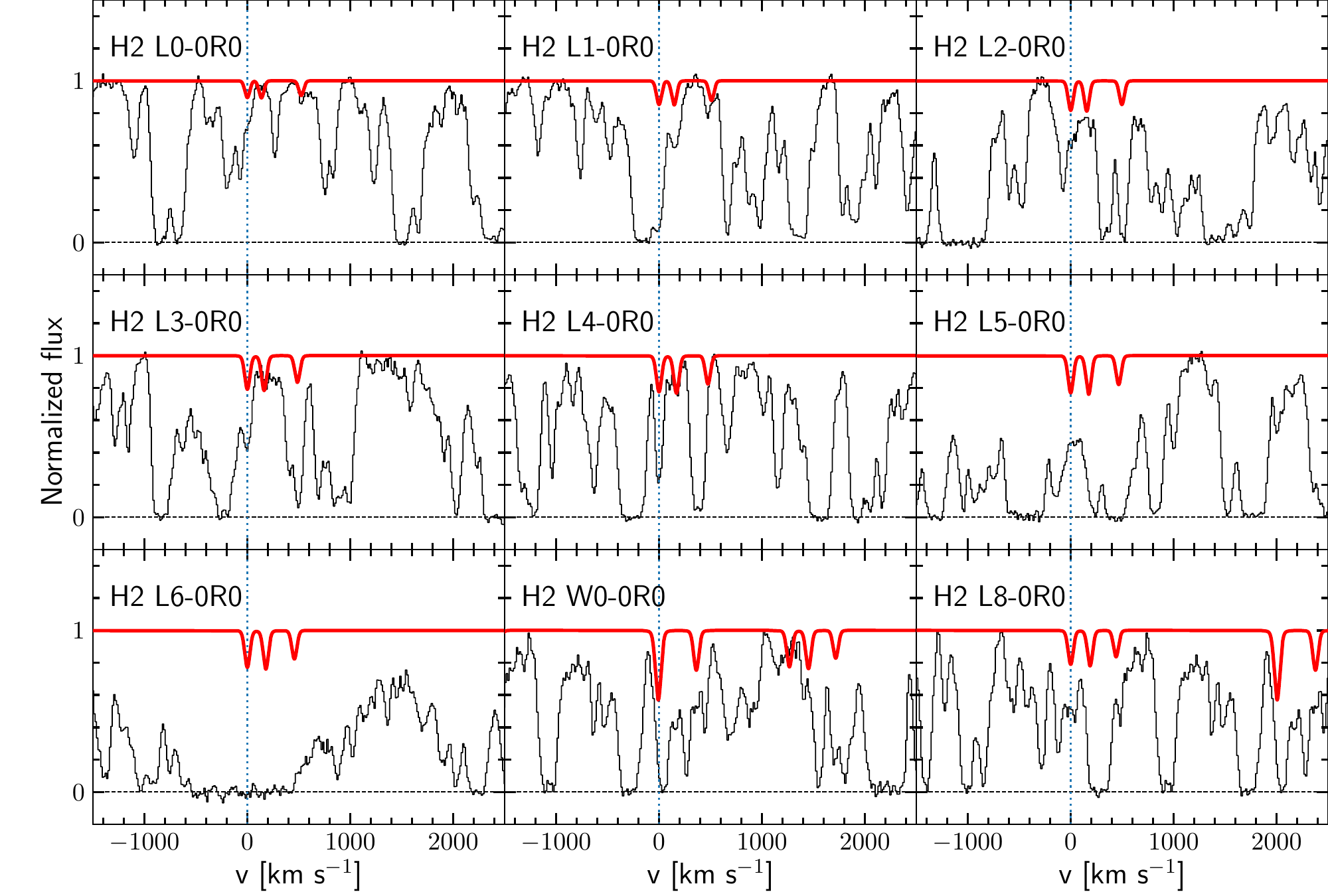}
\caption{The regions of J\,1238$+$1620 spectrum corresponding to the expected position of H$_2$ absorption lines associated with ESDLA at z$\sim$3.209. Each panel corresponds to a particular band of H$_2$ transitions. The red line presents the profile of the H$_2$ absorption lines used to obtain an upper limit on the H$_2$ column density. The blue vertical lines indicate the positions of the R0 H$_2$ transition, which are in agreement with the positions of the metal lines.}
\label{fig:J1238_H2}
\end{figure*}

\begin{figure*}
\includegraphics [width=\textwidth]{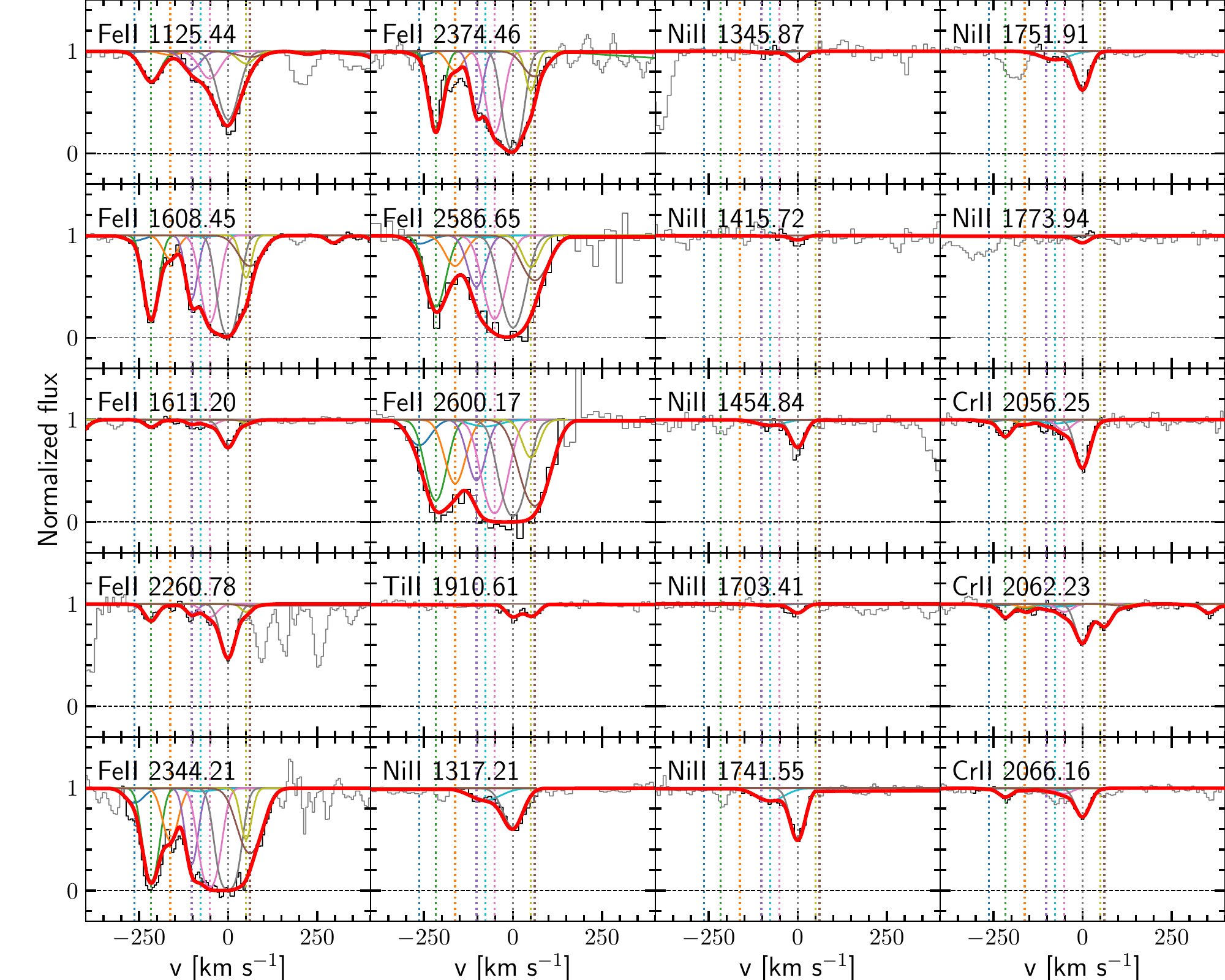}
\caption{Fit to metal absorption lines at z$\sim$3.209 towards J\,1238+1620. The red line presents the total profiles of the labeled metal transitions. The vertical dashed lines indicate the relative positions of individual components of the fit.} 
\label{fig:J1238_me}
\end{figure*}

The fitted \HI\ Ly-series absorptions are shown in Fig.~\ref{fig:J1238_HI}. The \HH\ Voigt profile, corresponded to the upper limit on \HH\ column density, is show in Fig.~\ref{fig:J1238_H2}. 
For this system, the \SiII\ $\lambda$1808\AA\ line is blended and all other \SiII\ lines are saturated, therefore, we did not include \SiII\ lines in the metal fit. 
The details of the fit to the metal lines are presented in Table~\ref{tab:fit_me_J1238}. The best-fit metal line Voigt-profiles are shown in Fig.~\ref{fig:J1238_me}.

\begin{table*}
\caption{Fit results of metal lines at z$\sim$3.209 towards J\,1238$+$1620.}
\label{tab:fit_me_J1238}
\begin{tabular}{ccccccccccccc}
\hline 
comp & $z$ & $b$, km~s$^{-1}$ & $\log N$(FeII) & $\log N$(CrII) & $\log N$(ZnII) & $\log N$(NiII) & $\log N$(TiII) \\
\hline
1 & $3.20550(^{+21}_{-10})$ & $25.6^{+24.0}_{-11.5}$ & $13.11^{+0.22}_{-0.13}$ & $12.50^{+0.25}_{-0.49}$ & $11.27^{+0.48}_{-0.26}$ &  &  \\
2 & $3.206279(^{+11}_{-10})$ & $14.8^{+0.8}_{-1.2}$ & $14.74^{+0.02}_{-0.04}$ & $13.13^{+0.06}_{-0.05}$ & $12.26^{+0.18}_{-0.41}$ &  &  \\
3 & $3.20706(^{+4}_{-5})$ & $21.8^{+7.9}_{-4.3}$ & $13.75^{+0.07}_{-0.05}$ & $12.66^{+0.20}_{-0.24}$ & $11.25^{+0.50}_{-0.19}$ &  &  \\
4 & $3.207904(^{+16}_{-22})$ & $9.6^{+0.8}_{-1.4}$ & $14.55^{+0.07}_{-0.04}$ & $12.52^{+0.25}_{-0.77}$ & $11.88^{+0.18}_{-0.58}$ &  &  \\
5 & $3.20819(^{+12}_{-13})$ & $53.1^{+22.1}_{-15.2}$ & $11.54^{+1.09}_{-0.45}$ & $13.05^{+0.27}_{-0.39}$ & $10.74^{+0.72}_{-0.22}$ & $13.83^{+0.12}_{-0.11}$ &  \\
6 & $3.20861(^{+8}_{-4})$ & $19.8^{+8.9}_{-4.2}$ & $14.70^{+0.11}_{-0.06}$ & $13.02^{+0.14}_{-0.16}$ & $11.16^{+0.42}_{-0.53}$ & $12.62^{+0.53}_{-0.82}$ &  \\
7 & $3.209333(^{+13}_{-11})$ & $19.7^{+1.4}_{-1.3}$ & $15.43^{+0.03}_{-0.03}$ & $13.76^{+0.02}_{-0.03}$ & $12.95^{+0.05}_{-0.06}$ & $14.26^{+0.01}_{-0.05}$ & $13.11^{+0.03}_{-0.07}$ \\
8 & $3.210062(^{+29}_{-36})$ & $4.1^{+2.2}_{-1.0}$ & $14.38^{+0.09}_{-0.24}$ & $11.20^{+0.55}_{-0.20}$ & $11.53^{+0.29}_{-0.45}$ &  &  \\
9 & $3.21011(^{+24}_{-22})$ & $45.8^{+7.3}_{-11.3}$ & $14.15^{+0.18}_{-0.28}$ & $11.18^{+0.70}_{-0.16}$ & $12.19^{+0.28}_{-0.45}$ &  &  \\
$\log N_{\rm tot}$ &  &  & $15.66^{+0.01}_{-0.01}$ & $13.99^{+0.02}_{-0.01}$ & $13.15^{+0.05}_{-0.05}$ & $14.39^{+0.01}_{-0.02}$ & $13.11^{+0.03}_{-0.07}$ \\
$\rm [X/H]$ &  &  & $-1.44^{+0.02}_{-0.02}$ & $-1.25^{+0.02}_{-0.02}$ & $-1.01^{+0.05}_{-0.05}$ & $-1.43^{+0.02}_{-0.02}$ & $-1.44^{+0.03}_{-0.07}$ \\
$\rm [X/ZnII]$ &  &  & $-0.43^{+0.06}_{-0.05}$ & $-0.24^{+0.06}_{-0.05}$ & $0$ & $-0.42^{+0.06}_{-0.05}$ & $-0.43^{+0.06}_{-0.08}$ \\
\hline
\end{tabular}
\end{table*}

\subsubsection{J\,1353$+$0956}
\begin{figure}
\includegraphics [width=\columnwidth]{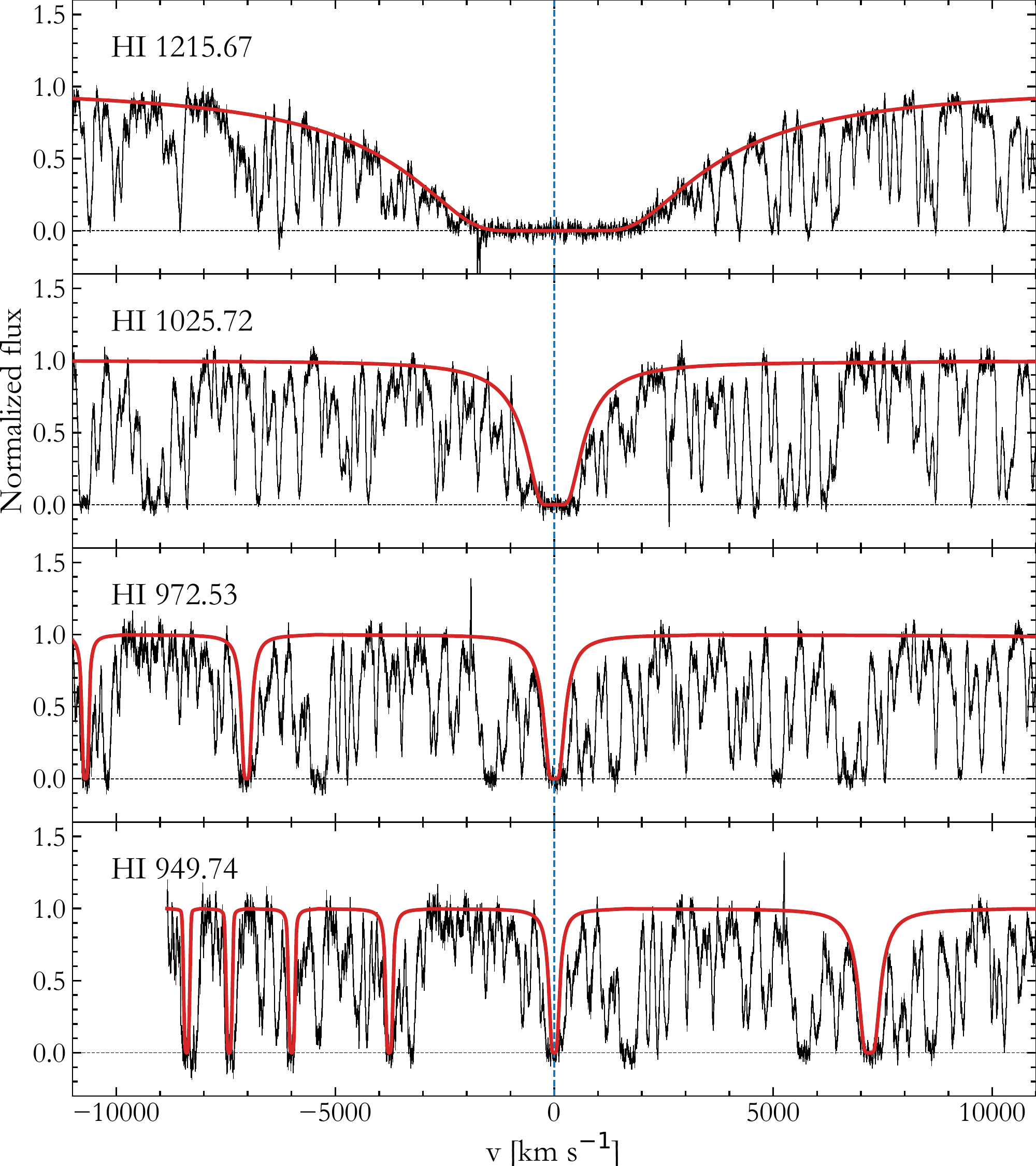}
\caption{Fit to \HI\ absorption lines at z$\sim$3.333 towards J\,1353+0956. The red line presents the profiles of the labeled \HI\ transitions. The vertical dashed line indicates the relative velocity of \HI\ absorption lines.}
\label{fig:J1353_HI}
\end{figure}

\begin{figure*}
\includegraphics [width=\textwidth]{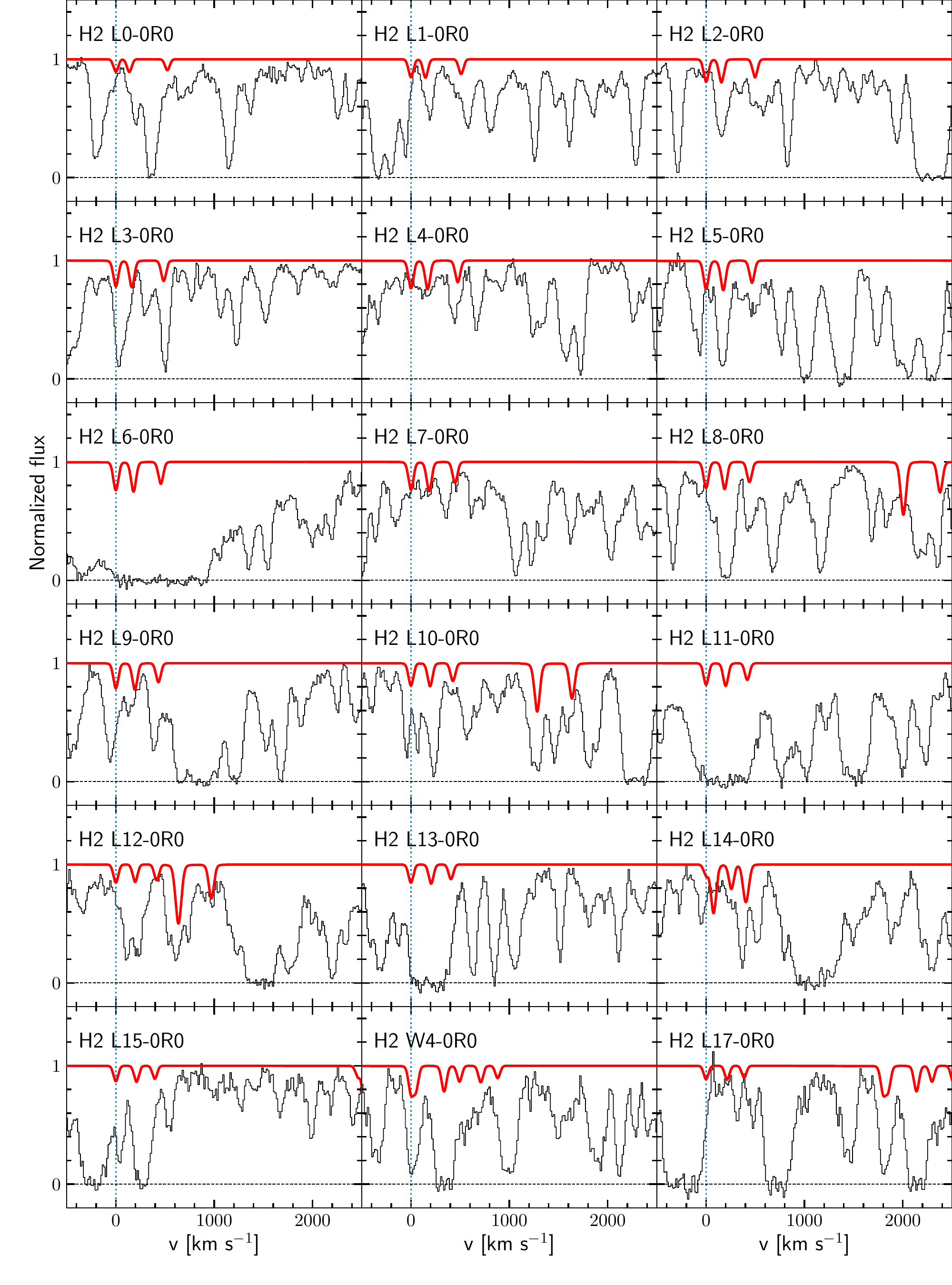}
\caption{The regions of J\,1353$+$0956 spectrum corresponding to the expected position of H$_2$ absorption lines associated with the ESDLA at z$\sim$3.333. Each panel corresponds to a particular band of H$_2$ transitions. The red line presents the profile of the H$_2$ absorption lines used to obtain an upper limit on the H$_2$ column density. The blue vertical lines indicate the positions of the R0 H$_2$ transition, which are in agreement with the positions of metal lines.}
\label{fig:J1353_H2}
\end{figure*}

\begin{figure*}
\includegraphics [width=\textwidth]{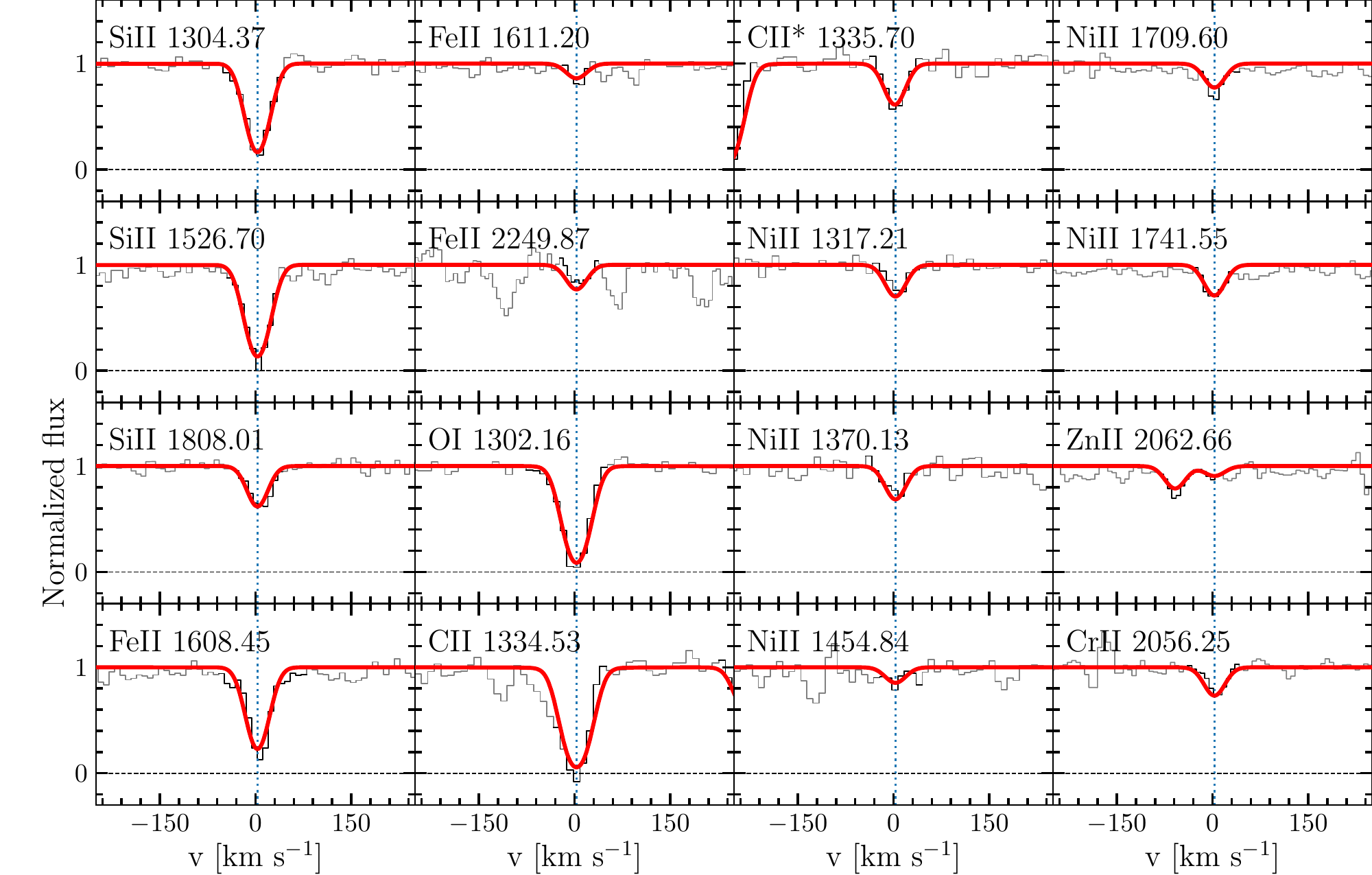}
\caption{Fit to metal absorption lines at z$\sim$3.3333 towards J\,1353+0956. The red line presents the total profile of the labeled metal transitions. The vertical dashed lines indicate the centers of the corresponding transitions.}
\label{fig:J1353_me}
\end{figure*}

The fitted \HI\ Ly-series absorptions are shown in Fig.~\ref{fig:J1353_HI}. The \HH\ Voigt profile, corresponded to the upper limit on \HH\ column density, is show in Fig.~\ref{fig:J1353_H2}.
The details of the fit to the metal lines are presented in Table~\ref{tab:fit_me_J1353}. The column densities on \OI\ and \CII\ are lower limits due the high saturation of the corresponding transitions. The best-fit metal line Voigt-profiles are shown in Fig.~\ref{fig:J1353_me}.

\begin{table*}
\caption{Fit results of metal lines at z$\sim$3.333 towards J\,1353$+$0956.}
\label{tab:fit_me_J1353}
\begin{tabular}{cccc}
\hline 
comp & 1 & $\rm [X/H]$ & $\rm [X/ZnII]$ \\
\hline
$z$ & $3.333463(^{+4}_{-6})$ &  &  \\
$b$, km~s$^{-1}$& $9.9^{+0.3}_{-0.3}$ &  &  \\
$\log N$(SiII) & $15.38^{+0.06}_{-0.06}$ & $-1.74^{+0.06}_{-0.06}$ & $-0.13^{+0.21}_{-0.14}$ \\
$\log N$(FeII) & $15.03^{+0.09}_{-0.08}$ & $-2.08^{+0.09}_{-0.08}$ & $-0.46^{+0.22}_{-0.15}$ \\
$\log N$(CrII) & $13.39^{+0.07}_{-0.04}$ & $-1.86^{+0.07}_{-0.05}$ & $-0.24^{+0.21}_{-0.14}$ \\
$\log N$(ZnII) & $12.56^{+0.13}_{-0.20}$ & $-1.61^{+0.13}_{-0.20}$ & $0$ \\
$\log N$(NiII) & $13.91^{+0.03}_{-0.04}$ & $-1.92^{+0.03}_{-0.04}$ & $-0.31^{+0.20}_{-0.14}$ \\
$\log N$(OI) & $16.50^{+0.20}_{-0.20}$ & $-1.80^{+0.20}_{-0.20}$ & $-0.19^{+0.28}_{-0.24}$ \\
$\log N$(CII*) & $13.74^{+0.09}_{-0.10}$ & $-4.30^{+0.09}_{-0.10}$ & $-2.68^{+0.22}_{-0.16}$ \\
$\log N$(CII) & $16.62^{+0.38}_{-0.24}$ & $-1.42^{+0.38}_{-0.24}$ & $0.20^{+0.43}_{-0.27}$ \\
\hline
\end{tabular}
\end{table*}

\subsubsection{J\,1418$+$0718}
\begin{figure}
\includegraphics [width=\columnwidth]{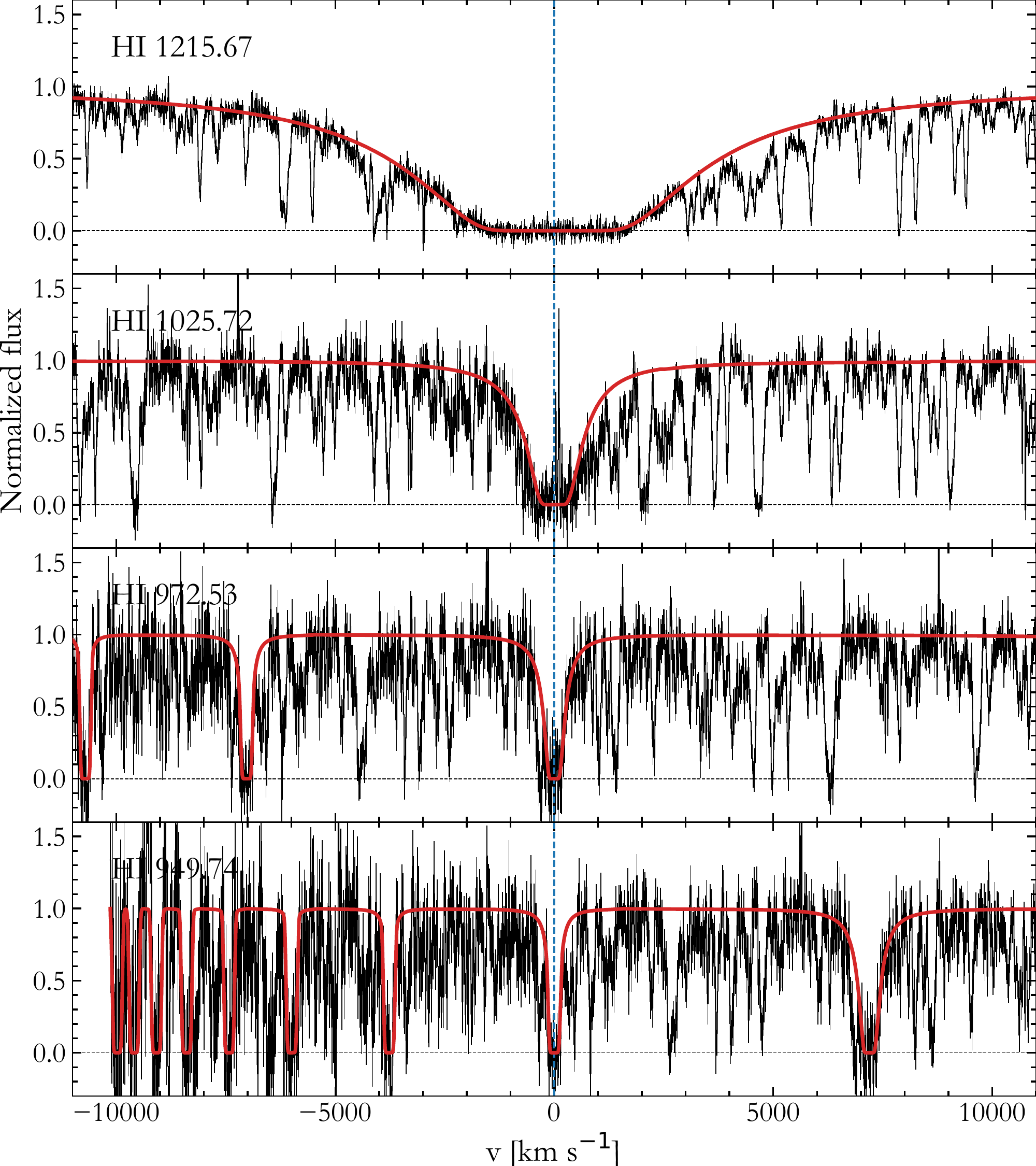}
\caption{Fit to \HI\ absorption lines at z$\sim$2.392 towards J\,1418+0718. The red line presents the profile of the labeled \HI\ transitions. The vertical dashed line indicates the center of \HI\ absorption lines.}
\label{fig:J1418_HI}
\end{figure}

\begin{figure*}
\includegraphics [width=\textwidth]{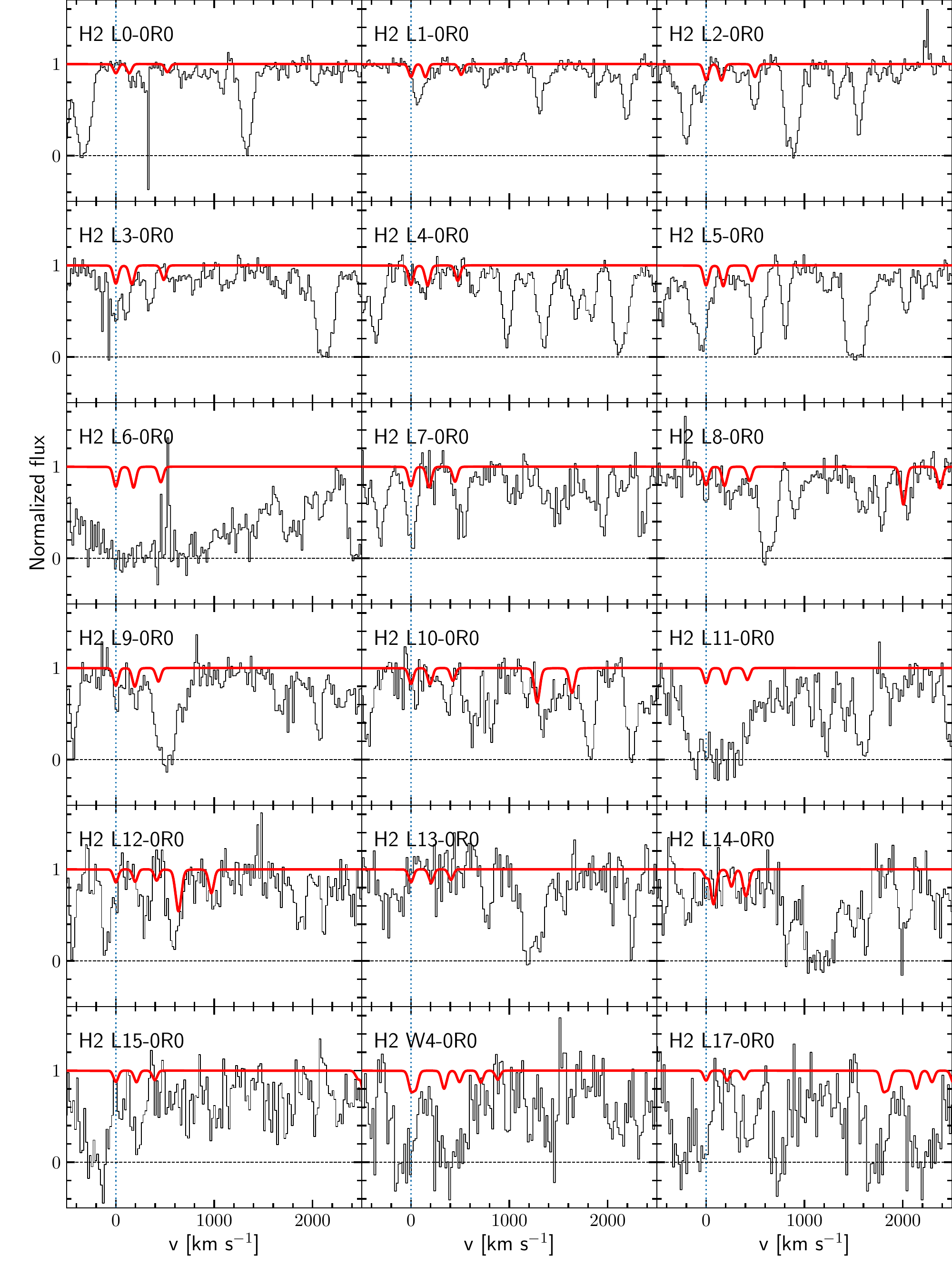}
\caption{The regions of J\,1418$+$0718 spectrum corresponding to the expected position of H$_2$ absorption lines associated with the ESDLA at z$\sim$2.392. Each panel corresponds to a particular band of H$_2$ transitions. The red line presents the profile of the H$_2$ absorption lines used to obtain an upper limit on the H$_2$ column density. The blue vertical lines indicate the positions of the R0 H$_2$ transition, which are in agreement with the positions of metal transitions.
}
\label{fig:J1418_H2}
\end{figure*}

\begin{figure*}
\includegraphics [width=\textwidth]{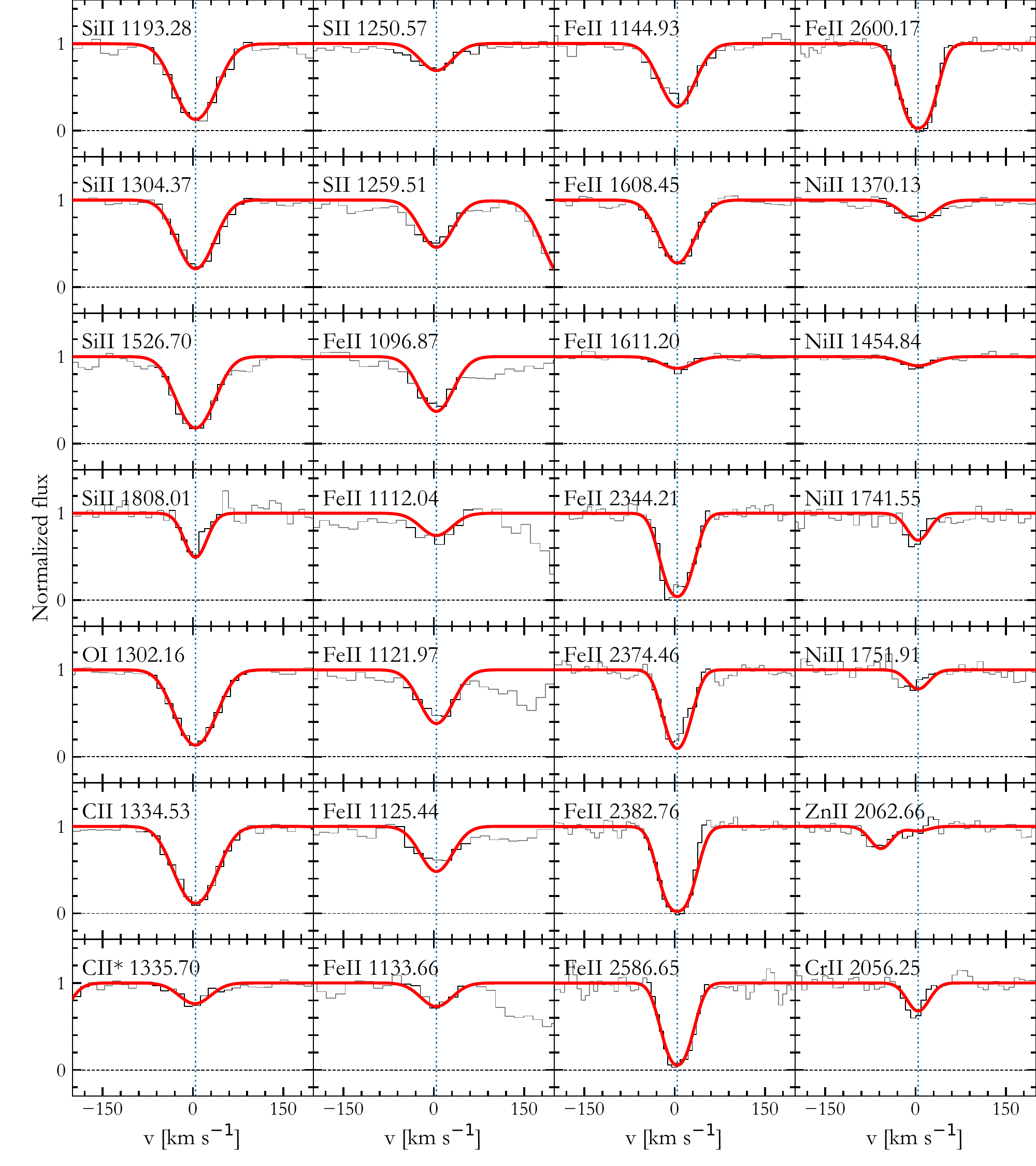}
\caption{Fit to metal absorption lines at z$\sim$2.392 towards J\,1418+0718. The red line presents the total profiles of the labeled metal transitions. The vertical dashed lines indicate the centers of the corresponding transitions.}
\label{fig:J1418_me}
\end{figure*}

The fitted \HI\ Ly-series absorptions are shown in Fig.~\ref{fig:J1418_HI}. The \HH\ Voigt profile, corresponded to the upper limit on \HH\ column density, is show in Fig.~\ref{fig:J1418_H2}. 
The details of the fit to the metal lines are presented in Table~\ref{tab:fit_me_J1418}. The column densities on \OI\ and \CII\ are lower limits due the high saturation of the corresponding transitions. The best-fit metal line Voigt-profiles are shown in Fig.~\ref{fig:J1418_me}.
 
\begin{table*}
\caption{Fit results of metal lines  at z$\sim$2.392 towards J\,1418$+$0718.}
\label{tab:fit_me_J1418}
\begin{tabular}{cccc}
\hline 
comp & 1 & $\rm [X/H]$ & $\rm [X/SII]$ \\
\hline
$z$ & $2.3919315(^{+31}_{-27})$ &  &  \\
$b$, km~s$^{-1}$ & $14.6^{+0.2}_{-0.3}$ &  &  \\
$\log N$(SiII) & $15.56^{+0.06}_{-0.05}$ & $-1.54^{+0.06}_{-0.06}$ & $-0.04^{+0.07}_{-0.07}$ \\
$\log N$(FeII) & $15.23^{+0.04}_{-0.04}$ & $-1.86^{+0.04}_{-0.05}$ & $-0.36^{+0.05}_{-0.06}$ \\
$\log N$(SII) & $15.21^{+0.05}_{-0.04}$ & $-1.50^{+0.05}_{-0.04}$ & $0$ \\
$\log N$(CII) & $16.39^{+0.18}_{-0.16}$ & $-1.63^{+0.18}_{-0.16}$ & $-0.13^{+0.19}_{-0.16}$ \\
$\log N$(NiII) & $13.95^{+0.04}_{-0.04}$ & $-1.86^{+0.04}_{-0.05}$ & $-0.36^{+0.05}_{-0.06}$ \\
$\log N$(OI) & $16.57^{+0.13}_{-0.09}$ & $-1.71^{+0.13}_{-0.10}$ & $-0.21^{+0.13}_{-0.11}$ \\
$\log N$(CII*) & $13.64^{+0.06}_{-0.07}$ & $-4.38^{+0.07}_{-0.07}$ & $-2.88^{+0.07}_{-0.08}$ \\
$\log N$(ZnII) & $12.25^{+0.33}_{-1.06}$ & $-1.90^{+0.33}_{-1.06}$ & $-0.40^{+0.33}_{-1.06}$ \\
$\log N$(CrII) & $13.52^{+0.05}_{-0.05}$ & $-1.71^{+0.05}_{-0.05}$ & $-0.21^{+0.06}_{-0.07}$ \\
\hline
\end{tabular}
\end{table*}

\subsubsection{J\,2205$+$102}
\begin{figure}
\includegraphics [width=\columnwidth]{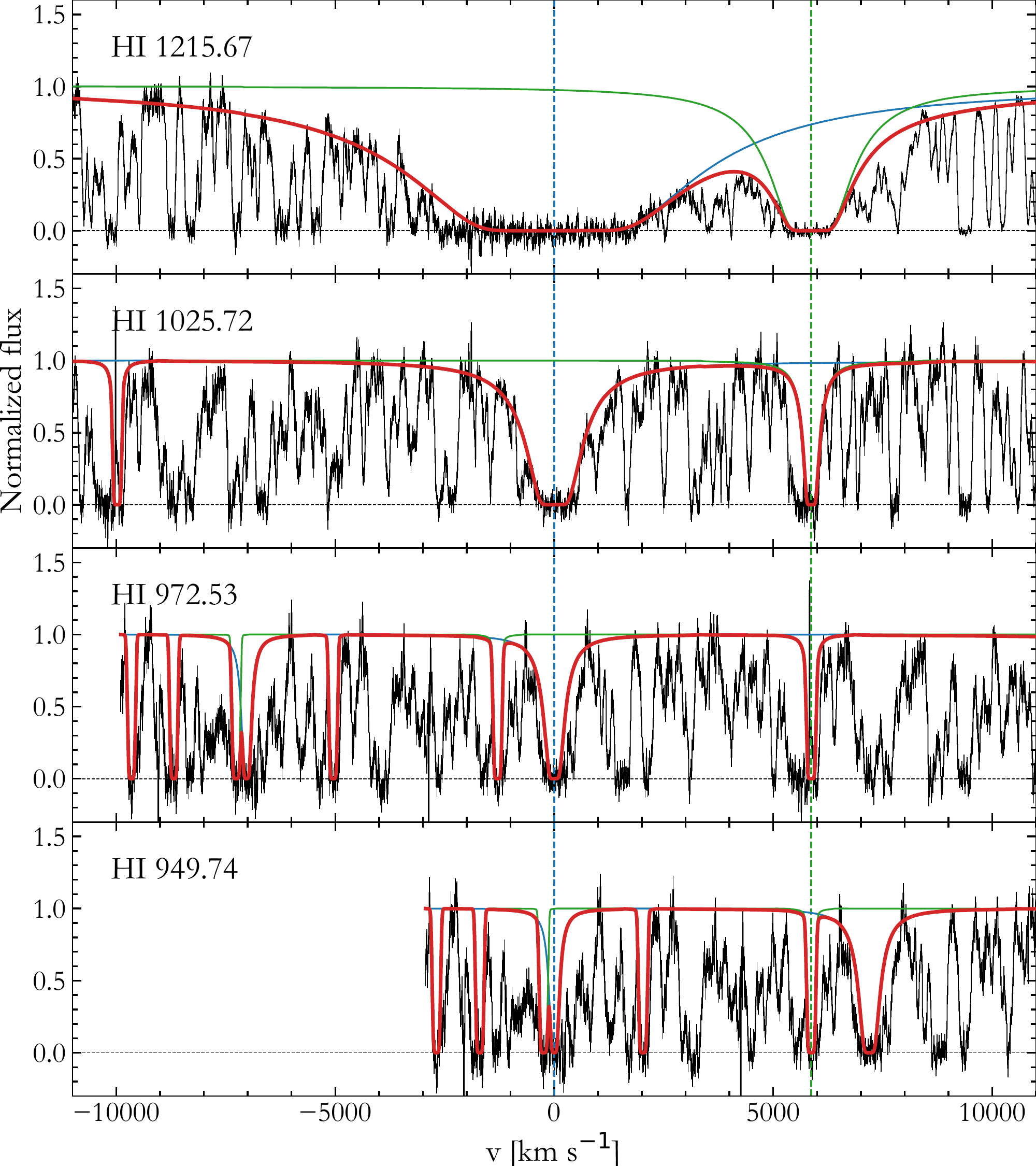}
\caption{Fit to \HI\ absorption lines at z$\sim$3.255 towards J\,2205+1021. The red line presents the total profile of the spectrum. The blue line indicates the ESDLA studied in this work, while the green line corresponds to adjacent and unrelated \HI\ systems. The vertical dashed lines indicate the relative positions of the two absorption systems.}
\label{fig:J2205_HI}
\end{figure}

\begin{figure*}
\includegraphics [width=\textwidth]{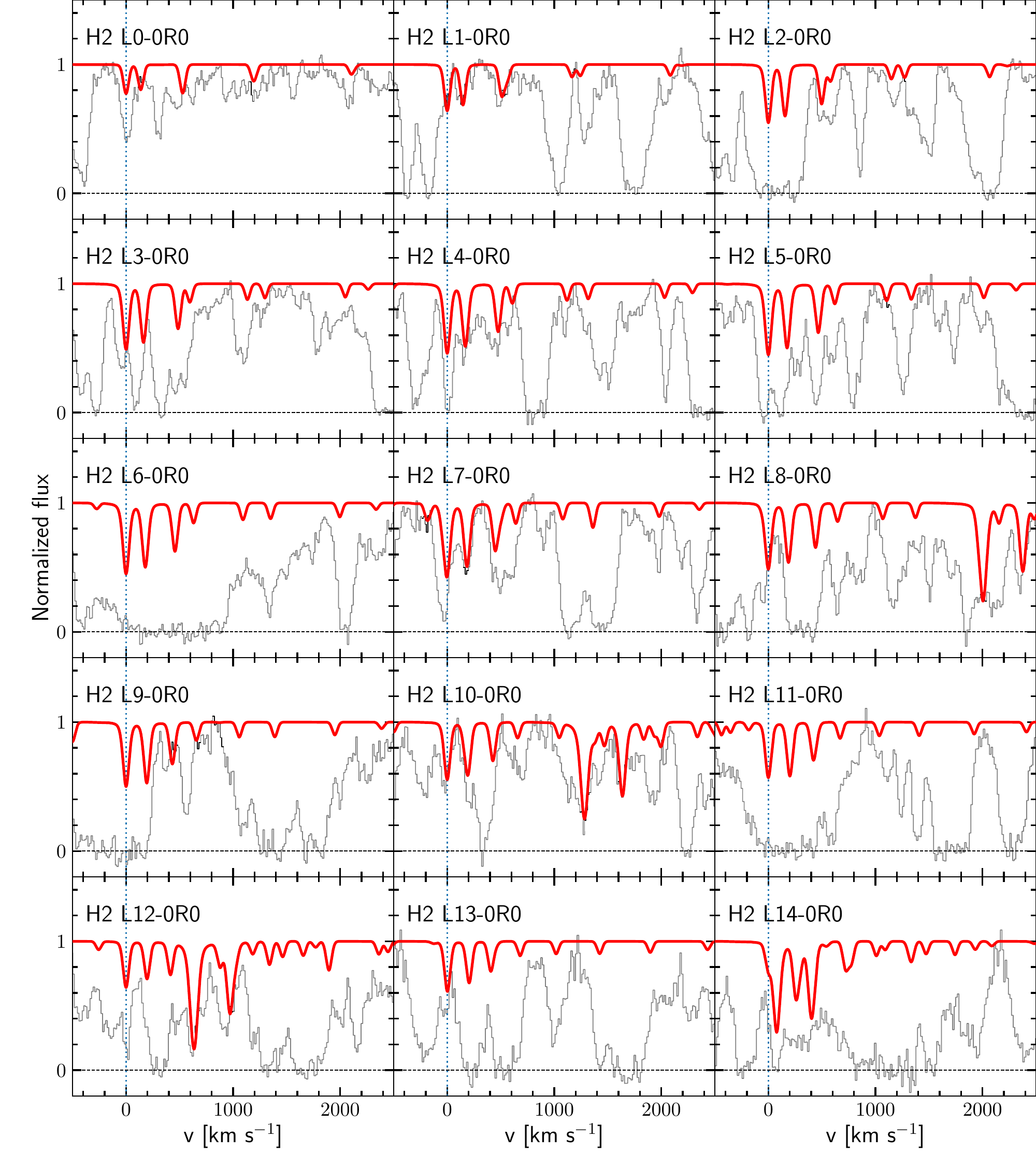}
\caption{The regions of J\,2205$+$1021 spectrum corresponding to the position of H$_2$ absorption lines associated with ESDLA at z$\sim$3.255. Each panel corresponds to a particular band of H$_2$ transitions. The red line presents the profile of the H$_2$ absorption lines plotted with respect to the best-fit value of \HH\ column density. The blue vertical lines indicate the positions of the R0 H$_2$ transition.}
\label{fig:J2205_H2_fit}
\end{figure*}

\begin{figure*}
\includegraphics [width=\textwidth]{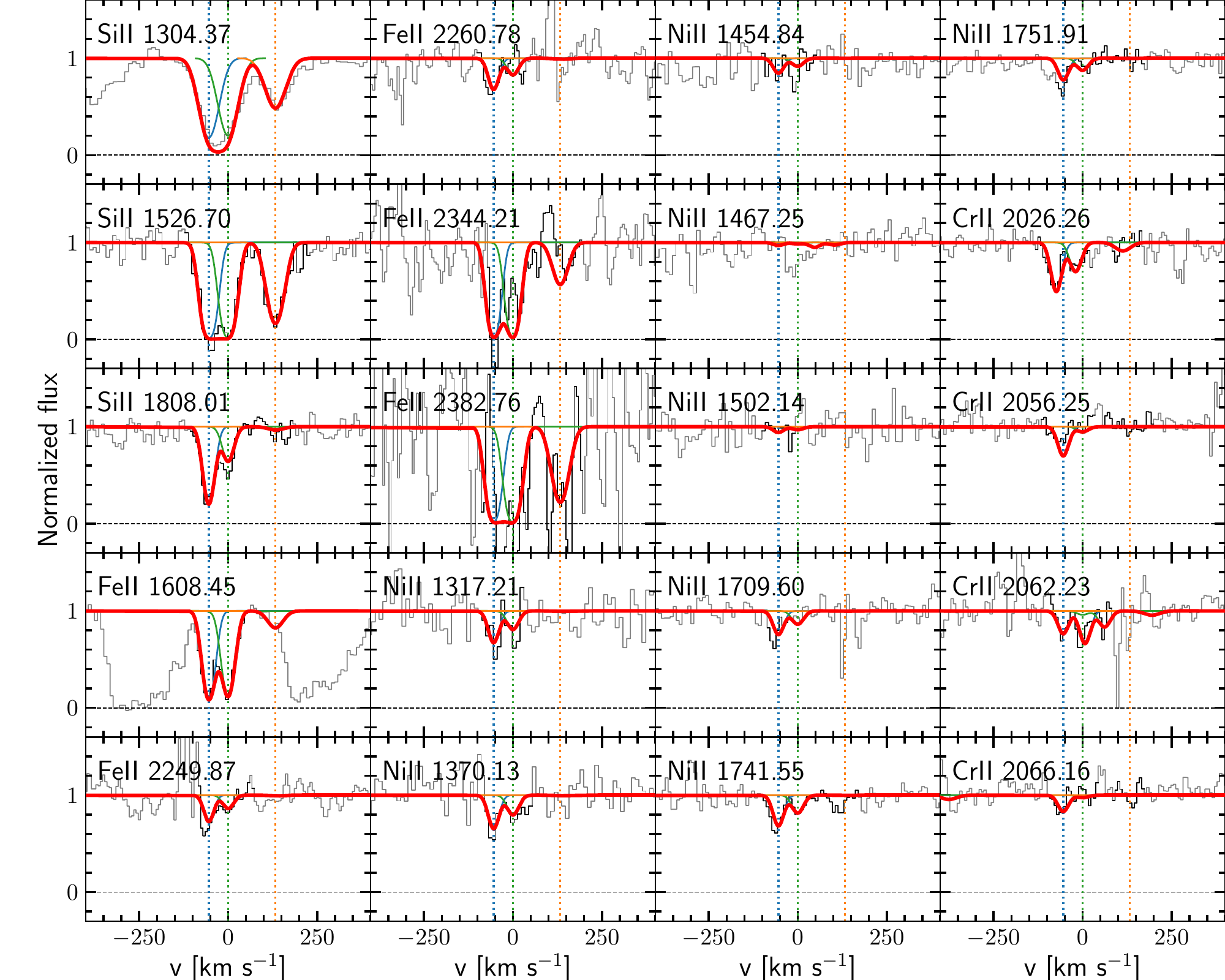}
\caption{Fit to metal absorption lines at z$\sim$3.255 towards J\,2205+1021. The red line presents the total profiles of the labeled metal transitions. The vertical dashed lines indicate the relative positions of individual components of the fit.}
\label{fig:J2205_me_sb}
\end{figure*}

The fitted \HI\ Ly-series absorptions are shown in Fig.~\ref{fig:J2205_HI}. The best-fit profile of \HH\ is shown in Fig.~\ref{fig:J2205_H2_fit}. The total column density has been determined from the $J=0-4$ rotational levels.
The details of the fit to the metal lines are presented in Table~\ref{tab:fit_me_J2205}. The best-fit metal line Voigt-profiles are shown in Fig.~\ref{fig:J2205_me_sb}.
For this target, the individual VIS exposures have different spectral resolutions. We have therefore additionally varied the resolution of these exposures during the metal line fitting.

\begin{table*}
\caption{Fit results of metal lines at z$\sim$3.255 towards J\,2205$+$1021.} 
\label{tab:fit_me_J2205}
\begin{tabular}{ccccccc}
\hline 
comp & 1 & 2 & 3 & $\log N_{\rm tot}$ & $\rm [X/H]$ & $\rm [X/ZnII]$ \\
\hline
$z$ & $3.2547756(^{+54}_{-3})$ & $3.255557(^{+17}_{-14})$ & $3.257465(^{+39}_{-30})$ &  &  &  \\
$b$, km~s$^{-1}$ & $10.6^{+0.8}_{-0.6}$ & $14.8^{+1.0}_{-1.2}$ & $33.3^{+5.3}_{-4.9}$ &  &  &  \\
$\log N$(SiII) & $16.20^{+0.18}_{-0.12}$ & $15.37^{+0.04}_{-0.08}$ & $14.21^{+0.05}_{-0.05}$ & $16.26^{+0.16}_{-0.10}$ & $-0.86^{+0.16}_{-0.10}$ & $0.07^{+0.16}_{-0.11}$ \\
$\log N$(NiII) & $13.98^{+0.04}_{-0.05}$ & $13.53^{+0.10}_{-0.10}$ & $13.21^{+0.28}_{-0.68}$ & $14.14^{+0.05}_{-0.05}$ & $-1.69^{+0.05}_{-0.05}$ & $-0.76^{+0.07}_{-0.07}$ \\
$\log N$(FeII) & $15.16^{+0.09}_{-0.07}$ & $14.70^{+0.09}_{-0.07}$ & $13.45^{+0.15}_{-0.12}$ & $15.30^{+0.07}_{-0.05}$ & $-1.81^{+0.07}_{-0.06}$ & $-0.87^{+0.09}_{-0.07}$ \\
$\log N$(ZnII) & $13.06^{+0.06}_{-0.06}$ & $12.73^{+0.06}_{-0.09}$ & $12.14^{+0.25}_{-1.15}$ & $13.24^{+0.05}_{-0.05}$ & $-0.93^{+0.05}_{-0.05}$ & $0$ \\
$\log N$(CrII) & $13.46^{+0.05}_{-0.06}$ & $12.83^{+0.24}_{-0.39}$ & $12.65^{+0.32}_{-1.35}$ & $13.56^{+0.07}_{-0.07}$ & $-1.69^{+0.07}_{-0.07}$ & $-0.76^{+0.08}_{-0.08}$ \\
\hline
\end{tabular}
\end{table*}

\subsubsection{J\,2351$-$0639}
\begin{figure}
\includegraphics [width=\columnwidth]{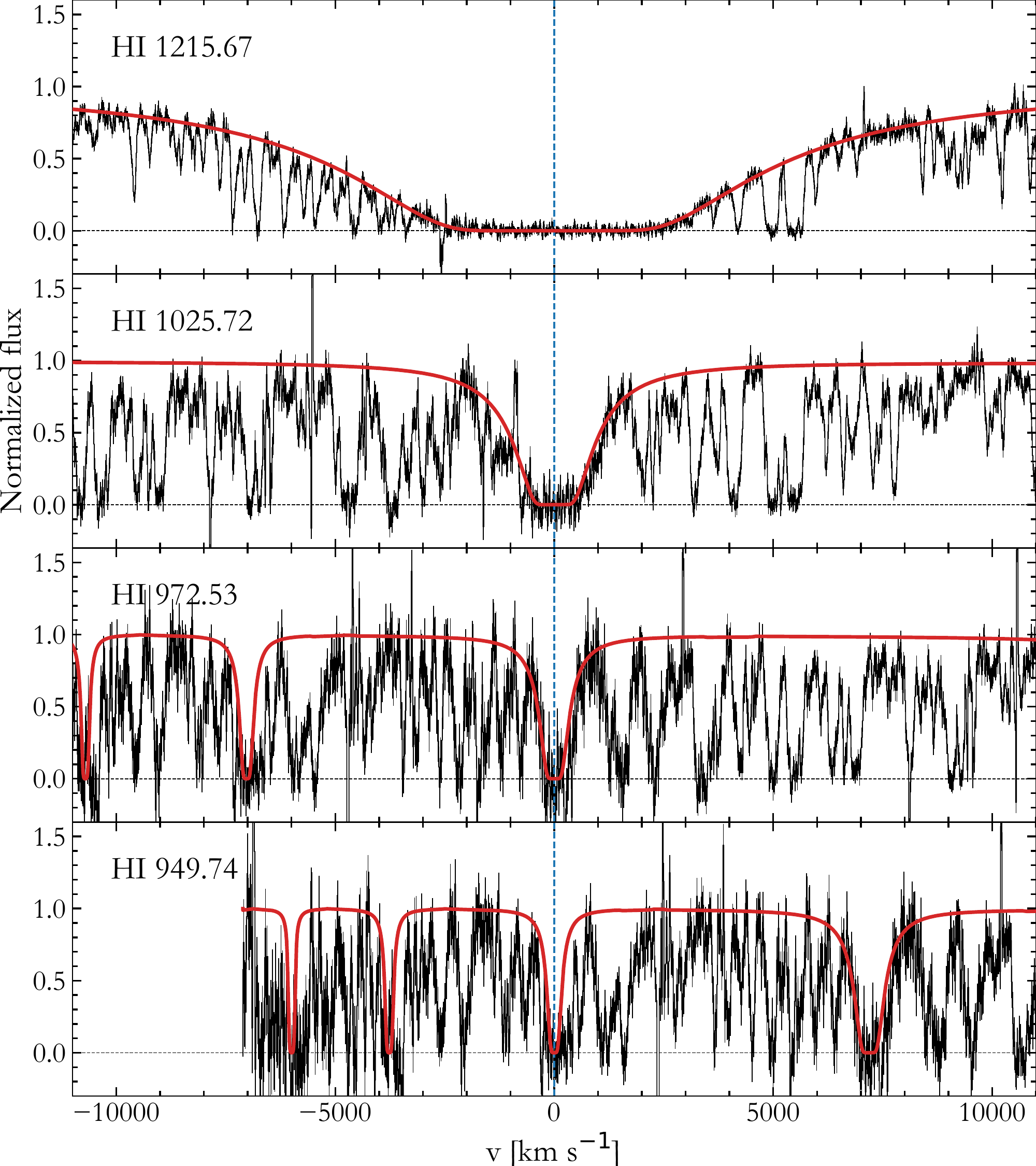}
\caption{Fit to \HI\ absorption lines at z$\sim$2.557 towards J\,2351-0639. The red line presents the profile of the labeled \HI\ transitions. The vertical dashed line indicates the center of the \HI\ absorption lines.}
\label{fig:J2351_HI}
\end{figure}

\begin{figure*}
\includegraphics [width=\textwidth]{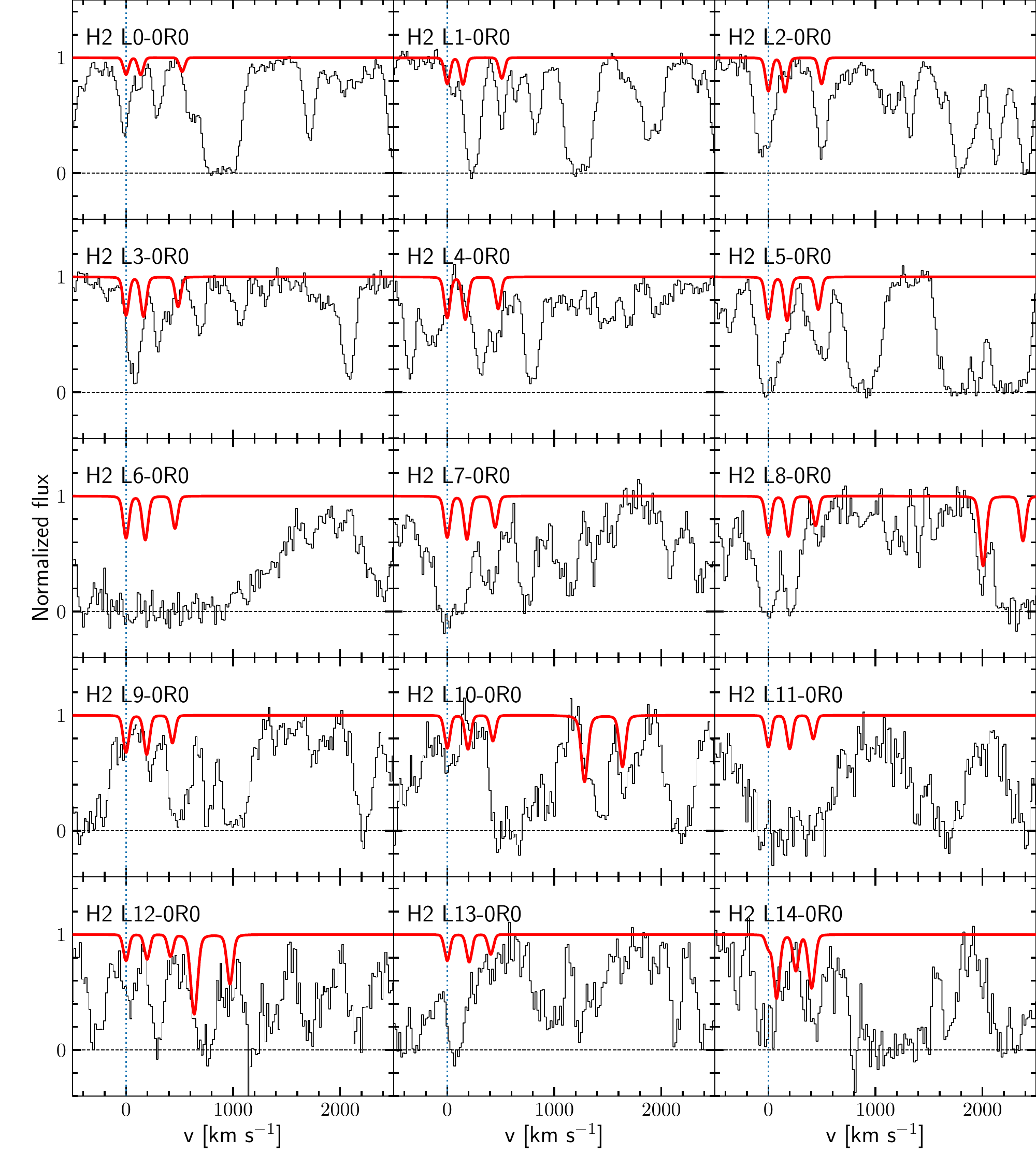}
\caption{The regions of J\,2351$-$0639 spectrum corresponding to the expected position of H$_2$ absorption lines associated with the ESDLA at $z=2.557$. Each panel corresponds to a particular band of H$_2$ transitions. The red line presents the profile of the H$_2$ absorption lines used to obtain an upper limit on the H$_2$ column density. The blue vertical lines indicate the positions of the R0 H$_2$ transition, which are in agreement with the positions of metal transitions.}
\label{fig:J2351_H2}
\end{figure*}

\begin{figure*}

\includegraphics [width=\textwidth]{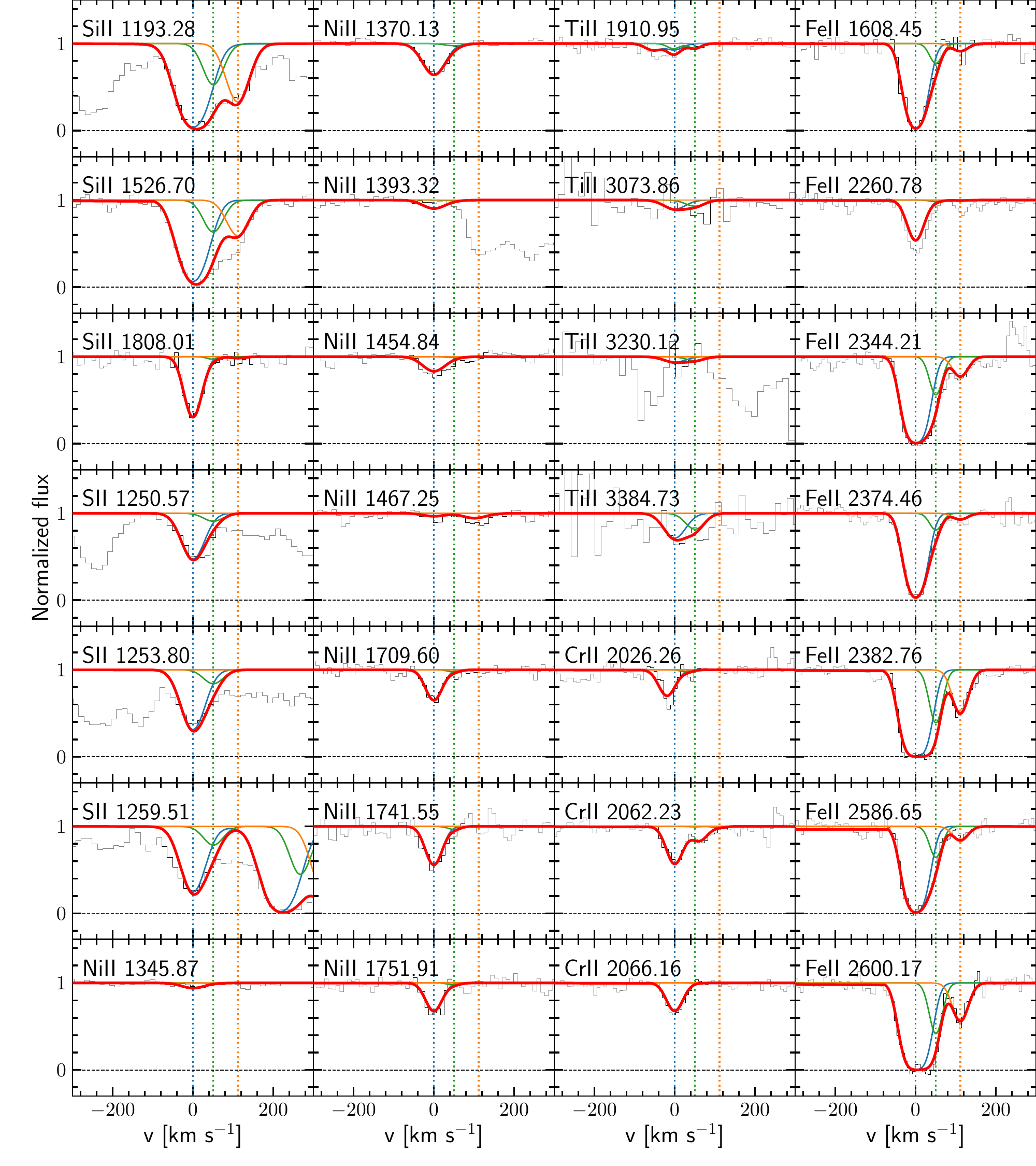}
\caption{Fit to metal absorption lines at $z\approx$2.557 towards J\,2351-0639. The red line presents the total profile of the labeled metal transitions. The vertical dashed lines indicate the relative positions of individual components of the fit. }
\label{fig:J2351_me}
\end{figure*}

The fitted \HI\ Ly-series absorptions are shown in Fig.~\ref{fig:J2351_HI}. The \HH\ Voigt profile, corresponded to the upper limit on \HH\ column density, is show in Fig.~\ref{fig:J2351_H2}.
The details of the fit to the metal lines are presented in Table~\ref{tab:fit_me_J2351}. The best-fit metal line Voigt-profiles are shown in Fig.~\ref{fig:J2351_me}.

\begin{table*}
\caption{Fit results of metal lines at z$\sim$2.557 towards J\,2351$-$0639.}
\label{tab:fit_me_J2351}
\begin{tabular}{ccccccc}
\hline 
comp & 1 & 2 & 3 & $\log N_{\rm tot}$ & $\rm [X/H]$ & $\rm [X/ZnII]$ \\
\hline
$z$ & $2.557846(^{+7}_{-9})$ & $2.558457(^{+25}_{-25})$ & $2.559162(^{+15}_{-17})$ &  &  &  \\
$b$, km~s$^{-1}$ & $20.1^{+1.0}_{-0.6}$ & $6.9^{+2.3}_{-1.1}$ & $15.5^{+4.3}_{-2.3}$ &  &  &  \\
$\log N$(SiII) & $15.80^{+0.06}_{-0.06}$ & $13.78^{+0.61}_{-0.49}$ & $13.81^{+0.36}_{-0.12}$ & $15.82^{+0.06}_{-0.05}$ & $-1.59^{+0.06}_{-0.05}$ & $-0.01^{+0.08}_{-0.07}$ \\
$\log N$(FeII) & $15.35^{+0.10}_{-0.11}$ & $13.61^{+0.09}_{-0.10}$ & $13.25^{+0.03}_{-0.03}$ & $15.36^{+0.10}_{-0.10}$ & $-2.04^{+0.10}_{-0.10}$ & $-0.46^{+0.11}_{-0.11}$ \\
$\log N$(NiII) & $14.19^{+0.02}_{-0.02}$ & $12.93^{+0.27}_{-0.66}$ &  & $14.21^{+0.02}_{-0.02}$ & $-1.91^{+0.02}_{-0.03}$ & $-0.33^{+0.05}_{-0.06}$ \\
$\log N$(CrII) & $13.85^{+0.02}_{-0.02}$ & $12.04^{+0.37}_{-0.65}$ &  & $13.85^{+0.03}_{-0.02}$ & $-1.69^{+0.03}_{-0.02}$ & $-0.11^{+0.06}_{-0.06}$ \\
$\log N$(ZnII) & $12.84^{+0.05}_{-0.05}$ & $11.85^{+0.22}_{-0.47}$ &  & $12.88^{+0.05}_{-0.05}$ & $-1.58^{+0.05}_{-0.05}$ & $0$ \\
$\log N$(SII) & $15.59^{+0.04}_{-0.06}$ & $14.56^{+0.21}_{-0.21}$ &  & $15.62^{+0.04}_{-0.04}$ & $-1.40^{+0.05}_{-0.04}$ & $0.18^{+0.07}_{-0.06}$ \\
$\log N$(TiII) & $12.81^{+0.17}_{-0.23}$ & $12.80^{+0.15}_{-0.24}$ &  & $13.06^{+0.10}_{-0.11}$ & $-1.79^{+0.10}_{-0.11}$ & $-0.20^{+0.12}_{-0.12}$ \\
\hline
\end{tabular}
\end{table*}

\subsubsection{J\,2359$+$1354}

\begin{figure}
\includegraphics [width=\columnwidth]{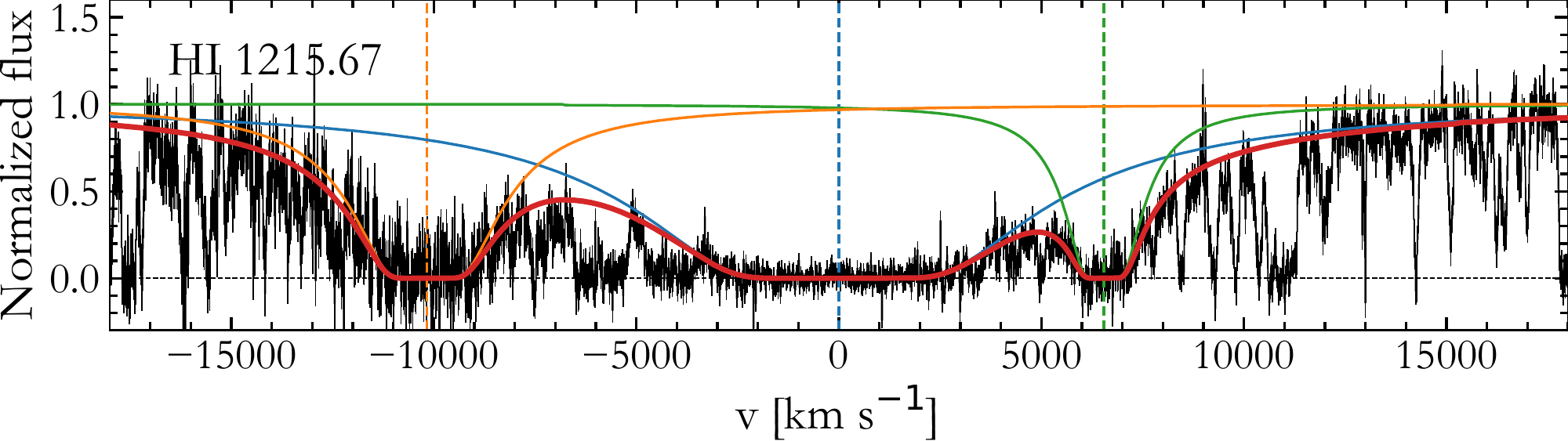}
\caption{Fit to \HI\ absorption lines at $z\approx$2.250 towards J\,2359$+$1354. The red line presents the total fitted profile. The blue line indicates the ESDLA targeted in this work, while the green and orange lines correspond to adjacent, unrelated \HI\ systems. The vertical dashed lines indicate the relative positions of the two absorption systems.}
\label{fig:J2359_HI}
\end{figure}

\begin{figure*}
\includegraphics [width=\textwidth]{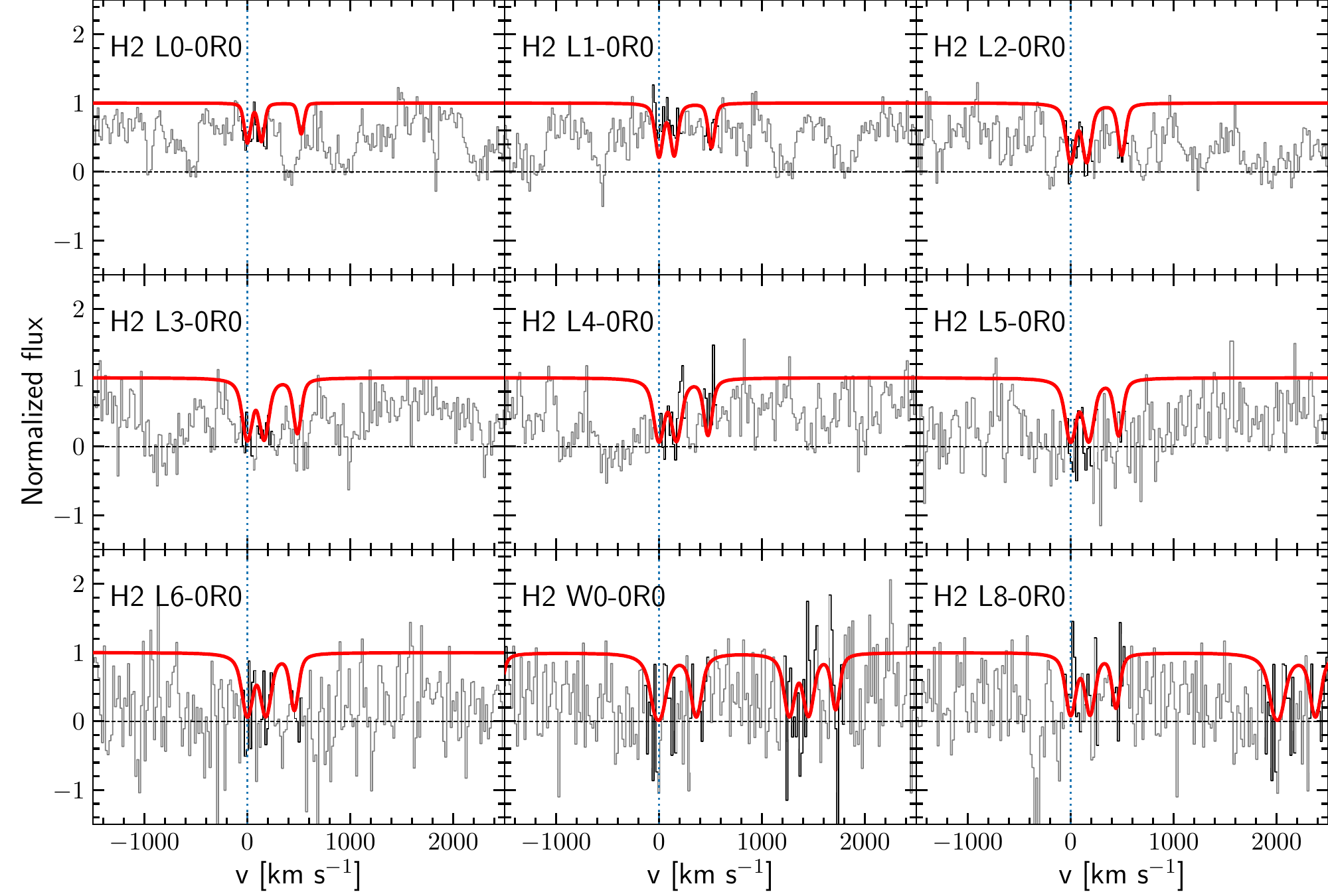}
\caption{The regions of J\,2359$+$1354 spectrum corresponding to the position of H$_2$ absorption lines associated with the ESDLA at $z\approx$2.249. Each panel corresponds to a particular band of H$_2$ transitions. The red line presents the profile of the H$_2$ absorption lines plotted with respect to the best-fit value of \HH\ column density. The blue vertical lines indicate the positions of the R0 H$_2$ transition.}
\label{fig:J2359_H2_fit}
\end{figure*}

\begin{figure*}
\includegraphics [width=\textwidth]{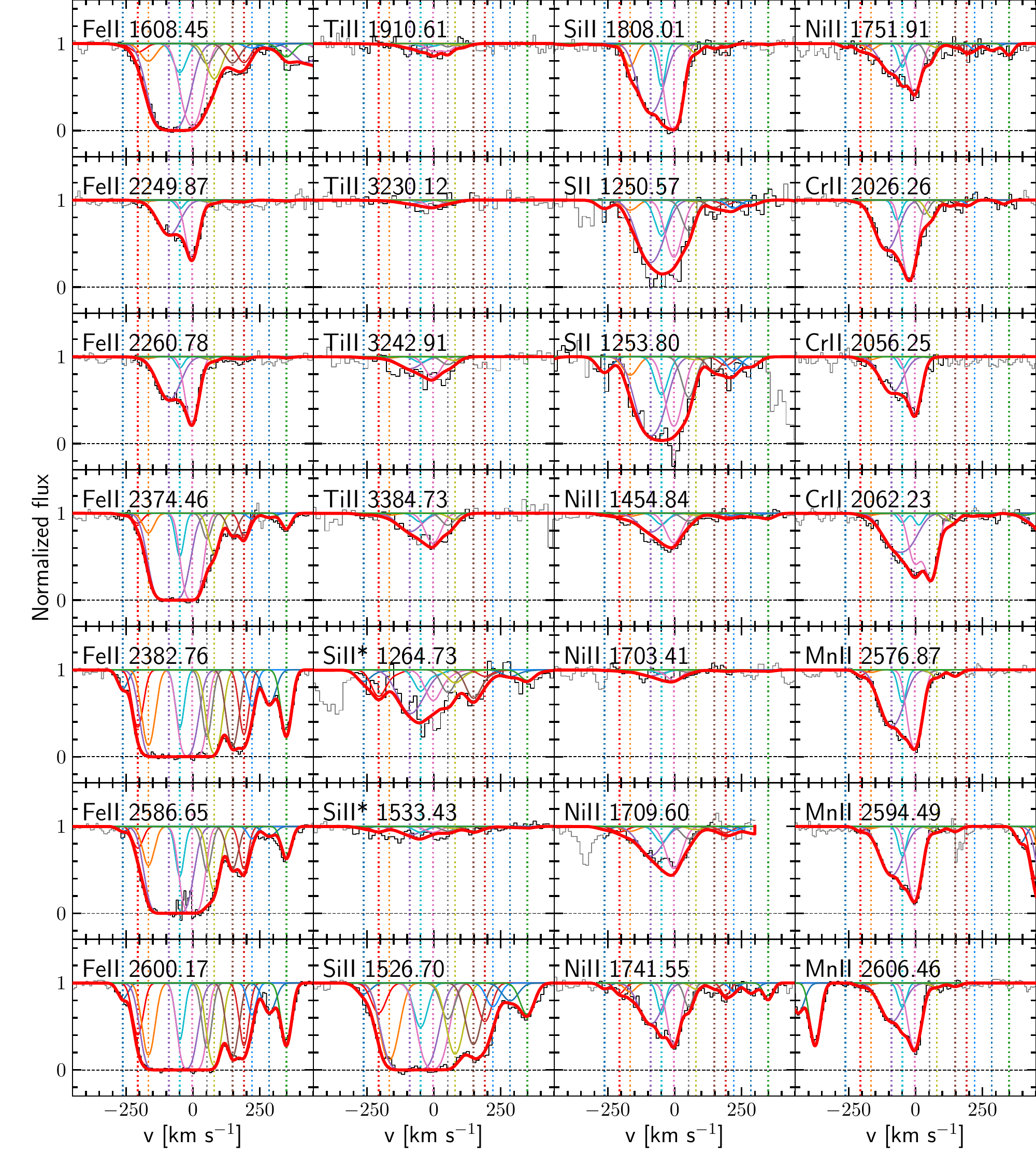}
\caption{Fit to metal absorption lines at $z\approx$2.250 towards J\,2359+1354. The red line presents the total profile of the labeled metal transitions. The vertical dashed lines indicate the relative positions of individual components of the fit.}
\label{fig:J2359_me}
\end{figure*}

The fitted \HI\ Ly-series absorptions are shown in Fig.~\ref{fig:J2359_HI}. \HI\ Ly$\alpha$ absorption from the ESDLA system at $z=2.2499$ falls on top of the quasar Ly$\beta$ emission line. Although, this contamination does not significantly change the total \HI\ column density, we took it into account during the fitting of the quasar continuum. \HH\ total Voigt profile are shown in Fig.~\ref{fig:J2359_H2_fit}. The total column density of \HH\ has been measured from the $J=0-2$ rotational levels. 
The details of the fit to the metal lines are presented in Table~\ref{tab:fit_me_J2359}. The best-fit metal line Voigt-profiles are shown in Fig.~\ref{fig:J2359_me}.

\newgeometry{margin=0.5cm}
\begin{landscape}
\begin{table}
  \caption{Fit results of metal lines at z$\sim$2.250 towards J\,2359$+$1354.}\label{tab:fit_me_J2359}
\begin{tabular}{ccccccccccccccc}
\hline 
comp & $z$ & $b$, km~s$^{-1}$ & $\log N$(SiII) & $\log N$(FeII) & $\log N$(NiII) & $\log N$(SII) & $\log N$(TiII) & $\log N$(SiII*) & $\log N$(MnII) & $\log N$(ZnII) & $\log N$(CrII) \\
\hline
1 & $2.24577(^{+3}_{-5})$ & $16.4^{+1.0}_{-1.7}$ & $13.36^{+0.14}_{-0.10}$ & $12.78^{+0.06}_{-0.05}$ & $13.25^{+0.13}_{-0.23}$ & $14.82^{+0.12}_{-0.21}$ & $12.00^{+0.07}_{-0.29}$ & $12.51^{+0.11}_{-0.13}$ &  &  &  \\
2 & $2.24638(^{+3}_{-3})$ & $19.2^{+2.3}_{-3.3}$ & $13.77^{+0.17}_{-0.16}$ & $13.56^{+0.07}_{-0.10}$ & $13.52^{+0.10}_{-0.12}$ & $11.67^{+0.14}_{-0.15}$ & $11.45^{+0.14}_{-0.18}$ & $12.80^{+0.14}_{-0.11}$ & $12.23^{+0.08}_{-0.12}$ & $11.45^{+0.19}_{-0.13}$ & $12.48^{+0.10}_{-0.31}$ \\
3 & $2.24682(^{+8}_{-8})$ & $23.6^{+1.7}_{-4.6}$ & $15.17^{+0.14}_{-0.14}$ & $13.97^{+0.15}_{-0.17}$ & $13.31^{+0.15}_{-0.14}$ & $14.69^{+0.13}_{-0.21}$ & $11.75^{+0.16}_{-0.23}$ & $10.40^{+0.22}_{-0.12}$ & $11.82^{+0.10}_{-0.20}$ & $11.45^{+0.10}_{-0.21}$ & $12.46^{+0.16}_{-0.22}$ \\
4 & $2.247598(^{+40}_{-28})$ & $51.1^{+2.5}_{-2.4}$ & $16.18^{+0.05}_{-0.05}$ & $15.68^{+0.02}_{-0.04}$ & $14.53^{+0.04}_{-0.05}$ & $16.01^{+0.05}_{-0.17}$ & $12.93^{+0.07}_{-0.07}$ & $13.27^{+0.05}_{-0.12}$ & $13.60^{+0.04}_{-0.02}$ & $13.52^{+0.04}_{-0.05}$ & $13.98^{+0.04}_{-0.04}$ \\
5 & $2.247981(^{+41}_{-24})$ & $7.3^{+2.1}_{-1.2}$ & $15.81^{+0.17}_{-0.15}$ & $14.72^{+0.11}_{-0.18}$ & $13.83^{+0.18}_{-0.05}$ & $16.11^{+0.20}_{-0.16}$ & $12.44^{+0.14}_{-0.15}$ & $12.84^{+0.14}_{-0.12}$ & $13.11^{+0.10}_{-0.07}$ & $12.48^{+0.16}_{-0.14}$ & $13.00^{+0.18}_{-0.13}$ \\
6 & $2.248548(^{+20}_{-25})$ & $24.6^{+0.8}_{-1.6}$ & $16.66^{+0.10}_{-0.08}$ & $15.73^{+0.03}_{-0.02}$ & $14.61^{+0.04}_{-0.04}$ & $15.80^{+0.21}_{-0.08}$ & $13.00^{+0.05}_{-0.08}$ & $13.19^{+0.09}_{-0.12}$ & $13.82^{+0.02}_{-0.03}$ & $13.81^{+0.05}_{-0.04}$ & $14.07^{+0.04}_{-0.05}$ \\
7 & $2.24912(^{+6}_{-3})$ & $11.5^{+1.2}_{-3.2}$ & $14.14^{+0.11}_{-0.19}$ & $14.06^{+0.13}_{-0.13}$ & $13.70^{+0.09}_{-0.14}$ & $15.20^{+0.13}_{-0.17}$ & $12.59^{+0.08}_{-0.15}$ & $12.62^{+0.15}_{-0.17}$ & $11.67^{+0.18}_{-0.19}$ & $12.39^{+0.19}_{-0.05}$ & $11.90^{+0.21}_{-0.14}$ \\
8 & $2.24944(^{+3}_{-4})$ & $20.0^{+3.2}_{-1.5}$ & $14.51^{+0.14}_{-0.13}$ & $14.16^{+0.08}_{-0.09}$ & $13.41^{+0.18}_{-0.11}$ & $13.52^{+0.18}_{-0.17}$ & $12.34^{+0.07}_{-0.21}$ & $12.48^{+0.20}_{-0.09}$ & $12.13^{+0.11}_{-0.24}$ & $12.59^{+0.06}_{-0.14}$ & $11.42^{+0.14}_{-0.21}$ \\
9 & $2.25015(^{+3}_{-4})$ & $17.5^{+2.3}_{-2.7}$ & $14.47^{+0.11}_{-0.08}$ & $13.88^{+0.04}_{-0.06}$ & $13.23^{+0.22}_{-0.08}$ & $14.76^{+0.11}_{-0.16}$ &  & $12.85^{+0.13}_{-0.11}$ & $12.09^{+0.10}_{-0.14}$ & $12.23^{+0.13}_{-0.15}$ &  \\
10 & $2.25057(^{+4}_{-3})$ & $10.5^{+1.8}_{-1.4}$ & $14.38^{+0.06}_{-0.29}$ & $13.89^{+0.08}_{-0.07}$ & $13.40^{+0.18}_{-0.12}$ & $14.36^{+0.18}_{-0.17}$ &  & $12.06^{+0.08}_{-0.23}$ &  & $11.78^{+0.19}_{-0.14}$ &  \\
11 & $2.250911(^{+41}_{-28})$ & $9.4^{+2.9}_{-1.1}$ & $13.89^{+0.19}_{-0.07}$ & $13.35^{+0.10}_{-0.11}$ & $13.41^{+0.15}_{-0.10}$ & $14.59^{+0.10}_{-0.16}$ &  & $10.30^{+0.17}_{-0.11}$ &  & $11.84^{+0.15}_{-0.11}$ &  \\
12 & $2.25167(^{+3}_{-4})$ & $28.4^{+3.0}_{-1.5}$ & $13.54^{+0.08}_{-0.08}$ & $13.17^{+0.04}_{-0.04}$ & $13.56^{+0.12}_{-0.15}$ & $14.40^{+0.21}_{-0.10}$ &  & $12.21^{+0.21}_{-0.11}$ &  &  &  \\
13 & $2.252353(^{+13}_{-24})$ & $14.8^{+1.8}_{-1.6}$ & $13.90^{+0.09}_{-0.05}$ & $13.65^{+0.04}_{-0.05}$ & $13.68^{+0.06}_{-0.17}$ &  &  & $12.35^{+0.14}_{-0.19}$ &  & $12.17^{+0.13}_{-0.12}$ &  \\
$\log N_{\rm tot}$ &  &  & $16.83^{+0.09}_{-0.04}$ & $16.05^{+0.01}_{-0.01}$ & $15.05^{+0.02}_{-0.02}$ & $16.52^{+0.09}_{-0.06}$ & $13.46^{+0.02}_{-0.03}$ & $13.84^{+0.04}_{-0.03}$ & $14.10^{+0.02}_{-0.01}$ & $14.05^{+0.02}_{-0.03}$ & $14.36^{+0.02}_{-0.03}$ \\
$\rm [X/H]$ &  &  & $-0.64^{+0.10}_{-0.04}$ & $-1.41^{+0.02}_{-0.03}$ & $-1.13^{+0.03}_{-0.03}$ & $-0.56^{+0.10}_{-0.06}$ & $-1.45^{+0.03}_{-0.04}$ & $-3.63^{+0.04}_{-0.04}$ & $-1.29^{+0.03}_{-0.02}$ & $-0.47^{+0.03}_{-0.03}$ & $-1.24^{+0.03}_{-0.03}$ \\
$\rm [X/ZnII]$ &  &  & $-0.17^{+0.10}_{-0.04}$ & $-0.94^{+0.03}_{-0.03}$ & $-0.66^{+0.03}_{-0.03}$ & $-0.09^{+0.10}_{-0.06}$ & $-0.99^{+0.03}_{-0.04}$ & $-3.17^{+0.05}_{-0.04}$ & $-0.83^{+0.03}_{-0.03}$ & $0.00^{+0.04}_{-0.04}$ & $-0.77^{+0.03}_{-0.04}$ \\
\hline
\end{tabular}
\end{table}
\end{landscape}
\restoregeometry

\label{lastpage}
\end{document}